\documentclass[letterpaper, 12pt]{article}

\usepackage{algorithm}
\usepackage{algorithmic}
\usepackage{amsfonts}
\usepackage{amsmath}
\usepackage{amssymb}
\usepackage{array}
\usepackage{epsfig}
\usepackage{geometry}
\usepackage{graphicx}
\usepackage[colorlinks=true, citecolor=blue, filecolor=black, linkcolor=red, urlcolor=blue]{hyperref}
\usepackage{lgrind}
\usepackage{multirow}
\usepackage{slashbox}
\usepackage{subfigure}
\usepackage[textsize=tiny]{todonotes}

\renewcommand{\[}{\left[}
\renewcommand{\]}{\right]}
\renewcommand{\(}{\left(}
\renewcommand{\)}{\right)}

\newcommand{\norm}[2]{\left\|\, #1 \,\right\|_{#2}}

\newcommand{\vvvert}{|\kern-1pt|\kern-1pt|}

\newcommand{\hJ}{\widehat{J}}

\newcommand{\tJ}{\widetilde{J}}
\newcommand{\tQ}{\tilde{Q}}

\newcommand{\EE}{\mathbb{E}}

\newcommand{\CD}{\mathcal{D}}

\newcommand{\CF}{\mathcal{F}}
\newcommand{\CH}{\mathcal{H}}

\newcommand{\CN}{\mathcal{N}}

\newcommand{\CX}{\mathcal{X}}
\newcommand{\CY}{\mathcal{Y}}

\newcommand{\argmax}{\operatornamewithlimits{arg\,max}}

\newcommand{\DKL}{D_{\mathrm{KL}}}

\title{Sequential Bayesian optimal experimental design \\ via
  approximate dynamic programming}

\author{Xun Huan$^\ast$ and Youssef
  M.~Marzouk\footnote{Massachusetts Institute of
    Technology, Cambridge, MA 02139 USA;
    \href{mailto:xunhuan@mit.edu,ymarz@mit.edu}{\{xunhuan,ymarz\}@mit.edu}; 
    \newline\url{http://uqgroup.mit.edu}.}}

\begin{document}

\maketitle

\begin{abstract}
  The design of multiple experiments is commonly undertaken via
  suboptimal strategies, such as batch (open-loop) design that omits
  feedback or greedy (myopic) design that does not account for future
  effects. This paper introduces new strategies for the
  \textit{optimal} design of sequential experiments.  First, we
  rigorously formulate the general sequential optimal experimental
  design (sOED) problem as a dynamic program. Batch and greedy designs
  are shown to result from special cases of this formulation.
  We then focus on sOED for parameter inference, adopting a Bayesian
  formulation with an information theoretic design objective. To make
  the problem tractable, we develop new numerical approaches for
  nonlinear design with continuous parameter, design, and observation
  spaces. We approximate the optimal policy by using backward
  induction with regression to construct and refine value function
  approximations in the dynamic program. The proposed algorithm
  iteratively generates trajectories via exploration and exploitation
  to improve approximation accuracy in frequently visited regions of
  the state space. Numerical results are verified against analytical
  solutions in a linear-Gaussian setting. Advantages over batch and
  greedy design are then demonstrated on a nonlinear source inversion
  problem where we seek an optimal policy for sequential sensing.
\end{abstract}

%% \begin{keywords} sequential experimental design, Bayesian experimental
%%   design, approximate dynamic programming, feedback control policy,
%%   lookahead, % cost-to-go, 
%%   approximate value iteration, information gain\end{keywords}

%\begin{AMS}62K05, 62L05, 62C10, 62F15, 49L20, 90C39\end{AMS}

%% \pagestyle{myheadings} \thispagestyle{plain} \markboth{HUAN AND
%%   MARZOUK}{SEQUENTIAL BAYESIAN OPTIMAL EXPERIMENTAL DESIGN}

\section{Introduction}
\label{ch:introduction}

Experiments are essential to learning about the physical world.
%% As George
%% E. P. Box points out, ``\ldots science is a means whereby learning is
%% achieved, not by mere theoretical speculation on the one hand, nor by
%% the undirected accumulation of practical facts on the other, but
%% rather by a motivated \textit{iteration} between theory and practice
%% \ldots''~\cite{Box1976}.  As illustrated in
%% Figure~\ref{f:learning_process}, theory is used to deduce what is
%% expected to be observed in practice, and observations from experiments
%% are used in turn to induce how theory may be further improved. In
%% science and engineering, experiments are a fundamental building block
%% of the scientific method, and crucial in the continuing development
%% and refinement of models of physical systems.
%% \begin{figure}[htb]
%%   \centering
%%   \includegraphics[width=0.6\textwidth]{figures/general/learning_process.pdf}
%%   \caption{The learning process can be characterized as an iteration
%%     between theory and practice via deductive and inductive
%%     reasoning.}
%%   \label{f:learning_process}
%% \end{figure}
Whether obtained through field observations or controlled laboratory
experiments, however, experimental data may be time-consuming or
expensive to acquire. Also, experiments are not equally useful: some
can provide valuable information while others may prove irrelevant to the
goals of an investigation. It is thus important to navigate the tradeoff
between experimental costs and benefits, and to maximize the ultimate value of
experimental data---i.e., to design experiments that are \emph{optimal} by some
appropriate measure. 
%Doing so can accelerate the advancement of scientific understanding.  
Experimental design thus addresses questions such as
where and when to take measurements, which variables to probe, and
what experimental conditions to employ.
%%  In this
%% paper, we develop a systematic framework for experimental design that
%% can help answer these questions.

%\subsection{Literature review}

% History of experimental design in literature. Existing approaches,
% literature review, shortcoming.

The systematic design of experiments has received much attention in
the statistics community and in many science and engineering
applications. Basic design approaches include factorial, composite,
and Latin hypercube designs, based on notions of space filling and
blocking~\cite{Fisher1966, Box1987, Cox2000, Box2005}. While these
methods can produce useful designs in relatively simple situations
involving a few design variables, they generally do not take into
account---or exploit---knowledge of the underlying physical
process. Model-based experimental design uses the relationship between
observables, parameters, and design variables to guide the choice of
experiments, and \textit{optimal} experimental design (OED) further
incorporates specific and relevant metrics to design experiments for a
particular purpose, such as parameter inference, prediction, or model
discrimination \cite{Fedorov1972, Atkinson1992,Chaloner1995}.

The design of \textit{multiple} experiments can be pursued via two
broad classes of approaches:
\begin{itemize}
\item \textit{Batch} or \textit{open-loop} design involves the design of all
  experiments concurrently, such that the outcome of any
  experiment cannot affect the design of the others.
  % In some situations, this
  % approach may be necessary, such as under certain scheduling
  % constraints.
  
\item \textit{Sequential} or \textit{closed-loop} design allows
  experiments to be chosen and conducted in sequence, thus permitting
  newly acquired data to guide the design of future experiments. In
  other words, sequential design involves \textit{feedback}.
\end{itemize}
Batch OED for linear models is well established (see,
e.g.,~\cite{Fedorov1972, Atkinson1992}), and recent years have seen
many advances in OED methodology for nonlinear models and large-scale
applications~\cite{Huan2014, Huan2013, Alexanderian2014, Bisetti2016,
  Haber2010, Long2013, Loredo2010, Terejanu2012, Weaver2015}. In the
context of Bayesian design with nonlinear models and non-Gaussian
posteriors, rigorous information-theoretic criteria have been proposed
\cite{Lindley1956, Ginebra2007}; these criteria lead to design
strategies that maximize the expected information gain due to the
experiments, or equivalently, maximize the \emph{mutual information}
between the experimental observables and the quantities of interest
\cite{Ryan2003, Huan2013, Krause2008a}.

In contrast, sequential optimal experimental design (sOED) has seen
much less development and use.
% The value of feedback
% through sequential design has been recognized early on, with original
% approaches typically involving a \textit{heuristic} partitioning of
% experiments into batches. For instance, in the context of experimental
% design for improving chemical plant filtration rate~\cite{Box1992}, an
% initial ``empirical feedback'' stage involving space-filling designs
% is administered to ``pick the winner'' and find designs that best fix
% the problem, and a subsequent ``scientific feedback'' stage with
% adapted designs is followed to better understand the reasons for what
% went wrong or why a solution worked.
Many approaches for sequential design rely directly on batch OED, simply by
repeating it in a greedy manner for each next experiment; this
strategy is known as \textit{greedy} or \textit{myopic} design.
% Some work made use of linear design theory by iteratively
% alternating between parameter estimation and applications of linear
% optimality (e.g.,~\cite{Agarwal1985a}).
Since many physically realistic models involve output quantities that
depend nonlinearly on model parameters, these models yield
non-Gaussian posteriors in a Bayesian setting. The key challenge for greedy design
is then to represent and propagate these posteriors beyond the first
experiment.
Various inference methodologies and representations have been employed
within the greedy design framework, with a large body of research
based on sample representations of the posterior. For example,
posterior importance sampling has been used to evaluate variance-based
design utilities~\cite{Solonen2012} and in greedy augmentations of
generalized linear models~\cite{Dror2008}. Sequential Monte Carlo
methods have also been used in experimental design for parameter
inference~\cite{Drovandi2013} and for model
discrimination~\cite{Cavagnaro2010, Drovandi2014}. Even grid-based
discretizations/representations of posterior probability density
functions have shown success in adaptive design using hierarchical
models~\cite{Kim2014}.  While these developments provide a convenient
and intuitive avenue for extending existing batch OED tools,
greedy design is ultimately \textit{suboptimal}.  An optimal
sequential design framework must account for all relevant
\textit{future effects} in making each design decision.
% , but such considerations are dampened by challenges in
% computational feasibility. With recent advances in numerical
% algorithms and computing power, sOED can now be made practical.

% Dror and Steinberg~\cite{Dror2008} proposed a greedy
% augmentation of designs for generalized linear models by using samples
% to represent posteriors. Kim~\etal~\cite{Kim2014} employ kernel
% density estimates (KDEs) in an adaptive design optimization framework
% and applied them to hierarchical models of visual psychophysics
% applications.  Drovandi~\etal~\cite{Drovandi2013} and Cavagnaro
% \etal~\cite{Cavagnaro2010} both address sequential design using
% sequential Monte Carlo techniques, with the former focusing on
% parameter inference and the latter for model discrimination.  Solonen
% \etal~\cite{Solonen2012} propose a framework using a variance-based
% utility and importance sampling for inference. 

sOED is essentially a problem of sequential decision-making under
uncertainty, and thus it can rigorously be cast in a dynamic
programming (DP) framework.
%%%%%%%%%%%%%%
While DP approaches are widely used in control
theory~\cite{Bertsekas1996, Bertsekas2005, Bertsekas2007}, operations
research~\cite{Puterman1994, Powell2011}, and machine
learning~\cite{Kaelbling1996, Sutton1998}, their application to sOED
raises several distinctive challenges. In the Bayesian sOED context,
the state of the dynamic program must incorporate the current
posterior distribution or ``belief state.'' In many physical
applications, this distribution is continuous, non-Gaussian, and
multi-dimensional. The design variables and observations are typically
continuous and multi-dimensional as well. These features of the DP
problem lead to enormous computational demands.
%
%% YMM: I thought the sentence below does not fit so well into this paper,
%% since our model is *not* computationally intensive, and it's not
%% expressed as the result of a numerical simulation. (Also, see
%% earlier comment on 'simulation.'
%
% ... and the use of simulation-based and often computationally intensive physical models. 
%
% These models often involve continuous input and output spaces of
% many dimensions, which makes them very challenging and expensive to
% characterize numerically.
Thus, while the DP description of sOED has received some attention in
recent years~\cite{Muller2007, VonToussaint2011}, implementations and
applications of this framework remain limited. 

Existing attempts have focused mostly on optimal stopping
problems~\cite{Berger1985}, motivated by the design of clinical
trials.  For example, direct backward induction with tabular storage
has been used in~\cite{Brockwell2003, Wathen2006}, but is only
practical for discrete variables that can take on a few possible
outcomes. More sophisticated numerical techniques have been used for
sOED problems with other special structure.  For instance,
\cite{Carlin1998} proposes a forward sampling method that directly
optimizes a Monte Carlo estimate of the objective,
%% \todo{This is the first time we mention the notion of a `utility,'
%% and thus it may not be clear how it fits in. Can we just say
%% `objective?'}
%% XH: Sure.
but targets monotonic loss functions and certain conjugate priors that
result in threshold policies based on the posterior mean.
Computationally feasible implementations of backward induction have
also been demonstrated in situations where policies depend only on
low-dimensional sufficient statistics, such as the posterior mean and
standard deviation~\cite{Berry2002, Christen2003}.
%
% Continued development on backward induction also find feasible
% numerical implementations owing to policies that depend only on
% lower-dimensional sufficient statistics such as the posterior mean and
% standard deviation~\cite{Berry2002, Christen2003}. \todo{This sentence
%   is hard to understand; what does ``continued development on backward
%   induction'' mean? Do you mean something like: ``Numerical
%   implementations of backward induction can be made feasible with
%   policies that depend only on low-dimensional sufficient statistics,
%   such as the posterior mean and standard deviation~\cite{Berry2002,
%     Christen2003}.''}
% % make BI work on larger problems by taking advantage of situations
% % where the policies depends on 
%
% Other approaches directly model experimental outcomes using standard
% distributions and introduce additional approximations; for instance,
% \cite{Murphy2003} solves a dynamic treatment problem using Gaussian
% models and Q-factors approximated by regret functions of quadratic
% form.
%
Other DP approaches introduce alternative approximations: for
instance, \cite{Murphy2003} solves a dynamic treatment problem over a
countable decision space % using Gaussian statistical models,
using $Q$-factors approximated by regret functions of quadratic form.
%
%\todo{Maybe the sharper distinction is between drawing the likelihood
%from a standard family of distributions (e.g., a Gaussian) and not?}
%
% XH: Let me revisit this point in the next days.
% YMM: I am fine with it as is. Probably not worth sweating over further?
%
% Berry~\etal~\cite{Berry2002} has successful used DP by taking
% advantage of the decision being dependent only on the posterior mean
% and standard deviation.  Similarly, Christen and
% Nakamura~\cite{Christen2003} also applied backward induction by
% replying on the existence of similar sufficient
% statistics. 
% Murphy~\cite{Murphy2003} solved a dynamic treatment
% problem using only statistical models with distributions assigned to
% be Gaussian, and Q-factors approximated by regret functions of
% quadratic form.  
%
Furthermore, most of these efforts employ relatively simple design
objectives. Maximizing information gain leads to design
objectives that are much more challenging to compute, and thus has
been pursued for sOED only in simple situations. For instance,
\cite{Ben-Gal2002} finds near-optimal stopping policies in
multidimensional design spaces by exploiting submodularity \cite{
  Krause2008a,Fujishige2005} of the expected incremental information
gain. However, this is possible only for linear-Gaussian problems,
where mutual information does not depend on the realized values of the
observations.

Overall, most current efforts in sOED focus on problems with specialized
structure and consider settings that are partially or completely
discrete (i.e., with experimental outcomes, design variables, or
parameters of interest taking only a few values).
This paper will develop a mathematical and computational framework for
a much broader class of sOED problems. We will do so by developing
refinable numerical approximations of the solution to the exact
optimal sequential design problem.
%
% With the current state-of-the-art in sOED relying on special problem
% structures and often feasible only for discrete variables that can
% take on a few values, we seek to contribute to its development with a
% more general framework and numerical tools that can accommodate
% broader classes of problems, by taking an approach that concentrates
% on numerical techniques to \textit{approximately solve the exact
%   problem}.
%% % What we want to do to improve.
%% Current research in OED has seen rapid advances in the design of batch
%% experiments. Progress towards the \textit{optimal} design of
%% sequential experiments, however, remains in relatively early
%% stages. Direct applications of batch OED methods to sequential
%% settings are suboptimal, and initial explorations of the optimal
%% framework have been limited to problems with discrete spaces of very
%% few states and with special problem and solution structures. We aim to
%% extend the optimal sequential design framework to much more general
%% settings.
In particular, we will:
\begin{itemize}
\item Develop a rigorous formulation of the sOED problem for
  finite numbers of experiments, accommodating nonlinear models (i.e.,
  nonlinear parameter-observable relationships); continuous
  parameter, design, and observation spaces; a Bayesian treatment of
  uncertainty encompassing non-Gaussian distributions; and design
  objectives that quantify information gain.
%This includes
%  formulating the \textit{DP form} of the sOED problem that is central
%  to the subsequent development of numerical methods.

\item Develop numerical methodologies for solving such sOED problems
  in a computationally tractable manner, using approximate dynamic
  programming (ADP) techniques to find principled approximations of
  the optimal policy.
\end{itemize}
We will demonstrate our approaches first on a linear-Gaussian problem
where an exact solution to the optimal design problem is available,
and then on a contaminant source inversion problem involving a
nonlinear model of advection and diffusion. 
% XH: While our model is indeed nonlinear and describes convection and
% diffusion phenomena, would people expect to see the
% convection-diffusion PDE when they see this sentence, and feel
% misled? Should we say something along ``... involving a nonlinear
% model of convection and diffusion''?
% YMM: done.
In the latter examples, we will
explicitly contrast the sOED approach with batch and greedy design
methods.

This paper focuses on the formulation of the optimal design problem
and on the associated ADP methodologies. The sequential design setting
also requires repeated applications of Bayesian inference, using data
realized from their prior predictive distributions. A companion paper
will describe efficient strategies for performing the latter; our
approach will use transport map representations
\cite{Villani2008, ElMoselhy2012, Parno2015a,Marzouk2016} of the prior and
posterior distributions, constructed in a way that allows for fast
Bayesian inference tailored to the optimal design problem. A full
exploration of such methods is deferred to that paper. To keep the
present focus on DP issues, here we will simply discretize the prior
and posterior density functions on a grid and perform Bayesian
inference via direct evaluations of the posterior density, coupled with
a grid adaptation procedure.

The remainder of this paper is organized as follows.
Section~\ref{ch:formulation} formulates the sOED problem as a dynamic
program, and then shows how batch and greedy design strategies result
from simplifications of this general formulation.
Section~\ref{ch:approximate_dynamic_programming} describes ADP
techniques for solving the sOED problem in dynamic programming form.
%% , including the development of an adaptive strategy to refine
%% the policy induced state space. 
% Section~\ref{ch:transport_maps} describes the use of transport map as
% belief state, along with the framework for using joint maps to enable
% fast and approximate Bayesian inference.
%% The full algorithm is then applied to
%% several numerical examples in Section~\ref{ch:numerical_results}. 
%% We first illustrate the solution on a simple linear-Gaussian problem to
%% provide intuitive insights and establish comparisons with analytic
%% references. 
Section~\ref{ch:numerical_results} provides numerical demonstrations
of our methodology, and Section~\ref{ch:conclusions} includes
concluding remarks and a summary of future work.

\section{Formulation}
\label{ch:formulation}

An optimal approach for designing a collection of experiments
conducted in sequence
%% XH: I want to make sure we don't accidentally give the impression
%% that our method is a sequential algorithm/method for solving a
%% batch OED problem.
should account for all sources of uncertainty occurring during the
experimental campaign, along with a full description of the system
state and its evolution. We begin by formulating an optimization
problem that encompasses these goals, then cast it as a dynamic
program. We next discuss how to choose certain elements of the
formulation in order to perform Bayesian OED for parameter inference.

\subsection{Problem definition}

\begin{figure}[htb]
  \centering
  \mbox{\subfigure[Batch (open-loop) design]
    {\includegraphics[width=0.52\textwidth]{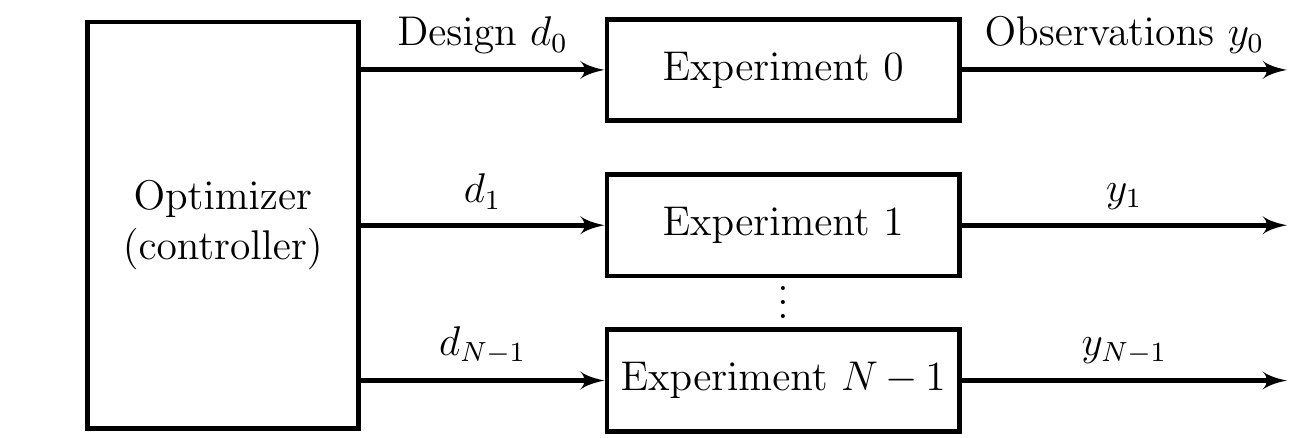}}
  }
  \mbox{\subfigure[Sequential (closed-loop) design]
    {\includegraphics[width=0.45\textwidth]{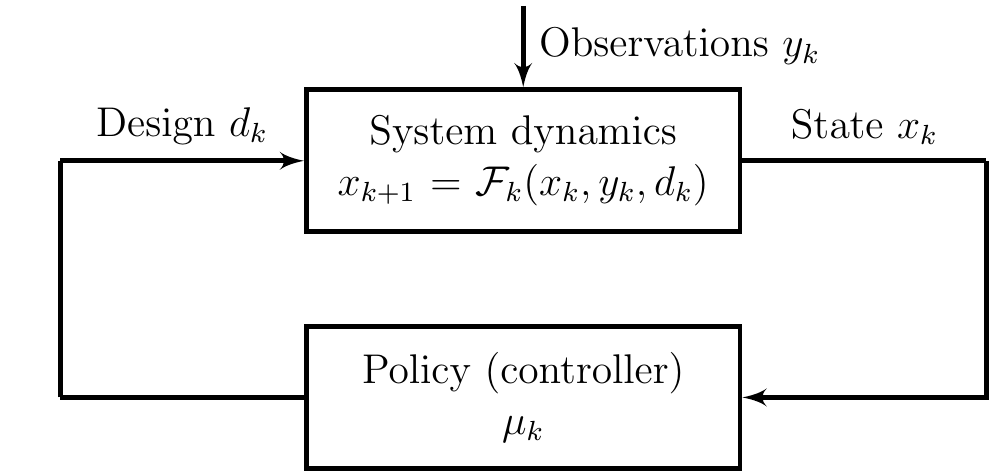}}
  }
  \caption{Batch design is an \textit{open-loop} strategy with no
    feedback of information, in that the observations $y_k$ from any
    experiment do not affect the design of any other
    experiments. Sequential design encodes a \textit{closed-loop}
    strategy, where feedback of information takes place, and the data
    $y_k$ from an experiment are used to guide the design of subsequent
    experiments.}
  \label{f:batch_sequential_designs}
\end{figure}

The core components of a general sOED formulation are as follows:

% (currently not assume designing for inference), and follow with
%stating the sOED problem.
\begin{itemize}

%% \item \textbf{Parameters:} $\theta\in \RR^{n_{\theta}}$. A finite
%%   number of uncertain model parameters to be inferred from noisy and
%%   indirect data.

\item \textbf{Experiment index:} $k=0,\ldots,N-1$. The experiments are
  assumed to occur at discrete times, ordered by the integer index
  $k$, for a total of $N < \infty$  experiments.

\item \textbf{State:} $x_k = \[x_{k,b},\,x_{k,p}\] \in \mathcal{X}_k$.
  The state contains information necessary to make optimal decisions
  about the design of future experiments. Generally, it comprises the
  \textit{belief state} $x_{k,b}$, which reflects the current state of
  uncertainty, and the \textit{physical state} $x_{k,p}$, which
  describes deterministic decision-relevant variables. We consider
  continuous and possibly unbounded state variables. Specific state
  choices will be discussed later.

\item \textbf{Design:} $d_k \in \CD_k$. The design $d_k$ represents the
  conditions under which the $k$th experiment is to be performed. We seek a
  \textit{policy} $\pi \equiv \left\{\mu_0, \mu_1, \ldots, \mu_{N-1} \right\}$
  consisting of a set of \textit{policy functions}, one for each
  experiment, that specify the design as a function of the
  current state: i.e., $\mu_k(x_k) = d_k$. We consider continuous
  real-valued design variables.
  
  Design approaches that produce a policy are \textit{sequential
    (closed-loop)} designs because the outcomes of the previous
  experiments are necessary to determine the current state, which in
  turn is needed to apply the policy.  These approaches contrast with
  \textit{batch (open-loop)} designs, where the designs are determined
  only from the initial state and do not depend on subsequent
  observations (hence, no feedback).
  Figure~\ref{f:batch_sequential_designs} illustrates these two
  different strategies.

\item \textbf{Observations:} $y_k \in \CY_k$. The observations from
  each experiment are endowed with uncertainties representing both
  measurement noise and modeling error. Along with prior uncertainty on the model
  parameters, these are assumed to be the only sources of uncertainty
  in the experimental campaign. Some models might also have internal
  stochastic dynamics, but we do not study such cases here. We
  consider continuous real-valued observations.

\item \textbf{Stage reward:} $g_k(x_k,y_k, d_k)$. The stage reward
  reflects the immediate reward associated with performing a
  particular experiment. This quantity could depend on the state,
  observations, or design. Typically, it reflects the cost of
  performing the experiment (e.g.,
  money and/or time), as well as any additional
  benefits or penalties.

\item \textbf{Terminal reward:} $g_N(x_N)$. The terminal reward
  reflects the value of the final state $x_N$ that is reached after
  all experiments have been completed.
 % serves
 %  as a mechanism to end the system dynamics by providing a reward
 %  value solely based on the final state $x_N$.

\item \textbf{System dynamics:} $x_{k+1} = \CF_k(x_k,y_k,d_k)$. The
  system dynamics describes the evolution of the system state from one
  experiment to the next, and includes dependence on both the current
  design and the observations resulting from the current
  experiment. This evolution includes the propagation of the belief
  state (e.g., statistical inference)
  %% (i.e., statistical inference)
  %% XH: at this point we have not introduced any notions of
  %% distributions/RV for the belief state yet, shall we hold off this
  %% ``i.e.'' for now?
  %% YM: How about e.g. instead? I wanted to make the notion of
  %% propagating the belief state more concrete, so that people are
  %% not left wondering.
  and of the physical state. The specific form
  of the dynamics depends on the choice of state variable, and
  will be discussed later.
\end{itemize}
\medskip

Taking a decision-theoretic approach, we seek a design policy that
maximizes the following \textit{expected utility} (also called an
\textit{expected reward}) functional:
\begin{eqnarray}
%  \pi^{\ast} 
%  = \left\{\mu_0^{\ast}(x_0),\ldots,
%    \mu_{N-1}^{\ast}(x_{N-1})\right\} 
  U(\pi) :=  \EE_{y_0,\ldots,y_{N-1}|\pi}\[\sum_{k=0}^{N-1}g_k\(x_k,y_k,\mu_k(x_k)\)  
  + g_N(x_N) \],
  \label{e:sOED_objective}
\end{eqnarray}
where the states must adhere to the system dynamics $x_{k+1} =
\CF_k(x_k,y_k,d_k)$. %, for $k=0,\ldots,N-1$.  %[okay to omit for simplicity]
The optimal policy is then
\begin{eqnarray}
  \pi^{\ast} 
  := \left\{\mu_0^{\ast},\ldots,
    \mu_{N-1}^{\ast}\right\} 
    = & \argmax_{\pi = \left\{\mu_0,
      \ldots, \mu_{N-1}\right\}}
    & \qquad U(\pi),
  \label{e:sOED} \\
& \rm{s.t.} & x_{k+1} = \CF_k(x_k,y_k,d_k), \nonumber \\
& & \mu_k\left ( \mathcal{X}_k \right ) \subseteq \CD_k, \ \ k=0,\ldots,N-1 . \nonumber
\end{eqnarray}
%subject to the system dynamics and the design space constraints
%$\mu_k(x_k) \in \CD_k$, $\forall x_k$, for $k=0,\ldots,N-1$. 
%
For simplicity, we will refer to \eqref{e:sOED} as ``the sOED
problem.''

\subsection{Dynamic programming form}

The sOED problem involves the optimization of the expected reward
functional \eqref{e:sOED_objective} over a set of policy functions,
which is a challenging problem to solve directly. Instead, we can
express the problem in an equivalent form using Bellman's principle of
optimality~\cite{Bellman1953, Bellman1956}, leading to a
finite-horizon dynamic programming formulation
(e.g.,~\cite{Bertsekas2005, Bertsekas2007}):
\begin{eqnarray}
  J_k(x_k) &=& \max_{d_k\in\CD_k}
  \EE_{y_k|x_k,d_k}\[g_k(x_k,y_k,d_k)+J_{k+1}\(\CF_k(x_k,y_k,d_k)\)\] \label{e:DP_1} \\
  J_N(x_N) &=& g_N(x_N), \label{e:DP_2}
\end{eqnarray}
for $k=0,\ldots,N-1$. The $J_k(x_k)$ functions are known as the
``cost-to-go'' or ``value'' functions. Collectively, these expressions
are known as Bellman equations. The optimal policies are now
implicitly represented by the arguments of each maximization: if
$d_k^{\ast} = \mu^*_k(x_k)$ maximizes the right-hand side of
\eqref{e:DP_1}, then the policy
$\pi^{\ast} = \left\{\mu_0^{\ast}, \mu_1^{\ast}, \ldots,
  \mu_{N-1}^{\ast} \right\}$
is optimal (under mild assumptions; see \cite{Bertsekas1996a} for more
detail on these verification theorems). 

The DP problem is well known to exhibit the ``curse of dimensionality,''
where the number of possible scenarios (i.e., sequences of design and
observation realizations) grows exponentially with the number of
stages $N$. It often can only be solved numerically and
approximately. We will develop numerical methods for finding an
approximate solution in
Section~\ref{ch:approximate_dynamic_programming}.

\subsection{Information-based Bayesian experimental design}
\label{s:info_sOED}

Our description of the sOED problem thus far has been somewhat general
and abstract. We now make it more specific, for the particular goal of
inferring uncertain model parameters $\theta$ from noisy and indirect
observations $y_k$. Given this goal, we can choose
appropriate state variables and reward functions.

We use a Bayesian perspective to describe uncertainty and
inference. Our state of knowledge about the parameters $\theta$ is
represented using a probability distribution. Moreover, if the $k$th experiment is
performed under design $d_k$ and yields the outcome $y_k$, then our
state of knowledge is updated via an application of Bayes' rule:
\begin{eqnarray}
  f(\theta|y_k,d_k, I_k) = %\frac{f(y_k|\theta,d_k, I_k) f(\theta|d_k, I_k)} {f(y_k|d_k, I_k)} = 
  \frac{f(y_k|\theta,d_k,I_k) f(\theta|I_k)} {f(y_k|d_k,
    I_k)}. \label{e:CL_Bayes}
\end{eqnarray}
Here, $I_k=\left\{d_0,y_0,\ldots,d_{k-1},y_{k-1} \right\}$ is the
information vector representing the ``history'' of previous
experiments, i.e., their designs and observations; the probability
density function $f(\theta|I_k)$ represents the prior for the $k$th
experiment; $f(y_k|\theta,d_k,I_k)$ is the likelihood function;
$f(y_k|d_k, I_k)$ is the evidence; and $f(\theta|y_k,d_k, I_k)$ is the
posterior probability density following the $k$th
experiment.\footnote{We assume that knowing the design $d_k$ of the
  current experiment (but not its outcome) does not affect our current belief about the
  parameters. In other words, the prior for the $k$th experiment does
  not change based on what experiment we \textit{plan} to do, and
  hence $f(\theta|d_k, I_k) = f(\theta|I_k)$.} Note that
$f(\theta|y_k,d_k, I_k) = f(\theta|I_{k+1})$. To keep notation
consistent, we define $I_0 = \emptyset$.

% XH: but the following assumption is violated from the 1D source
% inversion example with the rough and precise instruments.
%
% We also assume that observations from the $k$th experiment are
% independent from information of previous experiments when conditioned
% on the parameters and the current experimental design---hence $f(y_k
% |\theta,d_k ,I_k ) = f(y_k |\theta,d_k)$. While the first assumption
% is quite intuitive, the second assumption is also reasonable since the
% likelihood and forward models generally do not depend on the
% \textit{previous} experiments.
%

In this Bayesian setting, the ``belief state'' that describes the
state of uncertainty after $k$ experiments is simply the
posterior distribution. How to represent this distribution, and thus how to define
$x_{k,b}$ in a computation, is an important question. Options include:
(i) series representations (e.g., polynomial chaos expansions) of the
posterior random variable $\theta| I_k$ itself; (ii) numerical
%% XH: I have not been using upper case to denote RV (vs. lower case
%% for realization), it just gets very confusing while attempting to
%% do that everywhere. Shall we just consistently abuse this notation
%% and use lower case only? (for all the Theta in $\Theta|I_k$'s)
%% YM: OK, let's try this. If someone complains, then we can get picky
%% about uppercase.
discretizations of the posterior probability density function $f(\theta| I_k)$ or
cumulative distribution function $F(\theta| I_k)$; (iii) parameters of these
distributions, \textit{if} the priors and posteriors all belong to a simple
parametric family; or (iv) the prior $f(\theta | I_0)$ at $k=0$ plus the entire history of designs and observations from all previous
experiments. For example, if $\theta$ is a discrete random variable
that can take on only a finite number of distinct values, then it is
natural to define $x_{k,b}$ as the finite-dimensional vector
specifying the probability mass function of $\theta$. This is the
approach most often taken in constructing partially observable Markov
decision processes (POMDP)~\cite{Sigaud2010, Puterman1994}.
%% \todo{Are
%%   finite-dimensional random variables an intrinsic feature of POMDP? I
%%   tried to weaken this association somewhat.}
%% XH: POMDP does not preclude continuous variable
%% formulations. In practice, most POMDP textbooks and research have
%% been with discrete variables. There are some continuous works but
%% small in comparison. 
%
Since we are interested in continuous and possibly unbounded $\theta$,
an analogous perspective would yield in principle an
\textit{infinite-dimensional} belief state---unless, again, the
posteriors belonged to a parametric family (for instance, in the case
of conjugate priors and likelihoods).
%
%% \color{red}[[Furthermore, many state-of-the-art POMDP algorithms rely on the
%% piecewise linear and convex property of cost functions. However, as
%% shown in Section 4.5 of~\cite{Huan2015}, these algorithms would not be
%% suitable for OED that employs an information measure objective, as
%% practically useful information objectives lead to cost functions that
%% are not piecewise linear. We thus seek alternative
%% approaches.]] \normalcolor \todo{The bracketed discussion is useful, but seems
%%   entirely out of place here! We haven't even introduced the
%%   information theoretic objective yet, and we haven't talked about
%%   ADP. We brought up POMDP only because we wanted to talk about
%%   finite-dimensional belief states. Can this go elsewhere?}
%% XH: how about as a footnote after introducing (2.6)?
%% YM: I moved it after the Ginebra discussion.
%
 In this paper, we will not restrict our attention to standard
parametric families of distributions, however, and thus we will employ
finite-dimensional discretizations of infinite-dimensional belief
states. The level of discretization is a refinable numerical
parameter; details are deferred to
Section~\ref{ss:belief_state_representation}. We will also
use the shorthand $x_{k,b} = \theta|I_k$ to convey the underlying
notion that the belief state is just the current posterior distribution.

Following the information-theoretic approach suggested by Lindley
\cite{Lindley1956}, we choose the terminal reward to be
the Kullback-Leibler (KL) divergence from the final posterior (after
all $N$ experiments have been performed) to the prior (before any
experiment has been performed):
\begin{eqnarray}
  g_N(x_N) =
  \DKL\( f_{\theta \vert I_N}  || \, f_{\theta \vert I_0} \) =
  \int f_{\theta \vert I_N} (\theta)  \ln\[ \frac{  f_{\theta \vert I_N} (\theta) } {f_{\theta \vert I_0}(\theta) }\] \,d\theta \, . \label{e:cost1}
\end{eqnarray}
% where $\CH$ is the support of the prior. %YM: Why the prior? then you could have log of zero in the numerator. Thought it better to skip.
%% \todo{Notice that I changed notation above. The previous notation was
%%   confusing, I think.}
%% XH: agreed, though I worry the readers might ask where did the x's
%% go, so I added the reminder following the equation. We can remove
%% that if you feel it is unnecessary.
%% YM: the addition of the $x$s at the end seemed a little
%% unnecessary/confusing to me. We mention the shorthand/equivalence
%% at the end of the preceding paragraph.
The stage rewards $\{g_k\}_{k<N}$ can then be chosen to reflect all other immediate rewards or costs
associated with performing particular experiments.

We use the KL divergence in our design objective
\eqref{e:sOED_objective} for several reasons. First, as shown
in~\cite{Ginebra2007}, the expected KL divergence belongs to a broad
class of useful divergence measures of the information in a
statistical experiment; this class of divergences is defined by a
minimal set of requirements that must be satisfied to induce a total
information ordering on the space of possible experiments.\footnote{As
  shown in Section 4.5 of~\cite{Huan2015}, any reward function that
  introduces a non-trivial ordering on the space of possible
  experiments, according to the criteria formulated
  in~\cite{Ginebra2007}, cannot be linear in the belief state. (Note
  that this result precludes the use of non-centered posterior moments as
  reward functions.)  Consequently, the corresponding Bellman
  cost-to-go functions cannot be guaranteed to be piecewise linear and
  convex. Though many state-of-the-art POMDP algorithms have been
  developed specifically for piecewise linear and convex cost
  functions, they are generally not suitable for our sOED problem.}
Interestingly, these requirements do not rely on a Bayesian
perspective or a decision-theoretic formulation, though they can be
interpreted quite naturally in these settings.
Second, and perhaps more immediately, the KL divergence quantifies
information gain in the sense of Shannon
information~\cite{Cover2006,Mackay2005}. A large KL divergence from
posterior to prior implies that the observations $y_k$ decrease
entropy in $\theta$ by a large amount, and hence that the
observations are informative for parameter inference. Indeed, the
expected KL divergence is also equivalent to the \textit{mutual
  information} between the parameters $\theta$ and the observations
$y_k$ (treating both as random variables), given the design $d_k$.
Third, the KL divergence satisfies useful consistency conditions. It
is invariant under one-to-one reparameterizations of $\theta$. And
while it is directly applicable to non-Gaussian distributions and to
forward models that are nonlinear in the parameters $\theta$,
maximizing KL divergence in the linear-Gaussian case reduces to
Bayesian $D$-optimal design from linear optimal design
theory~\cite{Chaloner1995} (i.e., maximizing the determinant of the
posterior precision matrix, and hence of the Fisher information matrix
plus the prior precision).
%% \todo{KL is also invariant under
%%  reparameterization, so I moved that comment up.}
%% XH: sounds good.
%
Finally we should note that, as an alternative to KL divergence, it
is entirely reasonable to construct a terminal reward from some
other loss function tied to an alternative goal (e.g., squared error
loss if the goal is point estimation). But in the absence of such a
goal, the KL divergence is a general-purpose objective that
seeks to maximize learning about the uncertain environment represented
by $\theta$, and should lead to good performance for a broad range of
estimation tasks.

%  the use of
% information measure contrasts with a loss function in that, while the
% former does not target a particular task (such as estimation) in the
% context of a decision problem, it provides a general guidance of
% learning about the uncertain environment, and gaining information that
% performs well for a wide range of tasks albeit not best for any
% particular task. \todo{This paragraph seems much too wordy.}

\subsection{Notable suboptimal sequential design methods}
\label{s:notable_suboptimal_designs}

Two design approaches frequently encountered in the OED literature are
batch design and greedy/myopic sequential design. Both can be seen as
special cases or restrictions of the sOED problem formulated here, and
are thus in general suboptimal. We illustrate these relationships below.

Batch OED involves the concurrent design of all experiments, and hence
the outcome of any experiment cannot affect the design of the
others. Mathematically, the policy functions $\mu_k$ for batch design
do not depend on the states $x_k$, since no feedback is
involved. \eqref{e:sOED} thus reduces to an optimization problem over
the joint design space $\mathcal{D} := \CD_0 \times \CD_1 \times \cdots \CD_{N-1}$ rather
than over a space of policy functions, i.e.,
\begin{eqnarray}
  \left (d_0^{\ast},\ldots,d_{N-1}^{\ast}\right ) =
  \argmax_{ \left (d_0,\ldots,d_{N-1} \right ) \, \in \,  \CD}
  \EE_{y_0,\ldots,y_{N-1}|d_0,\ldots,d_{N-1}}\[\sum_{k=0}^{N-1}g_k(x_k,y_k,d_k) 
  + g_N(x_N) \],
\label{eq:batch}
\end{eqnarray}
subject to the system dynamics $x_{k+1} = \CF_k(x_k,y_k,d_k)$, for $k=0,\ldots,N-1$. Since
batch OED involves the application of stricter constraints to the sOED
problem than \eqref{e:sOED}, it generally yields suboptimal designs.

Greedy design is a particular sequential and closed-loop formulation
where only the next experiment is considered at each stage, without
taking into account the entire horizon of future experiments and
system dynamics.\footnote{The greedy experimental design strategies
  considered in this paper are instances of \textit{greedy
    experimental design with feedback}. This is a different notion
  than greedy \textit{optimization} of a batch experimental design
  problem, where no feedback of data occurs between experiments. The
  latter is simply a suboptimal solution strategy for
  \eqref{eq:batch}, wherein the $d_k$ are chosen in sequence.}
Mathematically, the greedy policy results from solving
\begin{eqnarray}
  J_k(x_k) &=& \max_{d_k\in\CD_k} \EE_{y_k|x_k,d_k}\[g_k(x_k,y_k,d_k)\] , \label{e:greedy}
\end{eqnarray}
where the states must obey the system dynamics
$x_{k+1} = \CF_k(x_k,y_k,d_k)$, $k=0,\ldots,N-1$. If
$d_k^{\textrm{gr}} = \mu^{\textrm{gr}}_k(x_k)$ maximizes the
right-hand side of \eqref{e:greedy} for all $k=0, \ldots, N-1$, then
the policy
$\pi^{\textrm{gr}} = \left\{\mu_0^{\textrm{gr}}, \mu_1^{\textrm{gr}},
  \ldots, \mu_{N-1}^{\textrm{gr}} \right\}$
is the greedy policy. Note that the terminal reward in \eqref{e:DP_2}
no longer plays a role in greedy design.\footnote{An information-based
  greedy design for parameter inference would thus require moving the
  information gain objective into the stage rewards $g_k$, e.g., an
  incremental information gain formulation \cite{Terejanu2012}.} 
Since greedy design involves truncating the DP form of the sOED
problem, it again yields suboptimal designs.
% While greedy design decouples the Bellman's equations,
% and may accommodate circumstances where the total number of
% experiments is unknown, but it is a truncation to the DP form of the
% sOED problem nonetheless suboptimal.

\section{Numerical approaches}
\label{ch:approximate_dynamic_programming}

%\subsection{Approximation approaches}

Approximate dynamic programming (ADP) broadly refers to numerical
methods for finding an approximate solution to a DP problem. The
development of such techniques has been the target of substantial
research efforts across a number of communities (e.g., control theory,
operations research, machine learning), targeting different variations
of the DP problem.
While a variety of terminologies are used in these fields,
there is often a large overlap among the fundamental spirits of their
solution approaches.  We thus take a perspective that groups many ADP
techniques into two broad categories:
\begin{enumerate}
\item \textbf{Problem approximation:} These are ADP techniques that do
  not provide a natural way to refine the approximation, or where
  refinement does not lead to the solution of the \textit{original}
  problem. Such techniques typically lead to suboptimal strategies
  (e.g., batch and greedy designs,
  % open-loop control,
  % YM: previously said 'open-loop feedback control' but this seemed
  % intrinsically contradictory!
  % XH: agreed it definitely does sound contradictory, but I swear
  % it's a real thing!  :)
  % Bertsekas DP book Vol1 p300: it is when you solve ALL remaining
  % experiments in an OL/batch setting, but only carry out the next
  % SINGLE experiment from them, update states, and repeat.
  % Maybe let's just leave it out altogether to avoid confusion?
  certainty-equivalent control, Gaussian approximations).

\item \textbf{Solution approximation:} Here there is some natural way
  to refine the approximation, such that the effects of approximation
  diminish with refinement. These methods have some notion of
  ``convergence'' and may be refined towards the solution of the
  \textit{original} problem. Methods used in solution approximation
  include policy iteration, value function and $Q$-factor
  approximations, numerical optimization, Monte Carlo sampling,
  regression, quadrature and numerical integration, discretization and
  aggregation, and rolling horizon procedures.
  %% \todo{Rolling horizon *what*? seems a follow-up word is needed.}
  %% XH: how about this ''procedure''?
\end{enumerate}
In practice, techniques from both categories are often combined in
order to find an approximate solution to a DP problem. The approach in
this paper will to try to preserve the original problem as much as
possible, relying more heavily on solution approximation techniques
to \textit{approximately solve the exact problem}.

Subsequent sections (Sections
\ref{s:policy_representation}--\ref{ss:belief_state_representation})
will describe successive building blocks of our ADP approach, and the
entire algorithm will be summarized in
Section~\ref{ss:algorithm_pseudocode}.

\subsection{Policy representation}
\label{s:policy_representation}

In seeking the optimal policy, we first must be able to represent a
(generally suboptimal) policy $\pi = \left\{\mu_0, \mu_1, \ldots,
  \mu_{N-1} \right\}$. 
One option is to represent a candidate policy function
$\mu_k(x_k)$ \textit{directly} (and approximately)---for example, by
brute-force tabulation over a finite collection of $x_k$ values
representing a discretization of the state space, or by using standard
function approximation techniques.
On the other hand, one can preserve the recursive structure of the
Bellman equations and ``parameterize'' the policy via approximations
% \todo{Should we say parameterization rather than approximations?}
%% XH: I put parameterize in quotes since value functions themselves
%% aren't parameters (though they could be furthermore
%% parameterized). The idea I want to convey is that the set of value
%% functions now represents/dictates/equivalently describes the
%% policy. Perhaps ``parameterize'' is misleading.
of the value functions appearing in \eqref{e:DP_1}.  We take this
approach here. In particular, we represent the policy using one step
of lookahead~\cite{Bertsekas2005}, thus retaining some structure from the original DP
problem while keeping the method computationally feasible. By looking
ahead only one step, the recursion between value functions is
broken and the exponential growth of computational cost with respect
to the horizon $N$ is reduced to linear growth.\footnote{Multi-step
  lookahead is possible in theory, but impractical, as the amount of
  online computation would be intractable given continuous state
  and design spaces.}  The one-step lookahead policy
representation\footnote{It is crucial to note that ``one-step
  lookahead'' is not greedy design, since future effects are
  still included (within the term $\tJ_{k+1}$). The name simply describes the
  \textit{structure} of the policy representation, indicating that 
  approximation is made after one step of looking ahead (i.e., in
  $\tJ_{k+1}$).} is:
\begin{eqnarray}
  \mu_k(x_k) = \argmax_{d_k\in\CD_k}
  \EE_{y_k|x_k,d_k}\[g_k(x_k,y_k,d_k)+\tJ_{k+1}\(\CF_k\(x_k,y_k,d_k\)\)\],
  \label{e:onestep_lookahead} 
\end{eqnarray}
for $k=0, \ldots, N-1$, and $\tJ_{N}(x_N) \equiv g_N(x_N)$. The policy
function $\mu_k$ is therefore indirectly represented via the
approximate value function $\tJ_{k+1}$, and one can view the policy
$\pi$ as implicitly parameterized by the set of value functions
$\tJ_{1}, \ldots, \tJ_{N}$.\footnote{A similar method is the use of
  $Q$-factors~\cite{Watkins1989, Watkins1992}: $\mu_k(x_k) =
  \argmax_{d_k\in\CD_k} \tQ_k(x_k,d_k)$, where the $Q$-factor
  corresponding to the optimal policy is $Q_k(x_k,d_k)\equiv
  \EE_{y_k|x_k,d_k}\[g_k(x_k,y_k,d_k)+J_{k+1}\(\CF_k(x_k,y_k,d_k)\)\]$. The
  functions $\tQ_k(x_k,d_k)$ have a higher input dimension than
  $\tJ_k(x_k)$, but once they are available, the corresponding policy
  can be applied without evaluating the system dynamics $\CF_k$, and is thus
  known as a ``model-free'' method. $Q$-learning via value iteration
  is a prominent method in reinforcement learning.} If
$\tJ_{k+1}(x_{k+1}) = J_{k+1}(x_{k+1})$, we recover the Bellman
equations \eqref{e:DP_1} and~\eqref{e:DP_2}, and hence we have 
$\mu_k=\mu_k^{\ast}$. Therefore we would like to find a collection of $\{ \tJ_{k+1}\}_k$
that is close to $\{J_{k+1}\}_k$.

We employ a simple parametric ``linear architecture'' for these value function approximations:
\begin{eqnarray}
  \tJ_{k}(x_{k}) = r_k^{\top}\phi_k(x_k) = \sum_{i=1}^m r_{k,i}
  \phi_{k,i}(x_k),\label{e:linear_arch}
\end{eqnarray}
where $r_{k,i}$ is the coefficient (weight) corresponding to the $i$th
feature (basis function) $\phi_{k,i}(x_k)$.  While more sophisticated
nonlinear or even nonparametric function approximations are possible
(e.g., $k$-nearest-neighbor~\cite{Gordon1995}, kernel
regression~\cite{Ormoneit2002}, neural networks~\cite{Bertsekas1996}),
the linear approximator is easy to use and intuitive to
understand~\cite{Lagoudakis2003}, and is often required for many
analysis and convergence results~\cite{Bertsekas2007}.  It follows
that the construction of $\tJ_k(x_k)$ involves the selection of
features and the training of coefficients.

The choice of features is an important but often difficult task. A
\textit{concise} set of features that is \textit{relevant} to the
actual dependence of the value function on the state can
substantially improve the accuracy and efficiency of
\eqref{e:linear_arch} and, in turn, of the overall
algorithm. Identifying helpful features, however, is non-trivial.  In
the machine learning and statistics communities, substantial research
has been dedicated to the development of systematic procedures for
both extracting and selecting features~\cite{Guyon2006,
  Liu1998}. Nonetheless, finding good features in practice often
relies on experience, trial and error, and expert knowledge of the
particular problem at hand.  We acknowledge the difficulty of this
process, but do not pursue a detailed discussion of general and
systematic feature construction here.
Instead, we employ a reasonable heuristic by choosing features that
are \textit{polynomial functions} of the mean and log-variance of the
belief state, as well as of the physical state. The main motivation
for this choice stems from the KL divergence term in the terminal
reward. The impact of this terminal reward is propagated
to earlier stages via the value functions, and hence the value
functions must represent the state-dependence of future information
gain. While the belief state is generally not Gaussian and the optimal
policy is expected to depend on higher moments, the analytic
expression for the KL divergence between two univariate Gaussian
distributions, which involves their mean and log-variance terms,
provides a starting point for promising features. Polynomials then
generalize this initial set.
We will provide more detail about our feature choices in
Section~\ref{ch:numerical_results}. For the present purpose of
developing our ADP method, we assume that the features are fixed. We now
focus on developing an efficient procedure for training the
coefficients.

\subsection{Policy construction via approximate value iteration}
\label{s:policy_construction}

Having decided on a way to represent candidate policies, we now aim to
construct policies within this representation class that are close to
the optimal policy. We achieve this goal by constructing and
iteratively refining value function approximations via regression over
targeted relevant states. 

Note that the procedure for policy construction described in this
section can be performed entirely offline. Once this process is
terminated and the resulting value function approximations are
available, \textit{applying} the policy as experimental data are
acquired is an online process, which involves evaluating
\eqref{e:onestep_lookahead}. The computational costs of these online
evaluations are generally much smaller than those of offline policy
construction.

\subsubsection{Backward induction with regression}
\label{ss:backward_induction_regression}

Our goal is to find value function approximations (implicitly, policy
parameterizations) $\tJ_{k}$ that are close to the value functions
$J_{k}$ of the optimal policy, i.e., the value functions that satisfy
\eqref{e:DP_1} and~\eqref{e:DP_2}. We take a direct approach, and
would in principle like to solve the following ``ideal'' regression
problem: minimize the squared error of the approximation under the
state measure induced by the optimal policy,
\begin{eqnarray}
  \min_{r_{k}, \forall k}  \int_{\CX_1 \times \cdots \times \CX_{N-1}}\[
  \sum_{k=1}^{N-1} \( J_{k}(x_k) - 
  r_{k}^{\top} \phi_{k}(x_k)\)^2\] f_{\pi^{\ast}}(x_1,\ldots,x_{N-1}) \,
  dx_1 \ldots dx_{N-1} .\label{e:ideal_regression_problem}
\end{eqnarray}
The  weighted $L^2$ norm above is also known as the $D$-norm in other
work~\cite{Tsitsiklis1997}; its
associated density function is denoted by
$f_{\pi^{\ast}}(x_1,\ldots,x_{N-1})$.
%% \todo{Did I mangle the
%%   description of the $D$-norm? Since it was a norm, and not just a
%%   density function, I introduced the notion of weighted $L^2$. Please
%%   correct as needed, and add a citation.}
%%
%% XH: From my understanding, this paper shows convergence proofs for
%% a situation of policy evaluation in an infinite-horizon Markov
%% chain setup (i.e., what is the value of a GIVEN policy), which is
%% often a crucial PART of policy-iteration-like algorithms (e.g.,
%% TD-lambda). So the D-norm it defines (starts in p.677 Section A;
%% and it is indeed a weighted L-2 norm) corresponds to the
%% steady-state probabilities of that GIVEN policy. But I
%% think the concept is still the same, whereas we more specifically
%% want our state measure to correspond to that of the OPTIMAL
%% policy. Maybe we don't put too much emphasis on D-norm here in case
%% of people nitpicking --- though I really think they are the same in
%% spirit.
Here we have imposed the linear architecture
$\tJ_{k}(x_{k}) = r_{k}^{\top} \phi_{k}(x_k)$ \eqref{e:linear_arch}.
$\CX_k$ is the support of $x_k$.

In practice, the integral above must be replaced by a sum over
discrete regression points, and the distribution of these points
reflects where we place more emphasis on the approximation being
accurate. Intuitively, we would like more accurate approximations in
regions of the state space that are more frequently visited under the
optimal policy, e.g., as captured by sampling from
$f_{\pi^{\ast}}$. 
%
% \todo{This notion is directly captured in $f_{\pi^{\ast}}$, right? The
%   next sentence seems to change what we want, though, because the
%   state measure induced by the numerical methods is not naturally part
%   of $f_{\pi^{\ast}}$, or is it? If we include the impact of numerical
%   methods on the policy-induced state measure, does it still
%   correspond to the $D$-norm above? I tried to rewrite to make this
%   change clearer.}
% \todo{Is the new version okay?}
%% XH: I don't think the D-norm concept takes into account anything
%% numerical, all theoretical analysis; this is a notion we
%% introduced. (The D-norm papers simply say it is the steady state of
%% probability of the Markov chain, maybe you can argue that for an
%% interpretation where it is the steady state under a particular
%% numerical implementation. I personally do not view that way.)
%% Agreed that then our ``practical D-norm'' would be
%% different. Perhaps this is getting more philosophical, but I think
%% while the original definition is good (and maybe required) for
%% theoretical analysis, but in practice the ``steady state'' that
%% will happen/observed will be different depending on numerical
%% methods/implementations, so we should build that in.
%
But we should actually consider a further desideratum: accuracy over
the state measure induced by the optimal policy \textit{and} by the
numerical methods used to evaluate this policy (whatever they may
be). Numerical optimization methods used to solve
\eqref{e:onestep_lookahead}, for instance, may visit many intermediate
values of $d_k$ and hence of $x_{k+1} = \CF_k\(x_k,y_k,d_k\)$.
The accuracy of the value function approximation at these intermediate
states can be crucial; poor approximations can potentially mislead the
optimizer to arrive at completely different designs, and in turn
change the outcomes of regression and policy evaluation.  We thus
include the states visited \textit{within} our numerical methods (such
as iterations of stochastic approximation for solving
\eqref{e:onestep_lookahead}) as regression points too. For simplicity
of notation, we henceforth let $f_{\pi^{\ast}}$ represent the state
measure induced by the optimal policy and the associated numerical methods.

In any case, as we have neither $J_{k}(x_k)$ nor
$f_{\pi^{\ast}}(x_1,\ldots,x_{N-1})$, we must solve
\eqref{e:ideal_regression_problem} approximately. First, to sidestep the need
for $J_k(x_k)$, we will construct the value function approximations
via an \textit{approximate value iteration}, specifically using
\textit{backward induction} with \textit{regression}.
% The resulting $\tJ_k$'s are
% then used as parameterization of the one-step lookahead policy in
% \eqref{e:onestep_lookahead}. 
Starting with $\tJ_{N}(x_N) \equiv
g_N(x_N)$, we proceed backwards from $k=N-1$ to $k=1$ and form
\begin{eqnarray}
  \tJ_{k}(x_k) &=& r_{k}^{\top} \phi_{k}(x_k)  \label{e:induction_projection} \\ &=& \mathcal{P}\,
  \left\{\max_{d_k\in\CD_k}
    \EE_{y_k|x_k,d_k}\[g_k(x_k,y_k,d_k)+\tJ_{k+1}\(\CF_k\(x_k,y_k,d_k\)\)\]\right\}
  \nonumber\\ &=& \mathcal{P}\,\hJ_k(x_k), \nonumber
\end{eqnarray}
where $\mathcal{P}$ is an approximation operator that here represents a regression procedure. This approach leads to a sequence of
ideal regression problems to be solved at each stage $k$:
\begin{eqnarray}
  \min_{r_{k}}  \int_{\CX_k}
  \( \hJ_{k}(x_k) - r_{k}^{\top} \phi_{k}(x_k)\)^2 f_{\pi^{\ast}}(x_k) \,
  dx_k, \label{e:stage_k_regression_problem}
\end{eqnarray}
where $\hJ_k(x_k) \equiv \max_{d_k\in\CD_k}
\EE_{y_k|x_k,d_k}\[g_k(x_k,y_k,d_k)+\tJ_{k+1}\(\CF_k\(x_k,y_k,d_k\)\)\]$
and $f_{\pi^{\ast}}(x_k)$ is the marginal of
$f_{\pi^{\ast}}(x_1,\ldots,x_{N-1})$. 

First, we note that since $\tJ_{k}(x_k)$ is built from
$\tJ_{k+1}(x_{k+1})$ through backward induction and regression, the
effects of approximation error can accumulate, 
% YM: previously said 'aggregate' but I thought accumulate sounded better. Any change, though?
% XH: accumulate sounds good
potentially at an exponential rate~\cite{Tsitsiklis2001}. The accuracy of {all}
$\tJ_{k}(x_k)$ approximations (i.e., for \textit{all} $k$) is thus
important. Second, while we no longer need $J_k(x_k)$ to construct
$\tJ_k(x_k)$, we remain unable to select regression points according
to $f_{\pi^{\ast}}(x_k)$. This issue is addressed next.

\subsubsection{Exploration and exploitation}

Although we cannot \textit{a priori} generate regression points from
the state measure induced by the optimal policy, it is possible to
generate them according to a \textit{given} (suboptimal) policy.
% This includes
% heuristic policies, and the current approximation to the optimal
% policy in the algorithm (we shall refer to this as the ``current
% policy'' throughout this section). 
We thus generate regression points via two main processes:
\textit{exploration} and \textit{exploitation}.  Exploration is
conducted simply by randomly selecting designs (i.e., applying a
random policy). For example, if the feasible design space is
bounded, the random policy could simply be uniform sampling. In
general, however, and certainly when the design spaces
$\{ \mathcal{D}_k \}_{k=0}^{N-1}$ are unbounded, a design
measure for exploration needs to be prescribed, often selected from experience and an
understanding of the problem.  The purpose of exploration is to allow
a positive probability of probing regions that can potentially lead to
good reward. Exploration states are generated from a 
design measure as follows: we sample $\theta$ from the prior, sample designs
$\{ d_k \}_{k=0}^{N-1}$ from the design measure, generate a $y_k$
from the likelihood $p(y_k \vert \theta, d_k, I_k)$ for each design,
and then perform inference to obtain states $x_k = \theta \vert y_k,
d_k, I_k$. 

Exploitation, on the other hand, involves using the current
understanding of a good policy to visit regions that are also likely
to be visited under the optimal policy.
Specifically, we will perform exploitation by exercising the one-step
lookahead policy based on the currently available approximate value
functions $\tJ_k$. In practice, a mixture of both exploration and
exploitation is used to achieve good results, and various strategies
have been developed and studied for this purpose (see, e.g.,
\cite{Powell2011}).  In our algorithm, the states visited from both
exploration and exploitation are used as regression points for the
least-squares problems in \eqref{e:stage_k_regression_problem}. Next,
we describe exactly how these points are obtained.

\subsubsection{Iteratively updating approximations of the optimal policy}
\label{ss:ell}

Exploitation in the present context involves a dilemma of sorts:
generating exploitation points for regression requires the
availability of an approximate optimal policy, but the construction of
such a policy requires regression points. To address this issue, we
introduce an iterative approach to update the approximation of the
optimal policy and the state measure induced by it. We refer to this
mechanism as ``policy update'' in this paper, to avoid confusion with
approximate value iteration introduced previously.

\begin{figure}[htb]
  \centering
  \includegraphics[width=0.9\textwidth]{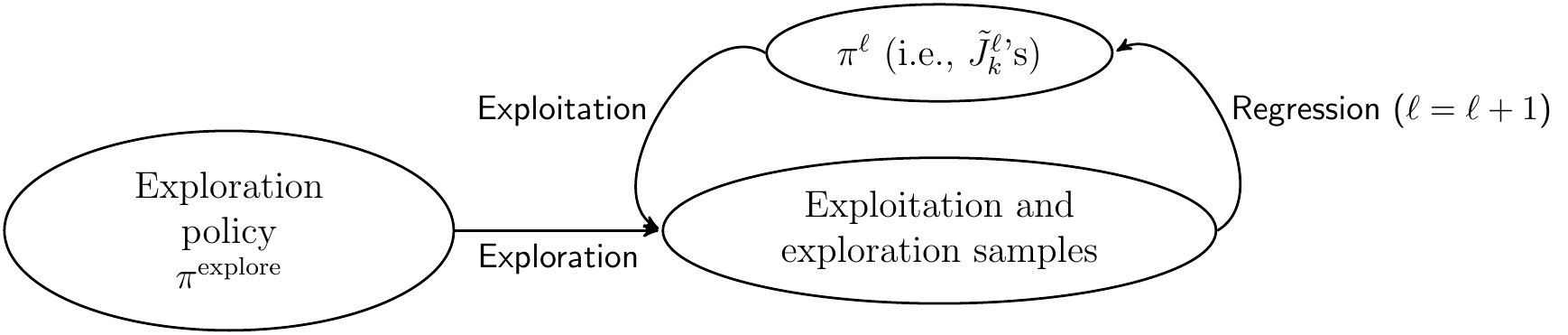}
  \caption{Iterative procedure for policy updates.}
  \label{f:state_measure_update}
\end{figure}

At a high level, our algorithm alternates between generating
regression points via exploitation and then constructing an
approximate optimal policy using those regression points. The
algorithm is initialized with only an exploration heuristic, denoted
by $\pi^{\textrm{explore}}$. States visited by exploration
trajectories generated from $\pi^{\textrm{explore}}$ are then used as
initial regression points to discretize
\eqref{e:stage_k_regression_problem}, producing a collection of value
functions $\{\tJ^1_k\}_{k=1}^{N-1}$ that parameterize the policy
$\pi^1$. The new policy $\pi^1$ is then used to generate exploitation
trajectories via \eqref{e:onestep_lookahead}. These states are mixed
with a random selection of exploration states from
$\pi^{\textrm{explore}}$, and this new combined set of states is used
as regression points to again discretize and solve
\eqref{e:stage_k_regression_problem}, yielding value functions
$\{\tJ^2_k\}_{k=1}^{N-1}$ that parameterize an updated policy
$\pi^2$. The process is repeated. As these iterations continue, we
expect a cyclical improvement: regression points should move closer to
the state measure induced by the optimal policy, and with more
accurate regression, the policies themselves can further improve.
% \todo{Adjusted slightly. REVISIT on skype.}
% XH: I agree with the description, but perhaps it might mislead that
% the regression points ``somehow'' improve over iters, and the
% policies subsequently improve because the regression points
% improved---in other words, can we clarify that it is a cycle, that
% regression improves because policy improves, and policy improves
% because regression improves, and repeat?
The largest change is expected to occur after the first iteration,
when the first exploitation policy $\pi^1$ becomes available; smaller
changes typically occur in subsequent iterations. A schematic of the
procedure is shown in Figure~\ref{f:state_measure_update}.

In this paper, we will focus on empirical numerical investigations of
these iterations and their convergence. A theoretical analysis of this
iterative procedure presents additional challenges, given the mixture of
exploration and exploitation points, along with the generally
unpredictable state measure induced by the numerical methods used to
evaluate the policy. We defer such an analysis to future work.
%% \todo{Take a look at this paragraph in our final pass, to see if it's
%%   calibrated correctly and not too negative.}
%\todo{I agree with what is said here, but wonder if this opens
  % us up to immediate criticism. We could defer these statements to the
  % conclusions? Or maybe not. Or otherwise reword, like ``Establishing
  % rigorous convergence theory for this iterative procedure is
  % difficult\ldots''}
%% XH: I agree this is a weak statement, gives the feeling that we are
%% finding an excuse. Is there a more graceful way to say: convergence
%% proof is important, we acknowledge it is difficult, but empirical
%% study is also important, so we do that for now?
%\todo{CHECK revised version to see if we like it.}

Combining the regression problems \eqref{e:stage_k_regression_problem} from all stages $k=1\ldots N-1$,
the overall problem that is solved approximates the original ``ideal'' regression problem of \eqref{e:ideal_regression_problem}:
\begin{eqnarray}
  \min_{r_{k},\forall k}  \int_{\CX_1 \times \cdots \times \CX_{N-1}}\[
  \sum_{k=1}^{N-1} \( \hJ^{\ell+1}_{k}(x_k) - 
  \(r_{k}^{\ell+1}\)^{\top} \phi_{k}(x_k)\)^2\] f_{\pi^{\textrm{explore}}+\pi^{\ell}}(x_1,\ldots,x_{N-1}) \,
  dx_1 \ldots dx_{N-1} \nonumber
%\label{e:actual_regression_problem}
\end{eqnarray}
where $f_{\pi^{\textrm{explore}}+\pi^{\ell}}(x_1,\ldots,x_{N-1})$ is
the joint density corresponding to the mixture of exploration and
exploitation from the $\ell$th iteration, and
$\(r_{k}^{\ell+1}\)^{\top} \phi_{k}(x_k)$ is
$\tJ_k^{\ell+1}(x_k)$. Note that
$f_{\pi^{\textrm{explore}}+\pi^{\ell}}(x_1,\ldots,x_{N-1})$ lags one
iteration behind $\hJ^{\ell+1}_{k}(x_k)$ and $\tJ_k^{\ell+1}(x_k)$,
since we need to have constructed the policy before we can
sample trajectories from it.

Simulating exploitation trajectories, evaluating policies, and
computing the values of $\hJ$ for the purpose of linear regression all
involve maximizing an expectation over a continuous design space (see
both \eqref{e:onestep_lookahead} and the definition of $\hJ$ following
\eqref{e:stage_k_regression_problem}). While the expected value generally
cannot be found analytically, a robust and natural approximation may
be obtained via Monte Carlo estimation. As a result, the optimization
objective is effectively noisy. We use Robbins-Monro (or
Kiefer-Wolfowitz if gradients are not available analytically)
stochastic approximation algorithms to solve these stochastic
optimization problems; more details can be found in~\cite{Huan2014}.

\subsection{Belief state representation}
\label{ss:belief_state_representation}

As discussed in Section~\ref{s:info_sOED}, a natural choice of the
belief state $x_{k,b}$ in the present context is the posterior
$\theta | I_k$. An important question is how to represent this belief
state numerically. There are two major considerations. First, since we
seek to accommodate general nonlinear forward models with continuous
parameters, the posterior distributions are continuous, non-Gaussian,
and not from any particular parametric family; such distributions are
difficult to represent in a finite- or low-dimensional manner. Second,
sequential Bayesian inference, as part of the system dynamics $\CF_k$,
needs to be performed repeatedly under different realizations of
$d_k$, $y_k$, and $x_k$.

In this paper, we represent belief states numerically by discretizing
their probability density functions on a dynamically evolving grid. To
perform Bayesian inference, the grid needs to be \textit{adapted} in
order to ensure reasonable coverage and resolution of the posterior
density. Our scheme first computes values of the unnormalized
posterior density on the current grid, and then decides whether grid
expansion is needed on either side, based on a threshold for the ratio
of the density value at the grid endpoints to the value of the density
at the mode. Second, a uniform grid is laid over the expanded regions,
and new unnormalized posterior density values are computed. Finally, a
new grid encompassing the original and expanded regions is constructed
such that the probability masses between neighboring grid points are
equal; this provides a mechanism for coarsening the grid in regions
of low probability density.

While this adaptive gridding approach is suitable for one- or perhaps
two-dimensional $\theta$, it becomes impractical in higher
dimensions. In a companion paper, we will introduce a more flexible
technique based on transport maps
(e.g.,~\cite{Villani2008,Marzouk2016}) that can represent
multi-dimensional non-Gaussian posteriors in a scalable
way, and that immediately enables fast Bayesian inference from
multiple realizations of the data.

\subsection{Algorithm pseudocode}
\label{ss:algorithm_pseudocode}

The complete approximate dynamic programming approach developed over
the preceding sections is outlined in Algorithm~\ref{a:sOED_algorithm}.
%% \todo{I replaced the index $rt$ with $\alpha$. A different letter is
%%   fine too, if you prefer something else. But two letters might be misunderstood as multiplication!}
%% \todo{A few edits to the algorithm description: (i) features are $\phi_{k,i}$ or $\phi_k$
%%   earlier; (ii) the notion of exploration measure might be clearer if
%%   we call it the exploration policy $\pi^{\textrm{explore}}$, or if we
%%   write out the probability density $p^{\textrm{explore}}(d_0, d_1,
%%   \ldots, d_{N-1})$, then below say ``$d_k$ from exploration
%%   distribution $p^{\textrm{explore}}$'' (iii) should $N$ also be a parameter; (iv) do we
%%   want to briefly assign meaning to $L$, $R$, and $T$, as in ``number
%%   of value iterations $L$'' or ``number of policy updates $L$'';
%%   number of exploration trajectories $R$; number of exploitation
%%   trajectories $T$; (v) below we could be clearer in saying ``sampling
%%   \ldots $y_k$ from likelihood $p(y_k | \theta, d_k)$'' [or $p(y_k |
%%   \theta, d_k, I_k)$]; (vi) instead of $\forall k$, would it be good
%%   to say $k=1, \ldots, N-1$? This may be too wordy, but it's clearer }
\begin{algorithm}[htb]
  \caption{Pseudocode for solving the sOED problem.
%    , based
%    on one-step lookahead policy representation, backward induction
%    (with regression) policy construction, iteratively updated policy
%    induced state measure, and transport map belief state
%    representation and joint map-accelerated Bayesian inference.
  }
  \label{a:sOED_algorithm}
\begin{algorithmic}[1]
  %  \SetAlgoLined
  \STATE{\textbf{Set parameters:} Select number of experiments $N$, features
  $\{\phi_{k}\}_{k=1}^{N-1}$, 
  exploration policy $\pi^{\textrm{explore}}$, number of policy
  updates $L$, number of exploration trajectories $R$, number of
  exploitation trajectories $T$}
  % \textbf{Initial exploration:} Simulate $R_0$ exploration
  % trajectories by sampling $\theta$ from
  % prior, $d_k$ from exploration measure, $y_k$ from likelihood,
  % $\forall k$, without inference\;
  % \textbf{Make exploration joint map:} Make $T_{\mathrm{explore}}$
  % from these samples\;
  \FOR{$\ell=1,\ldots,L$}
%  {
  \STATE{\textbf{Exploration:} Simulate $R$ exploration trajectories by
    sampling $\theta$ from 
    the prior, $d_k$ from exploration policy $\pi^{\textrm{explore}}$,
    and $y_k$ from the likelihood $p(y_k | \theta, d_k, I_k)$, for
    $k=0,\ldots,N-1$}
  \STATE{Store all states visited: 
    $\CX_{k,\mathrm{explore}}^{\ell} = \{x_k^r\}_{r=1}^R$, $k=1,\ldots,N-1$}
  \STATE{\textbf{Exploitation:} If $\ell >1$,
    simulate $T$ exploitation trajectories by sampling $\theta$ from
    the prior, $d_k$ from the one-step lookahead policy
    $$\mu^{\ell-1}_k(x_k)=\argmax_{d'_k}
    \EE_{y_k|x_k,d'_k}\[g_k(x_k,y_k,d'_k)+\tJ_{k+1}^{\ell-1}(\CF_k(x_k,y_k,d'_k))\],$$
    and $y_k$ from the likelihood $p(y_k | \theta, d_k, I_k)$, $k=0,\ldots,N-1$}
  \STATE{Store all states visited: 
    $\CX_{k,\mathrm{exploit}}^{\ell} = \{x_k^t\}_{t=1}^T$, $k=1,\ldots,N-1$}
  \STATE{\textbf{Approximate value iteration:} Construct functions $\tJ_k^{\ell}$ via
    backward induction using new
    regression points $\{\CX_{k,\mathrm{explore}}^{\ell} \cup
    \CX_{k,\mathrm{exploit}}^{\ell}\}$,
    $k=1,\ldots,N-1$, as described in the loops below}
    \FOR{$k=N-1,\ldots,1$}
%    {
      \FOR{$\alpha = 1, \ldots, R+T$ where $x_k^{(\alpha)}$ are members of
        $\{\CX_{k,\mathrm{explore}}^{\ell} \cup
        \CX_{k,\mathrm{exploit}}^{\ell}\}$} 
%      {
      \STATE{Compute training values: $$\hJ^{\ell}_k \( x_k^{(\alpha)} \)=\max_{d'_k}
        \EE_{y_k|x_k^{(\alpha)},d'_k}\[g_k(x_k^{(\alpha)},y_k,d'_k)+\tJ_{k+1}^{\ell}(\CF_k(x_k^{(\alpha)},y_k,d'_k))\] $$
        Construct $\tJ_k^{\ell}=\mathcal{P}\,\hJ^{\ell}_k$ by regression on training values}
      \ENDFOR
      \ENDFOR
      \ENDFOR
      %      }
%    }
%  }
 \STATE{\textbf{Extract final policy parameterization:} $\tJ_k^{L}$,
  $k=1,\ldots,N-1$}
  \end{algorithmic}
\end{algorithm}

\section{Numerical examples}
\label{ch:numerical_results}

We present two examples to highlight different aspects of the
approximate dynamic programming methods developed in this paper. First
is a \textit{linear-Gaussian problem}
(Section~\ref{s:linear_gaussian}). This example establishes (a) the
ability of our numerical methods to solve an sOED problem, in a
setting where we can make direct comparisons to an exact solution
obtained analytically; and (b) agreement between results generated
using grid-based or analytical representations of the belief state,
along with their associated inference methods. Second is a nonlinear
\textit{source inversion problem}
(Section~\ref{s:source_inversion_1D}).  This problem has three cases:
Case 1 illustrates the advantage of sOED over batch
(open-loop) design; Case 2 illustrates the advantage
% XH: synonyms... ``asserts the superiority''?
of sOED over
greedy (myopic) design; and Case 3 demonstrates the ability to
accommodate longer sequences of experiments, as well as the effects of
policy updates.

\subsection{Linear-Gaussian problem}
\label{s:linear_gaussian}

%% \todo{I know we wanted to have some text before each new level of
%%   hierarchy, but getting rid of the ``problem setup'' heading makes
%%   the next ``results'' heading look orphaned. So I put it back.}

\subsubsection{Problem setup}
\label{ss:linear_gaussian_setup}

Consider a forward model that is linear in its parameters $\theta$,
with a scalar output corrupted by additive Gaussian noise
$\epsilon\sim \CN(0, \sigma_{\epsilon}^2)$:
\begin{eqnarray}
  y_k = G(\theta,d_k)+\epsilon =  \theta d_k +
  \epsilon.\label{e:linear_gaussian_model} 
\end{eqnarray}
The prior on $\theta$ is $\CN(s_0, \sigma_0^2)$ and the design
parameter is $d\in\[d_L,d_R\]$.  The resulting inference
problem on $\theta$ has a conjugate Gaussian structure, such that all
subsequent posteriors are Gaussian with mean and variance given by
\begin{eqnarray}
  \(s_{k+1},\sigma_{k+1}^2\) =
  \(\frac{\frac{y_k/d_k}{\sigma_{\epsilon}^2/d_k^2}+\frac{s_k}{\sigma_k^2}} 
  {\frac{1}{\sigma_{\epsilon}^2/d_k^2}+\frac{1}{\sigma_k^2}},
  \frac{1}{\frac{1}{\sigma_{\epsilon}^2/d_k^2}+\frac{1}{\sigma_k^2}}\).
  \label{e:linear_gaussian_analytic_inference}
\end{eqnarray}
Let us consider the design of $N=2$ experiments, with prior parameters
$s_0=0$ and $\sigma^2_0=9$, noise variance $\sigma^2_{\epsilon}=1$, and
design limits $d_L=0.1$ and $d_R=3$.
%% \todo{I changed these to
%%   $\sigma^2$ for better correspondence with the figures later.} 
%
The Gaussian posteriors in this problem---i.e., the belief
states---are completely specified by values of the mean and variance;
hence we may designate $x_{k,b}=(s_k,\sigma_k^2)$. We call this parametric
representation the ``analytical method,''
% XH: 'analytical', or 'analytic'?
as it also allows inference
to be performed exactly using
\eqref{e:linear_gaussian_analytic_inference}.  The analytical method
will be compared to the adaptive-grid representation of the belief
state (along with its associated inference procedure) described in
Section~\ref{ss:belief_state_representation}.
%% \todo{Should we say
%%   something about how inference is performed when using the grid
%%   representation? Or is it obvious?}
% XH: We described the entire grid approach, including adaptation in
% inference, in section 3.3. I feel we don't need to repeat it here
% and let the reader refer to Section 3.3? Slight wording change to
% try to be slightly more descriptive.
In this example, the adaptive
grids use 50 nodes. There is no physical state $x_{k,p}$; we
simply have $x_k = x_{k,b}$.

Our goal is to infer $\theta$. The stage and terminal reward functions are:
\begin{eqnarray}
  g_k(x_k, y_k, d_k) &=& 0, \ \ k \in \{0,1\} \nonumber \\
  g_N(x_N) &=& \DKL \(f_{\theta|I_N} \, || \, f_{\theta|I_0} \)
  -2 \( \ln \sigma_N^2 - \ln 2 \)^2.  \nonumber \label{e:linear_gaussian_terminal}
\end{eqnarray}
The terminal reward is thus a combination of information gain in
$\theta$ and a penalty for deviation from a particular log-variance target.
The latter term increases the difficulty of this problem by moving the
outputs of the optimal policy away from the design space boundary;
doing so helps avoid the fortuitous construction of successful
policies.\footnote{Without the second term in the terminal reward, the
  optimal policies will always be those that lead to the highest
  achievable signal, which occurs at the $d_k=3$ boundary. Policies
  that produce such boundary designs can be realized even when the
  overall value function approximation is poor. Nothing is wrong with
  this situation \textit{per se}, but adding the second term leads to
  a more challenging test of the numerical approach.}
%% \todo{Check that the footnote rewording is okay.}
Following the discussion in
Section~\ref{s:policy_representation}, we approximate the value
functions using features $\phi_{k,i}$ (in \eqref{e:linear_arch}) that
are polynomials of degree two or less in the posterior mean and
log-variance: 1, $s_k$, $\ln(\sigma_k^2)$, $s_k^2$,
$\ln^2(\sigma_k^2)$, and $s_k\ln(\sigma_k^2)$. When using the grid
representation of the belief state, the values of the features are
evaluated by trapezoidal integration rule. The terminal KL divergence
is approximated by first estimating the mean and variance, and then
applying the analytical formula for KL divergence between
Gaussians. 
% Since we know the posteriors \textit{should} be Gaussian in
% this example, these approximations are expected to be quite accurate.
% YM: This sentence may be confusing if readers haven't seen the
% second example yet. Better to skip for now?
% XH: Okay.
The ADP approach uses $L=3$ policy updates (described in Section~\ref{ss:ell});
%\todo{check language okay?}
% iterations of state measure update 
these updates are conducted using regression points that, for the
first iteration ($\ell=1$), are generated entirely via
exploration. The design measure for exploration is chosen to be $d_k\sim\CN(1.25,
0.5^2)$ in order to have a
wide coverage of the design space.\footnote{Designs proposed outside
  the design constraints are simply projected back to the nearest feasible
  design; thus the actual design measure is not exactly Gaussian.}
%% \todo{What is the design measure leading to the
%%   exploration policy?} 
Subsequent iterations use regression with a mix of 30\%
exploration samples and 70\% exploitation samples. At each iteration,
we use 1000 regression points for the analytical method and 500
regression points for the grid method.
%\todo{Why 1000 versus 500? Is this just because grid is more expensive?}
% XH: actually, yes :)

To compare the policies generated via different numerical methods, we
apply each policy to generate 1000 simulated trajectories. Producing a
trajectory of the system involves first sampling $\theta$ from its
prior, applying the policy on the inital belief state $x_0$ to
obtain the first design $d_0$, drawing a sample $y_0$ to simulate the outcome
of the first experiment, updating the belief state to
$x_1 = \theta | y_0, d_0, I_0$, applying the policy to $x_1$ in order
to obtain $d_1$, and so on. The mean reward over all these
trajectories is an estimate of the expected total reward
\eqref{e:sOED_objective}. But evaluating the reward requires some
care. Our procedure for doing so is summarized in
Algorithm~\ref{a:linear_gaussian_fair_evaluator}. Each policy is first
applied to belief state trajectories computed using the \textit{same} state
representation (i.e., analytical or grid) originally used to construct
that policy; this yields sequences of designs and observations.  Then,
to assess the reward from each trajectory, inference is performed on
the associated sequence of designs and observations using a
\textit{common} assessment framework---the analytical method in this
case---regardless of how the trajectory was produced. This process
ensures a fair comparison between policies, where the designs are
produced using the ``native'' belief state representation for which
the policy was originally created, but all final trajectories are
assessed using a common method.
%
% [[We should note that the final policies produced by each method are
% themselves random, due stochasticity in each numerical method (e.g.,
% stochastic approximation, Monte Carlo sampling of regression
% points). But this variation is overall quite small relative to other
% sources of uncertainty in the problem, and randomly sampling over
% this policy distribution would be an extremely expensive
% undertaking.]] \todo{Maybe omit this [[bracketed]] part? It is a
%   subtle point.}
% XH: agreed, let's leave it out for now.

% there is also a distribution for the final policy due to
% the randomness involved in the numerical methods (e.g., repeating the
% algorithm to construct the policy would not result in exactly the same
% policy, simply due to the different random numbers being used in
% simulations). Accounting for this policy distribution is an extremely
% expensive undertaking, and is presently not included in our numerical
% assessment.

% ; instead, only a single policy realization
% is used to generate all 1000 trajectories. A more comprehensive study
% by repeating the policy constructing algorithm many times may be
% conducted in the future, although such an undertaking would be
% extremely expensive.

\begin{algorithm}[h]
  \caption{Procedure for assessing policies by simulating
    trajectories.}
  \label{a:linear_gaussian_fair_evaluator}
  \begin{algorithmic}[1]
    %  \SetAlgoLined
    \STATE{\textbf{Select a ``native'' belief state representation} to generate
      policy (e.g., analytical or grid)}
    \STATE{\textbf{Construct policy} $\pi$ using this belief state representation}
    \FOR{$q=1,\ldots,n_{\textrm{trajectories}}$}
    \STATE{\textbf{Apply policy} using the native belief state
      representation: Sample $\theta$ from the prior. Then evaluate
      $d_k$ by applying the constructed policy $\mu_k(x_k)$, sample
      $y_k$ from the likelihood $p(y_k | \theta, d_k, I_k)$ to simulate
      the experimental outcome, and
      evaluate $x_{k+1} = \CF_k(x_k,y_k,d_k)$, for $k=0,\ldots,N-1$.}

    \STATE{\textbf{Evaluate rewards via a common assessment framework:}
      Discard the belief state sequence $\{x_k\}_k$. Perform inference on
      the $\{d_k\}_k$ and $\{y_k\}_k$ sequences from the current trajectory and evaluate the
      total reward using a chosen common assessment framework.}
    \ENDFOR
  \end{algorithmic}
\end{algorithm}

\subsubsection{Results}

\begin{figure}[htb]
  \centering
  \mbox{\subfigure[Analytical method]
    {\includegraphics[width=0.32\textwidth]{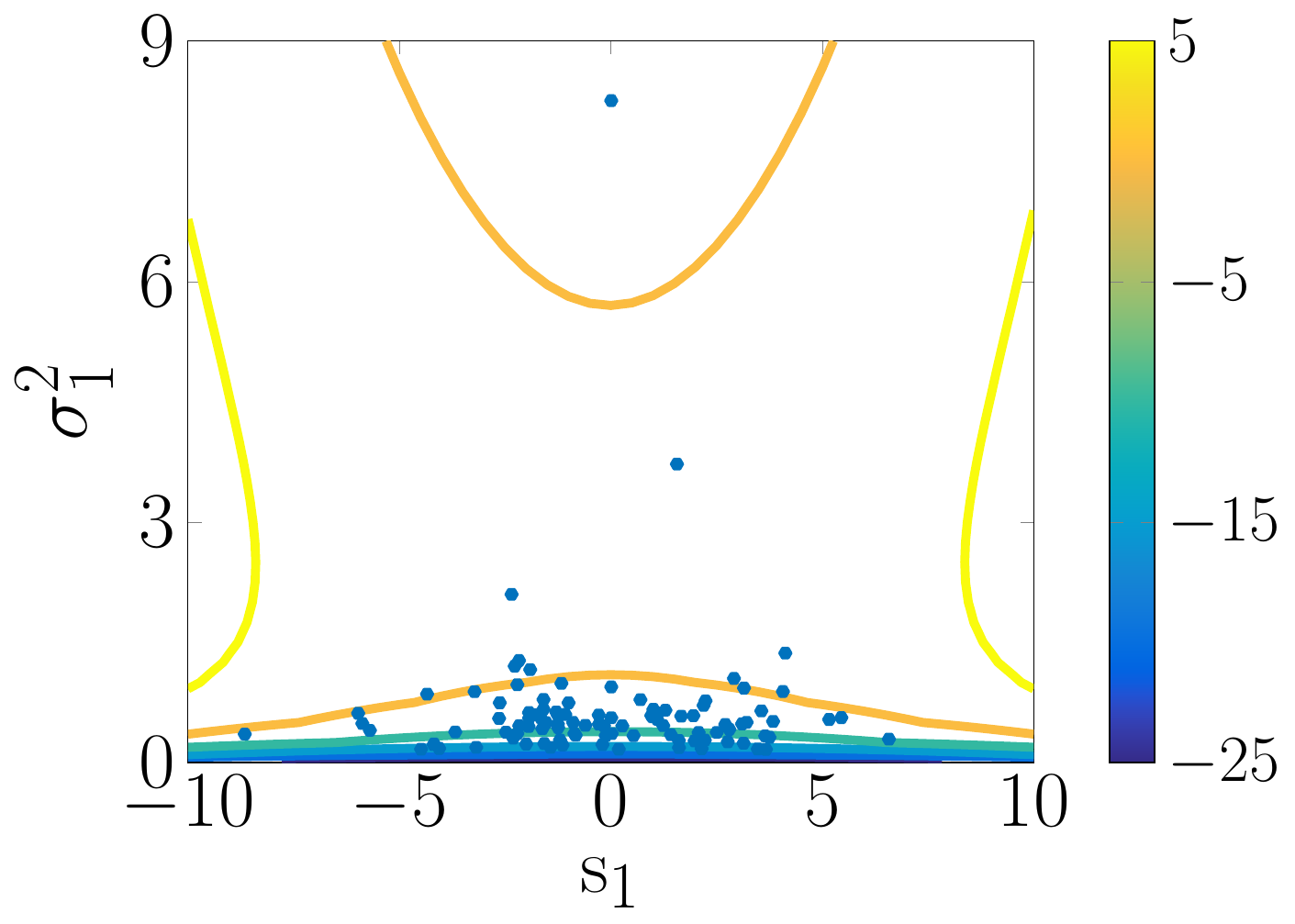}
      \includegraphics[width=0.32\textwidth]{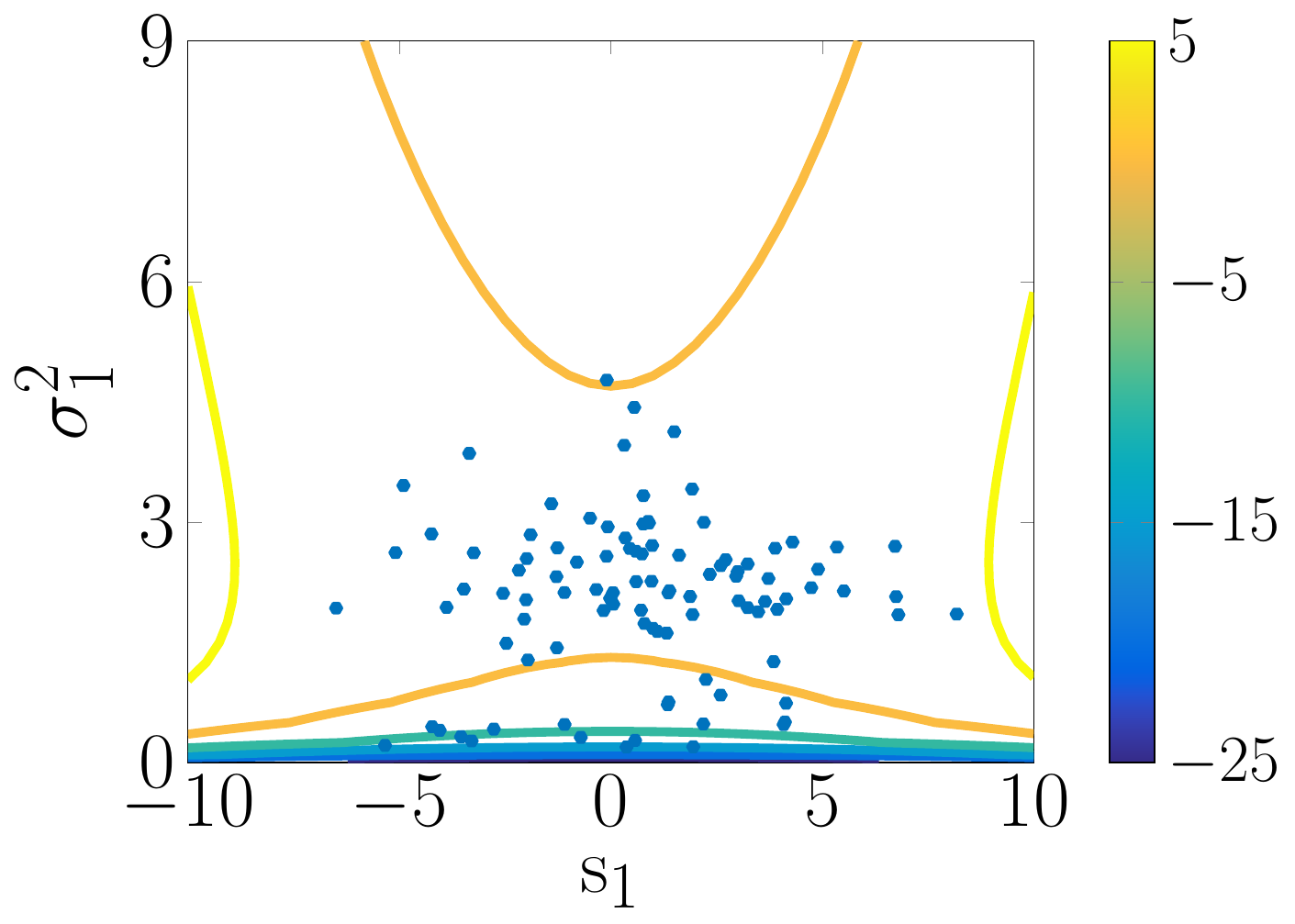}
      \includegraphics[width=0.32\textwidth]{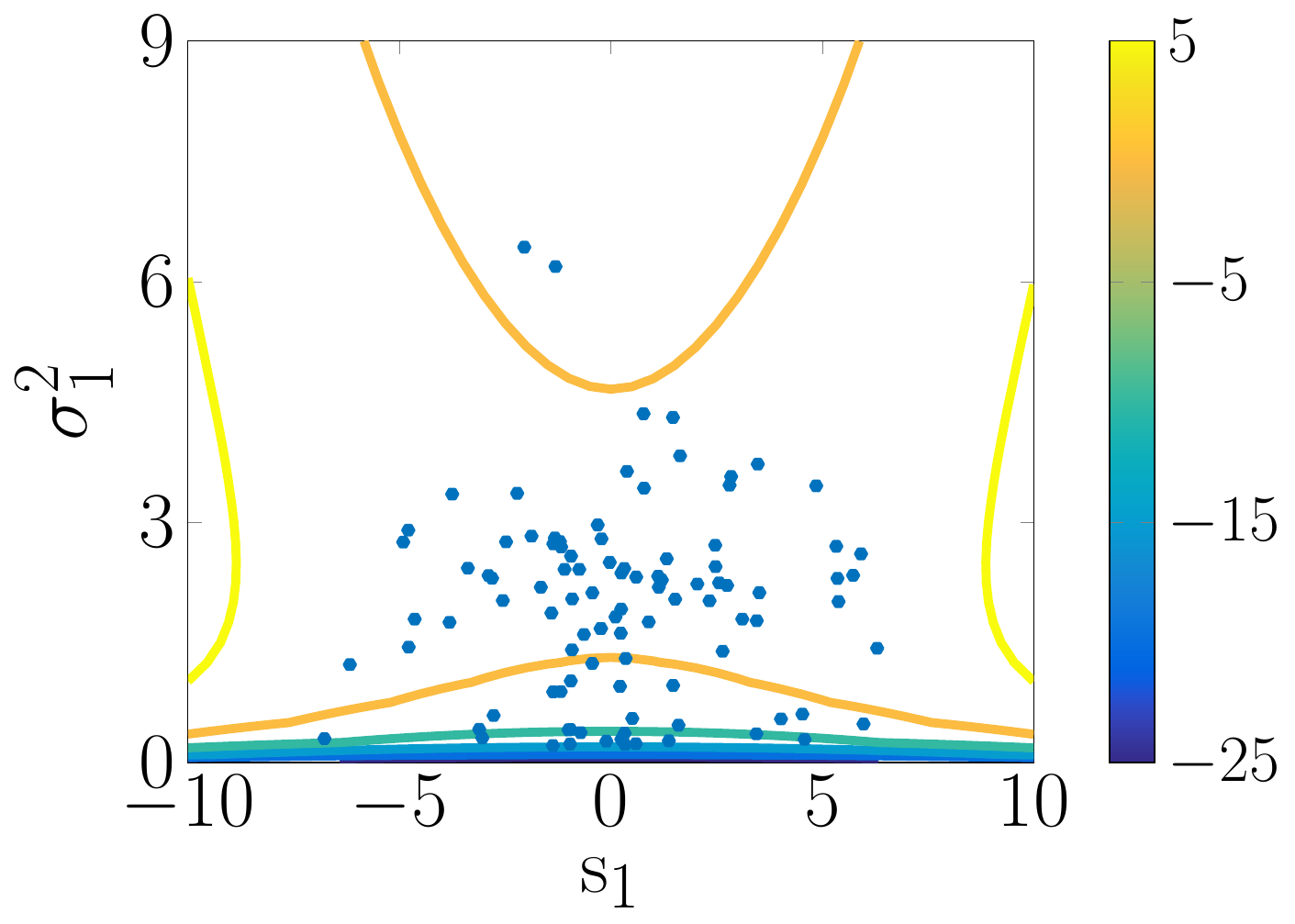}
    }
  }
  \mbox{\subfigure[Grid method]
    {\includegraphics[width=0.32\textwidth]{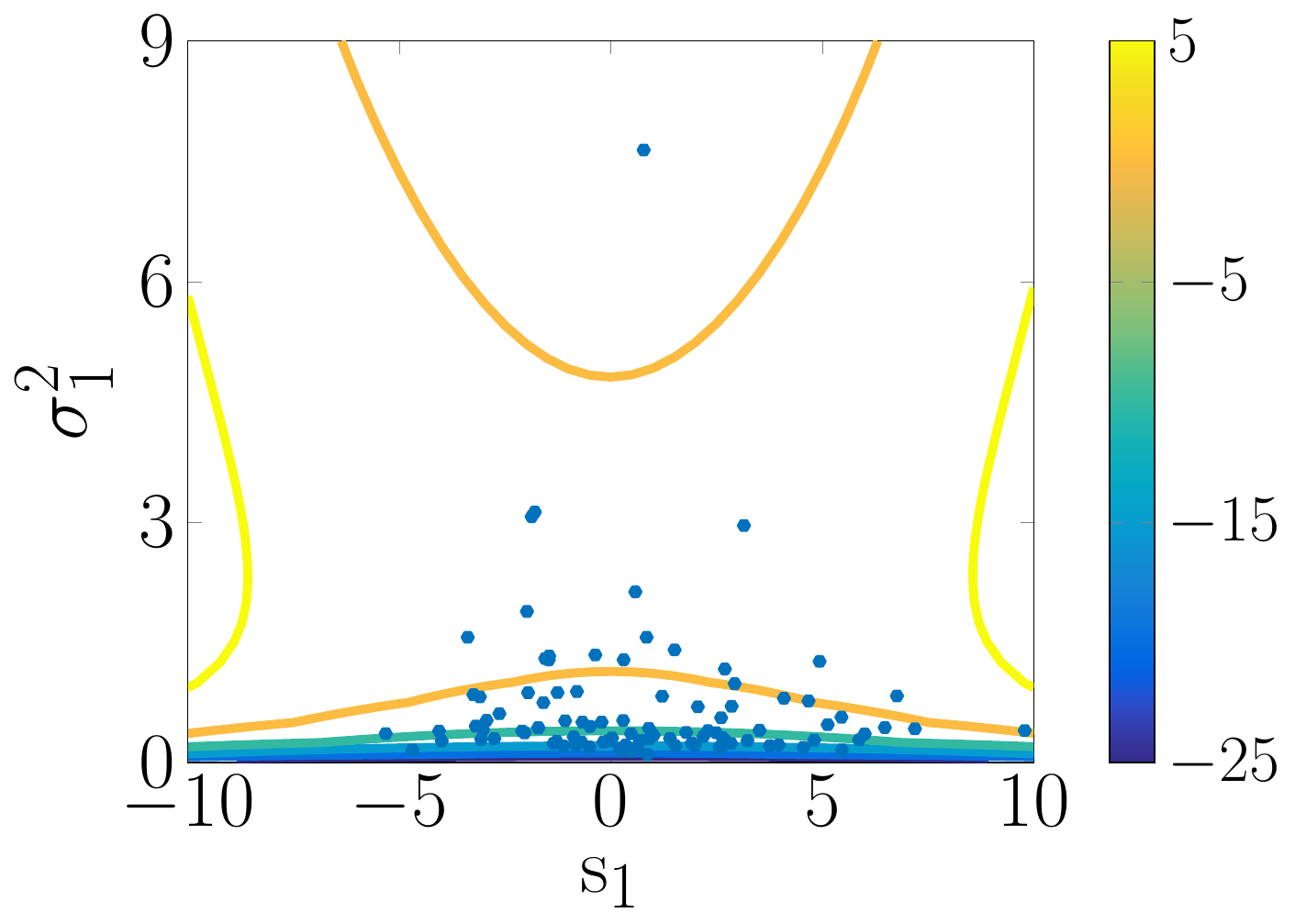}
      \includegraphics[width=0.32\textwidth]{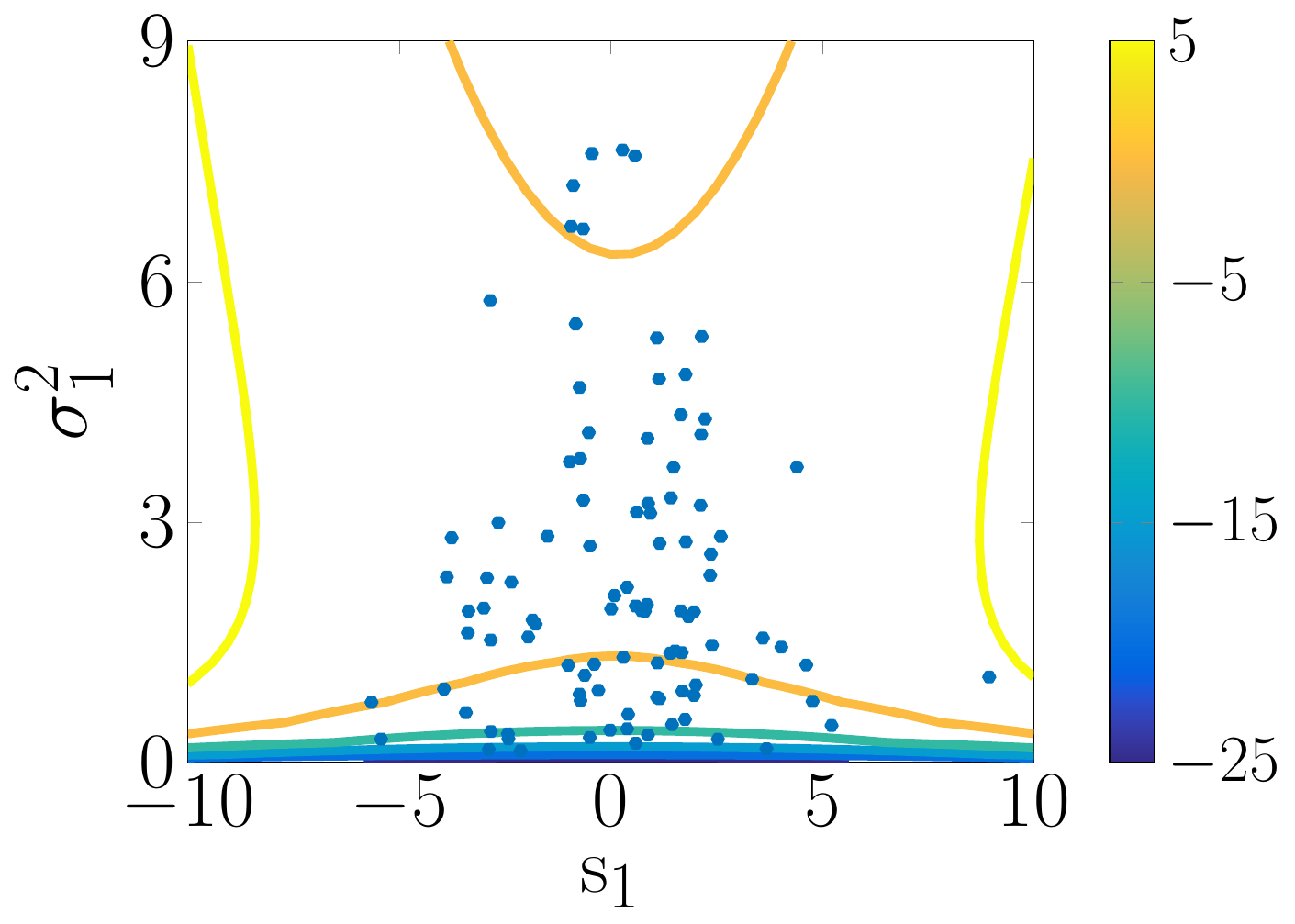}
      \includegraphics[width=0.32\textwidth]{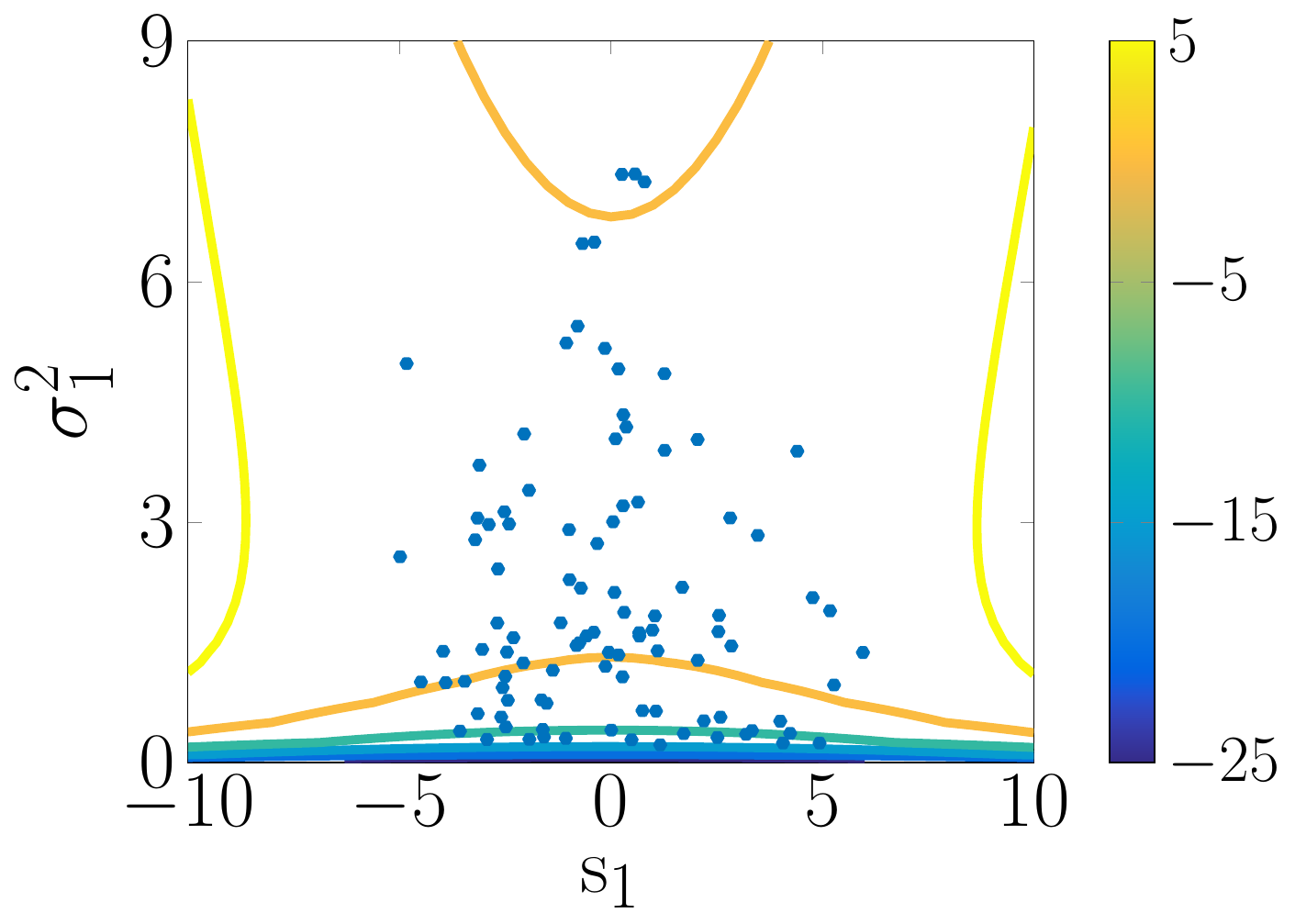}
    }
  }
  % \mbox{\subfigure[Map method]
  %   {\includegraphics[width=0.32\textwidth]{figures/linear_gaussian/state_samples_VFA_iter0_linear_gaussian_map_exploration_map_only_perturb.pdf}
  %     \includegraphics[width=0.32\textwidth]{figures/linear_gaussian/state_samples_VFA_iter1_linear_gaussian_map_exploration_map_only_perturb.pdf}
  %     \includegraphics[width=0.32\textwidth]{figures/linear_gaussian/state_samples_VFA_iter2_linear_gaussian_map_exploration_map_only_perturb.pdf}
  %   }
  %   }
  \caption{Linear-Gaussian problem: contours represent $\tJ_1(x_1)$
    and blue points are the regression points used to build these
    value function approximations. The left, middle, and right columns
    correspond to $\ell=1$, $2$, and $3$, respectively.}
  \label{f:linear_gaussian_tJ1}
\end{figure}

\begin{figure}[htb]
  \centering
  \mbox{\subfigure[Analytical method]
    {\includegraphics[width=0.32\textwidth]{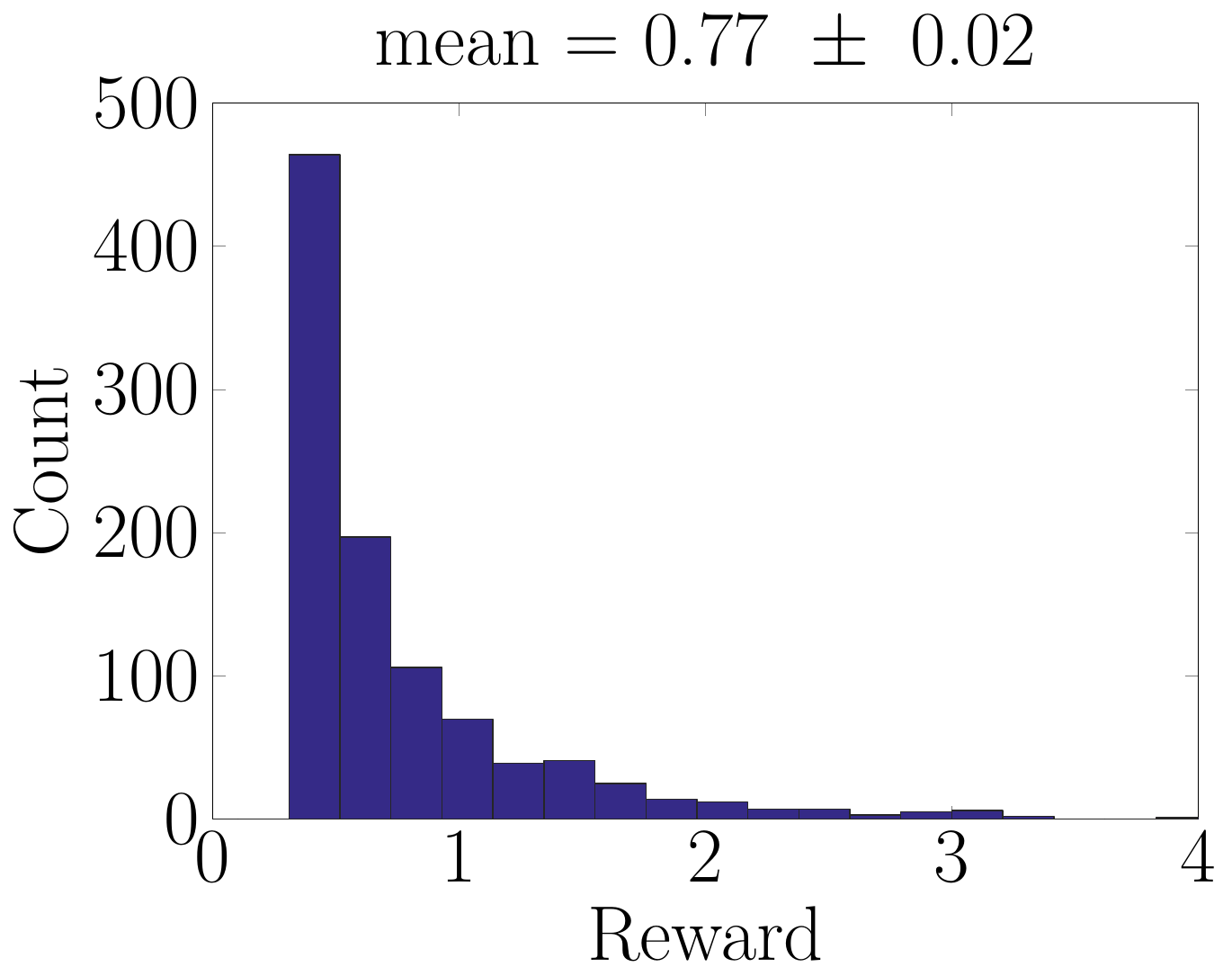}
      \includegraphics[width=0.32\textwidth]{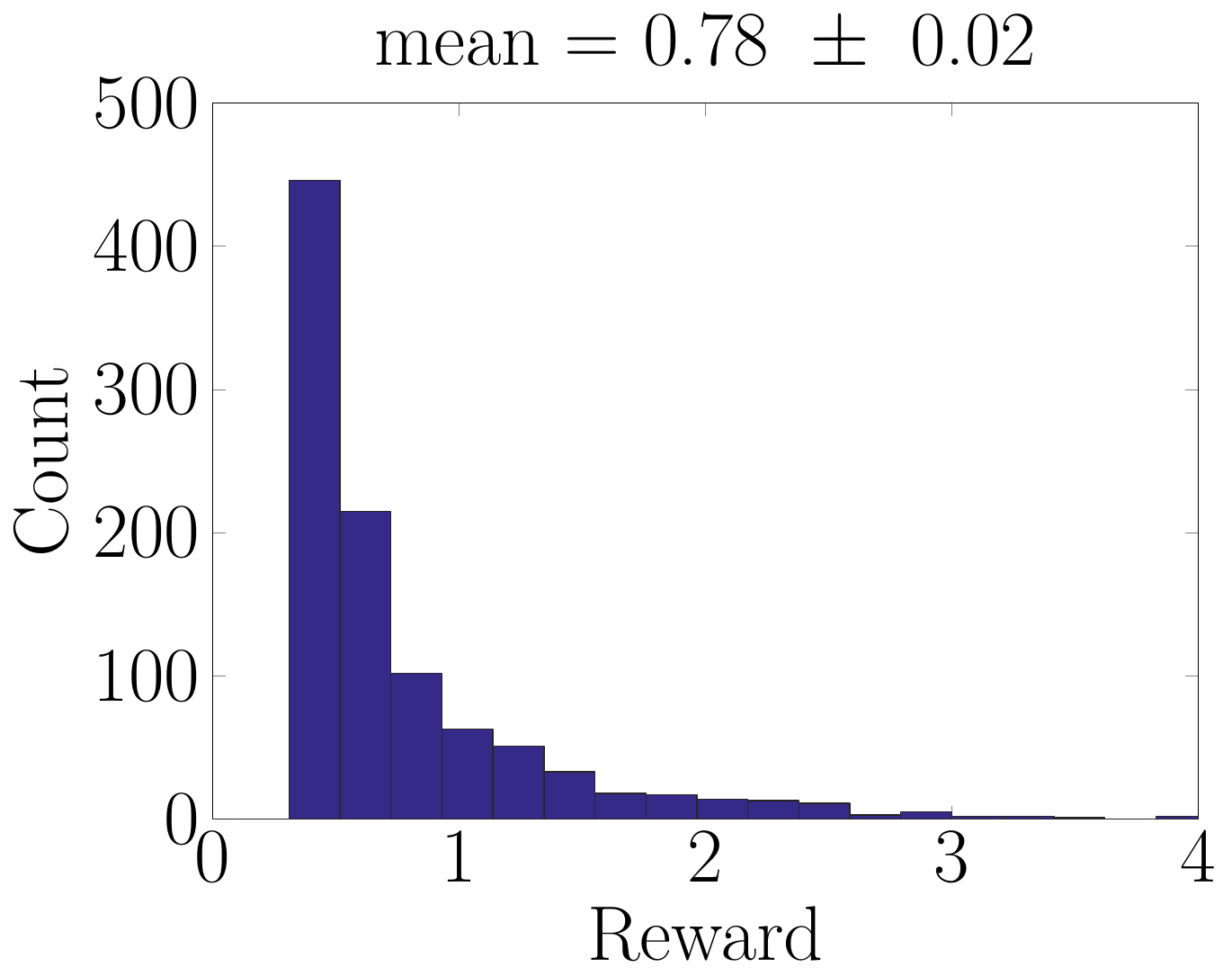}
      \includegraphics[width=0.32\textwidth]{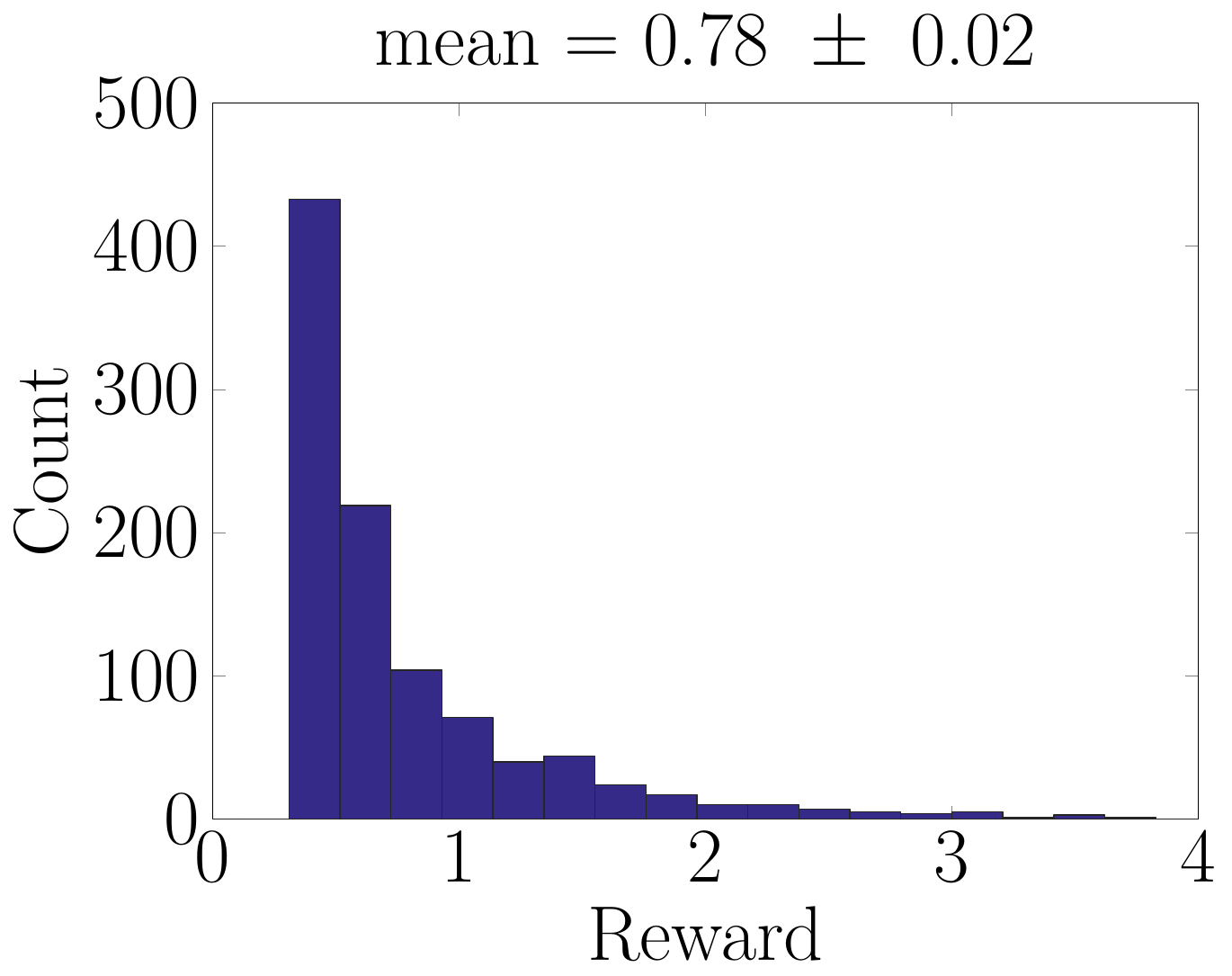}
    }
  }
  \mbox{\subfigure[Grid method]
    {\includegraphics[width=0.32\textwidth]{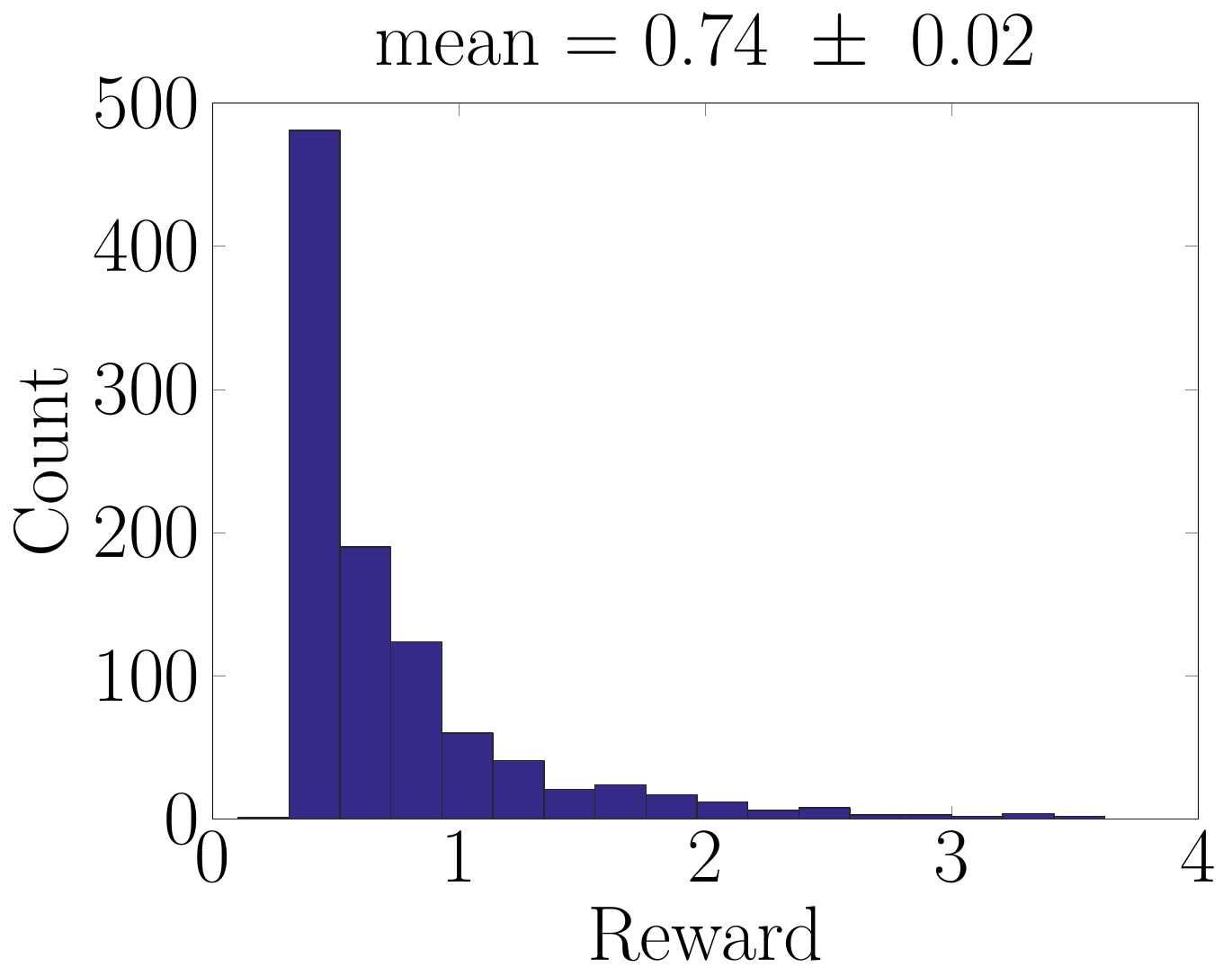}
      \includegraphics[width=0.32\textwidth]{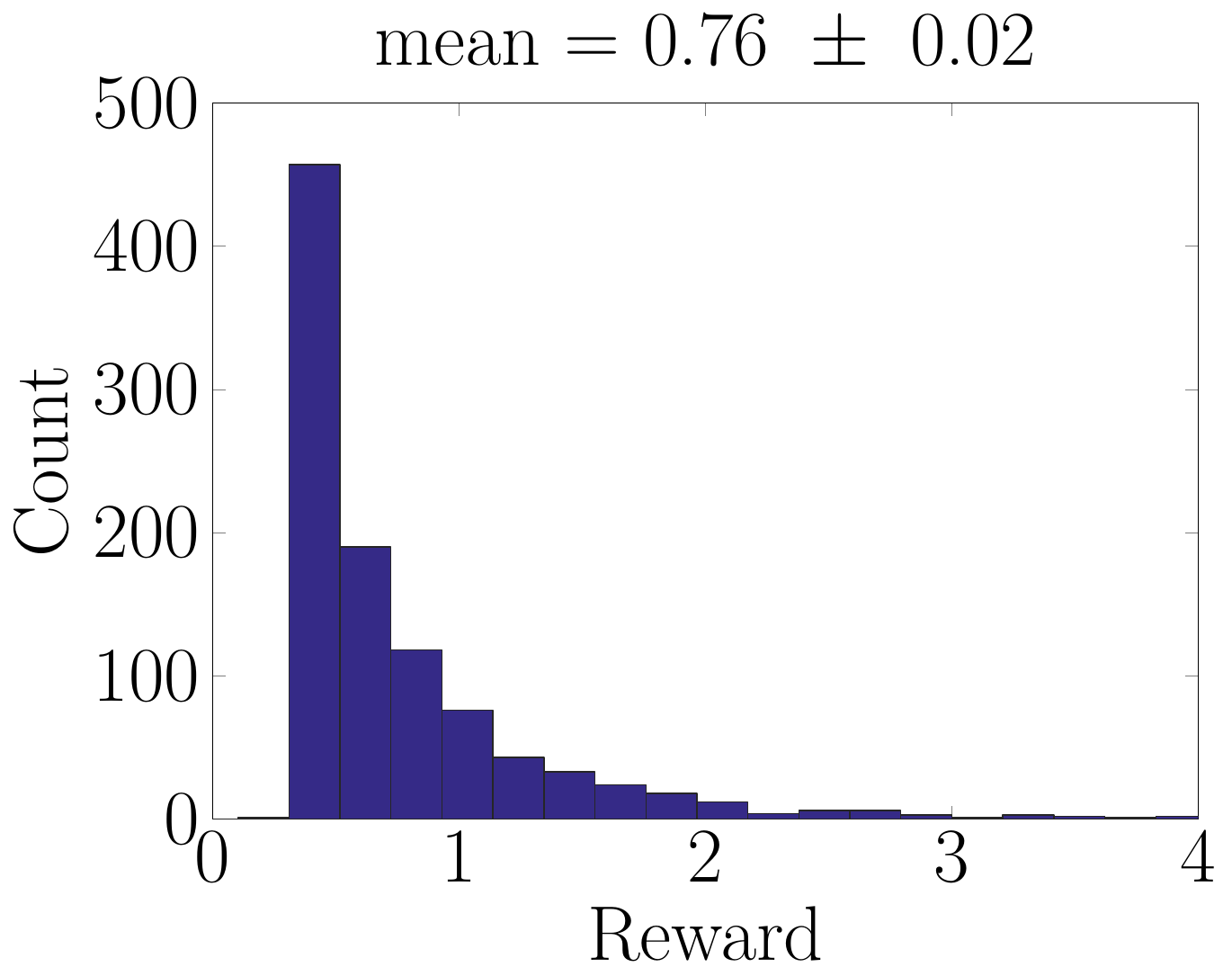}
      \includegraphics[width=0.32\textwidth]{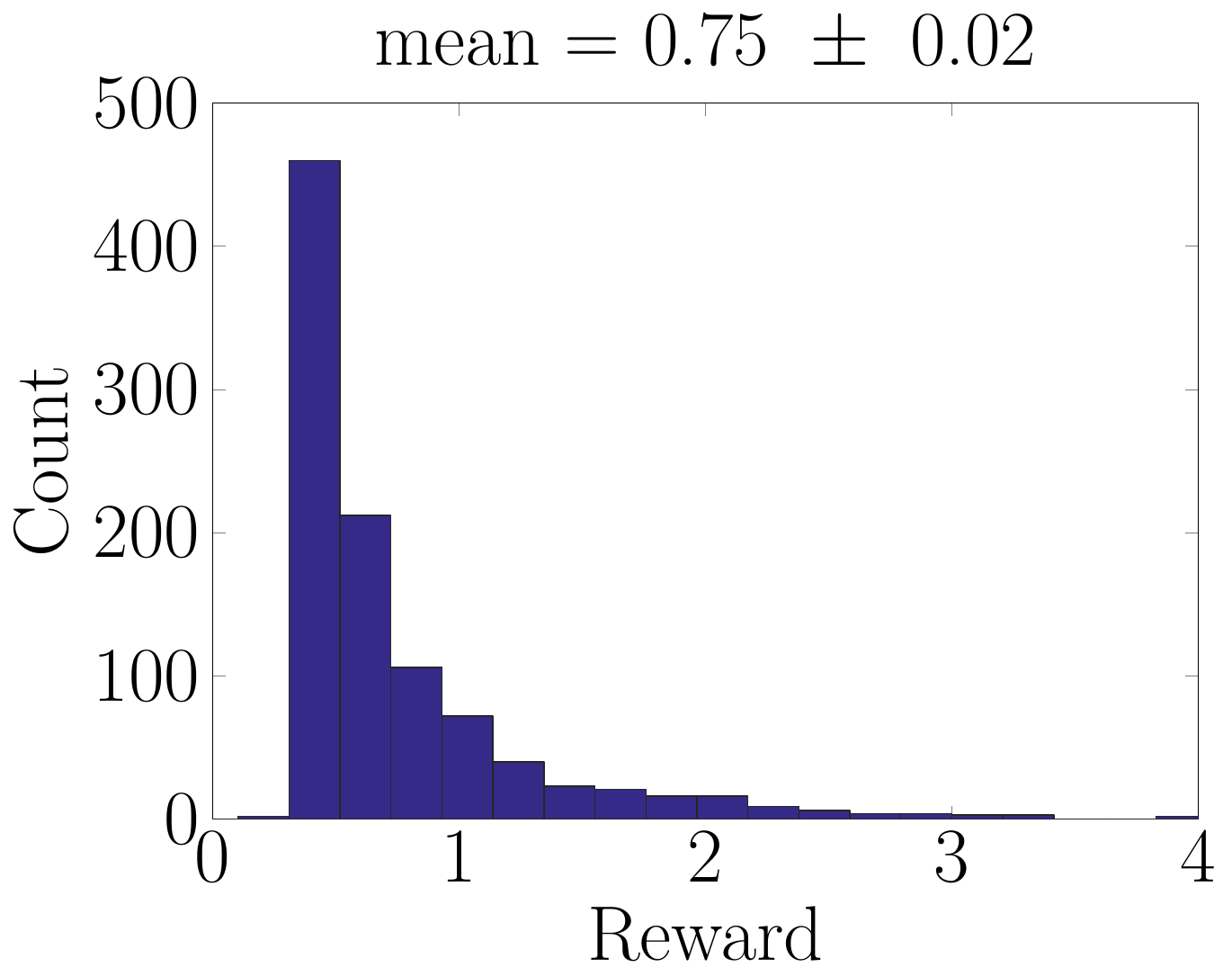}
    }
  }
  % \mbox{\subfigure[Map method]
  %   {\includegraphics[width=0.32\textwidth]{figures/linear_gaussian/rewards_VFA_iter0_linear_gaussian_map_exploration_map_only_perturb.pdf}
  %     \includegraphics[width=0.32\textwidth]{figures/linear_gaussian/rewards_VFA_iter1_linear_gaussian_map_exploration_map_only_perturb.pdf}
  %     \includegraphics[width=0.32\textwidth]{figures/linear_gaussian/rewards_VFA_iter2_linear_gaussian_map_exploration_map_only_perturb.pdf}
  %   }
  % }
  \caption{Linear-Gaussian problem: total reward histograms from 1000
    simulated trajectories. The left, middle, and right columns
    correspond to $\ell=1$, 2, and 3, respectively.} %The plus-minus
  % quantity is one standard error. The true optimal expected reward is $\approx 0.783289$.
  \label{f:linear_gaussian_rewards}
\end{figure}

Since this example has a horizon of $N=2$, we only need to construct
the value function approximation $\tJ_1$. $J_2 = g_2$ is the terminal
reward and can be evaluated
directly. Figure~\ref{f:linear_gaussian_tJ1} shows contours of
$\tJ_1$, along with the regression points used to build the
approximation, as a function of the belief state (mean and variance).
We observe excellent agreement between the analytical and grid-based
methods. We also note a significant change in the distribution of
regression points from $\ell=1$ (regression points obtained only from
exploration) to $\ell=2$ (regression points obtained from a mixture of
exploration and exploitation), leading to a better approximation of
the state measure induced by the optimal policy. The regression points
appear to be grouped more closely together for $\ell=1$ even though
they result from exploration; this is because exploration in fact
covers a large region of $d_k$ space that leads to small values of
$\sigma_k^2$. In this simple example, our choice of design
measure for exploration did not have a particularly negative impact on the expected
reward for $\ell=1$ (see
Figure~\ref{f:linear_gaussian_rewards}). However, this situation can
easily change for problems with more complicated value functions and
less suitable choices of the exploration policy.

%\todo{Small point, but vertical axis should be labled $\sigma^2_1$.}

% Histograms for $d_0$ and $d_1$ are shown in
% Figures~\ref{f:linear_gaussian_d0} and~\ref{f:linear_gaussian_d1}.
% While overall the agreement is quite good between the three methods,
% $d_0$ for the grid and map methods $\ell=2$ and $\ell=3$ are
% concentrated slightly to the left compared to those for the analytic
% method. The opposite is true for $d_1$, where the grid and map methods
% $\ell=2$ and $\ell=3$ are slightly to the right compared to those for
% the analytic method. This is because 

\begin{figure}[htb]
  \centering
  \mbox{\subfigure[Analytical method]
    {\includegraphics[width=0.32\textwidth]{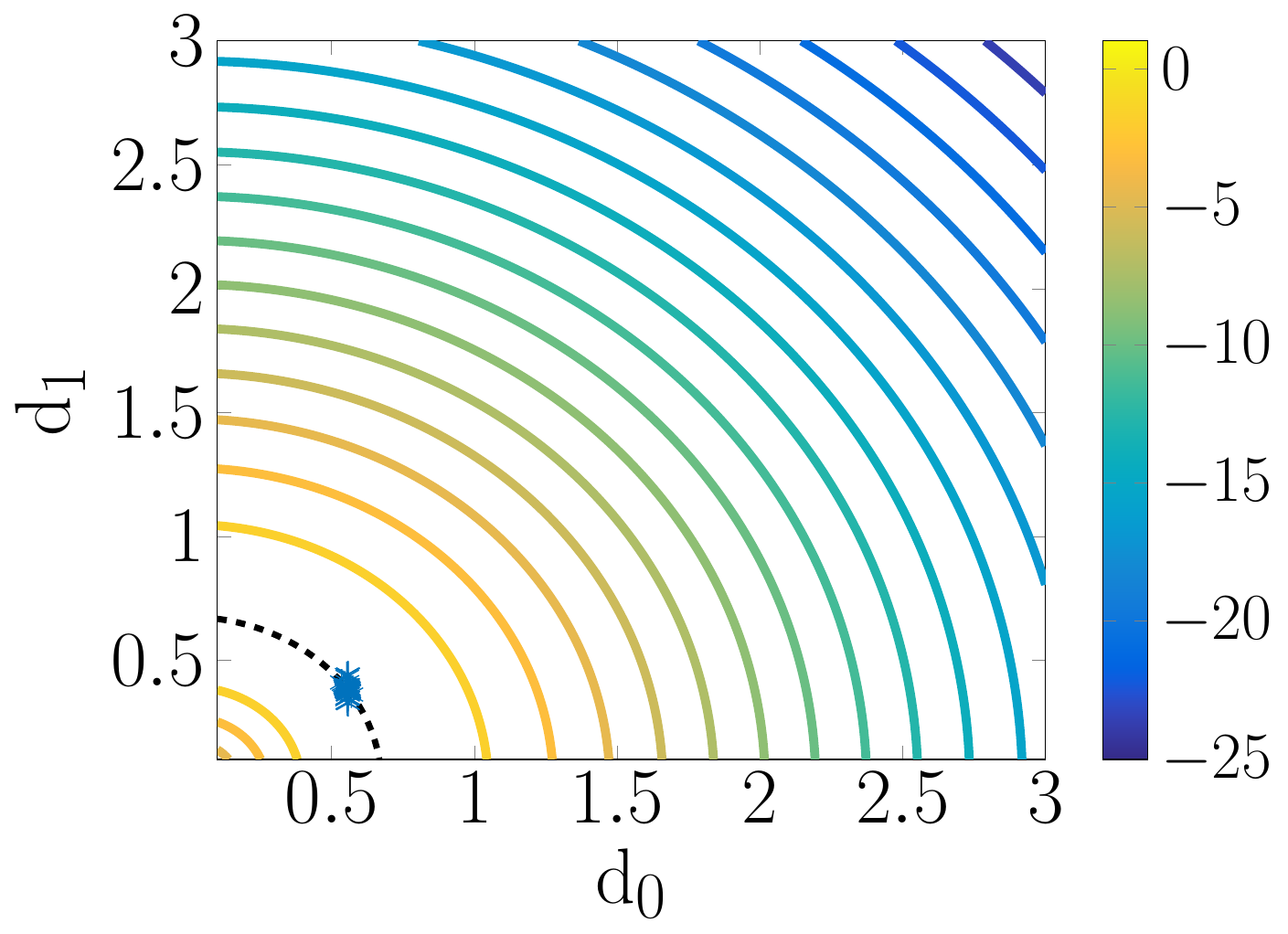}
      \includegraphics[width=0.32\textwidth]{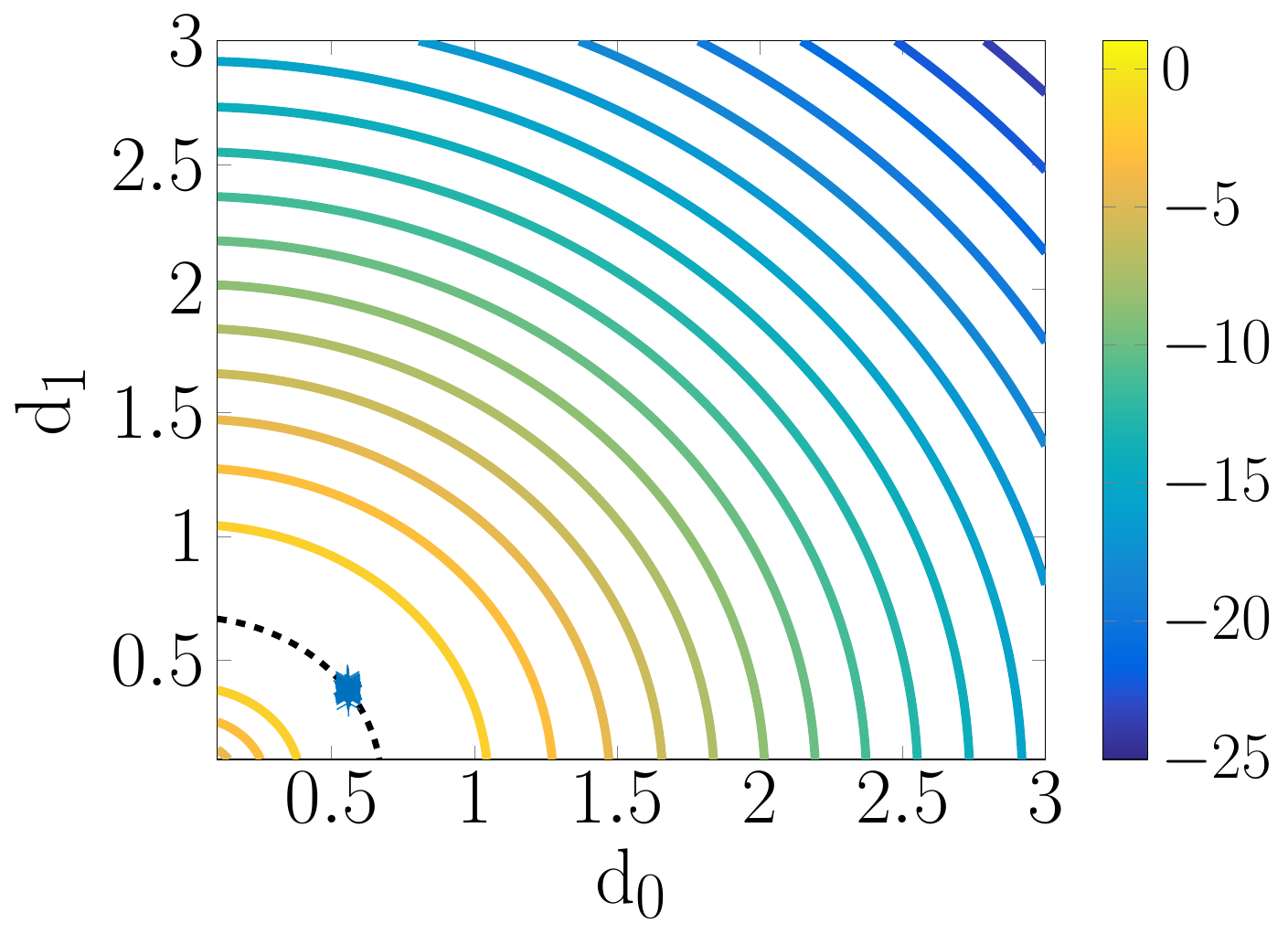}
      \includegraphics[width=0.32\textwidth]{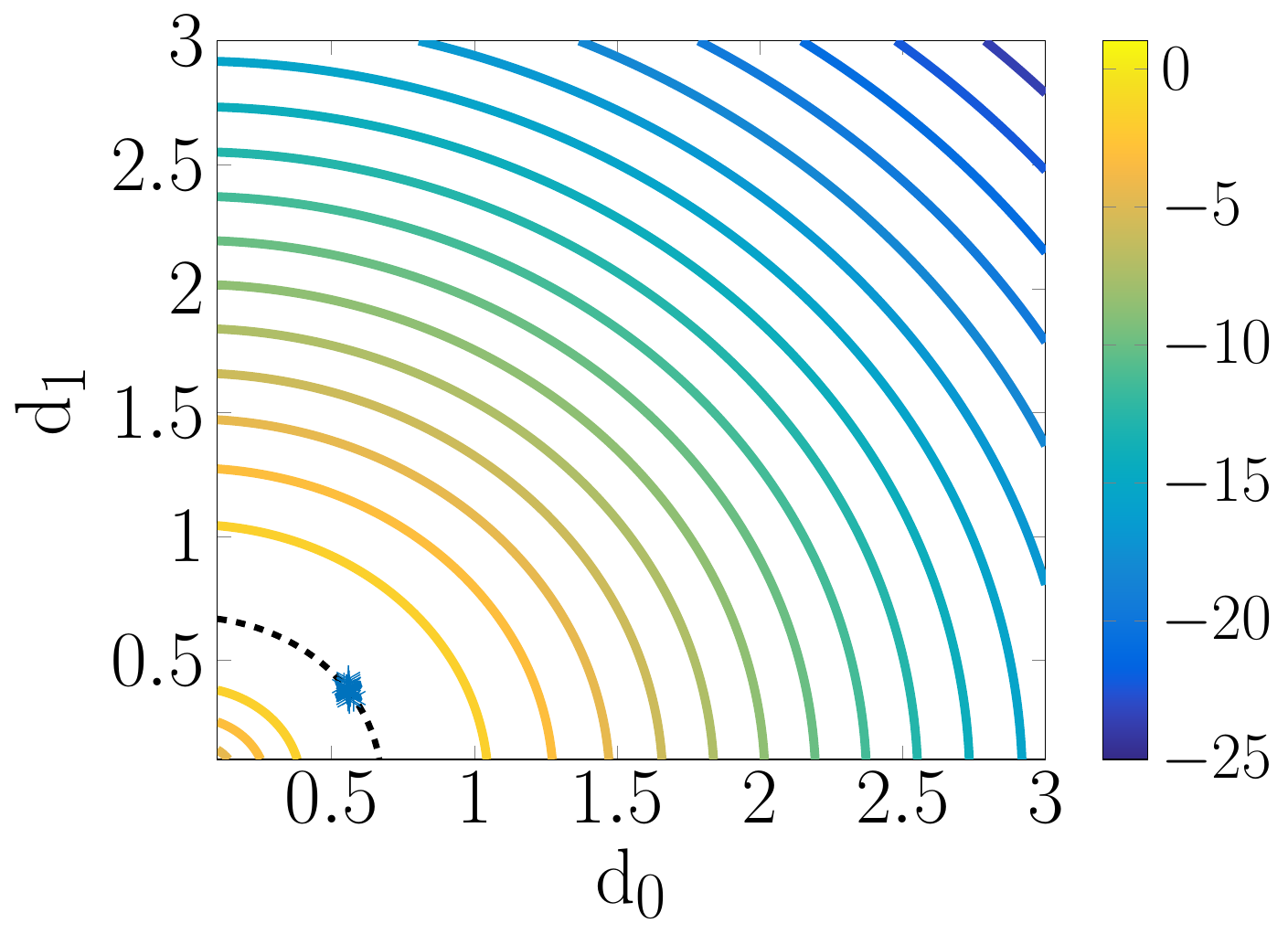}
    }
  }
  \mbox{\subfigure[Grid method]
    {\includegraphics[width=0.32\textwidth]{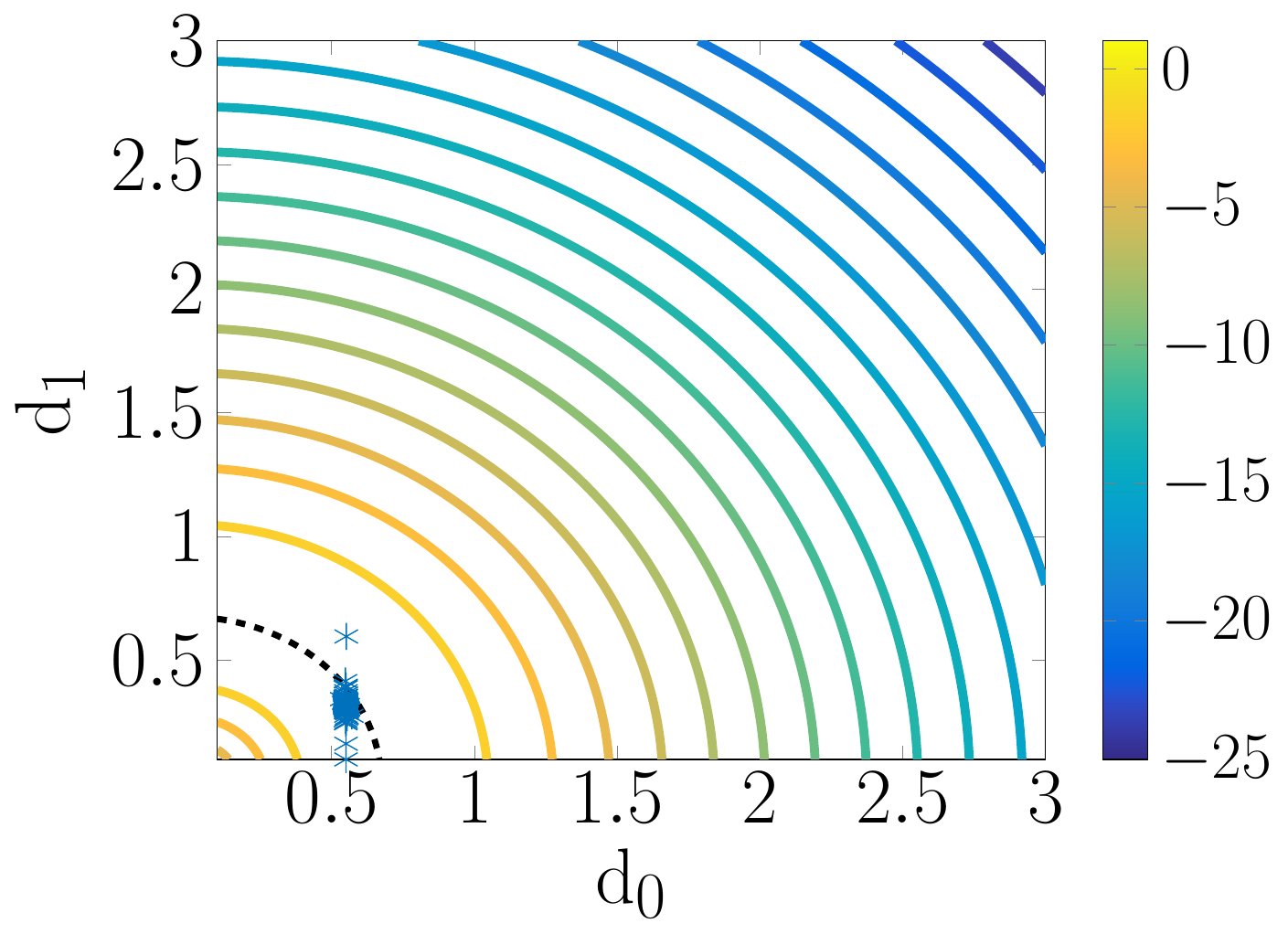}
      \includegraphics[width=0.32\textwidth]{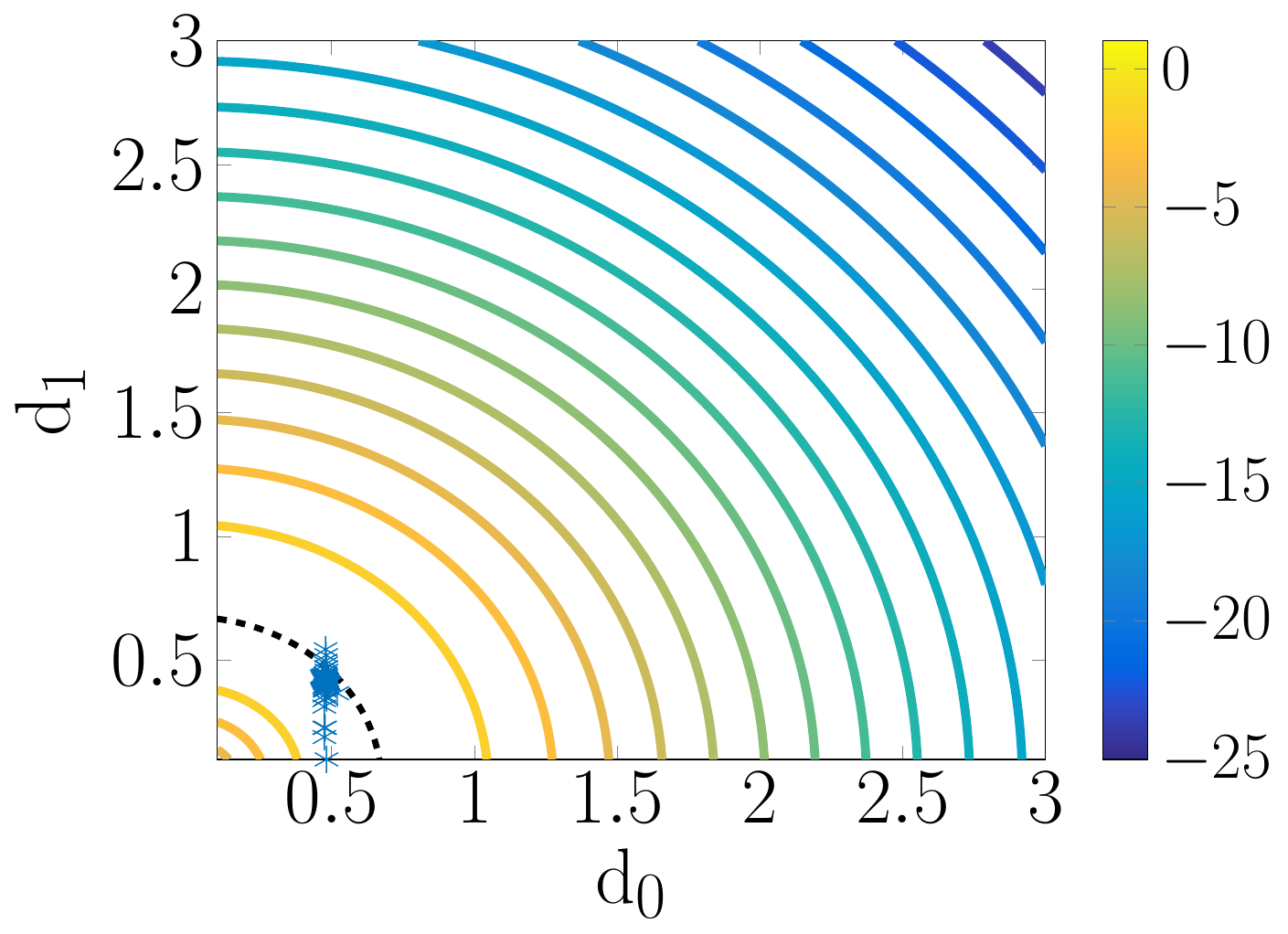}
      \includegraphics[width=0.32\textwidth]{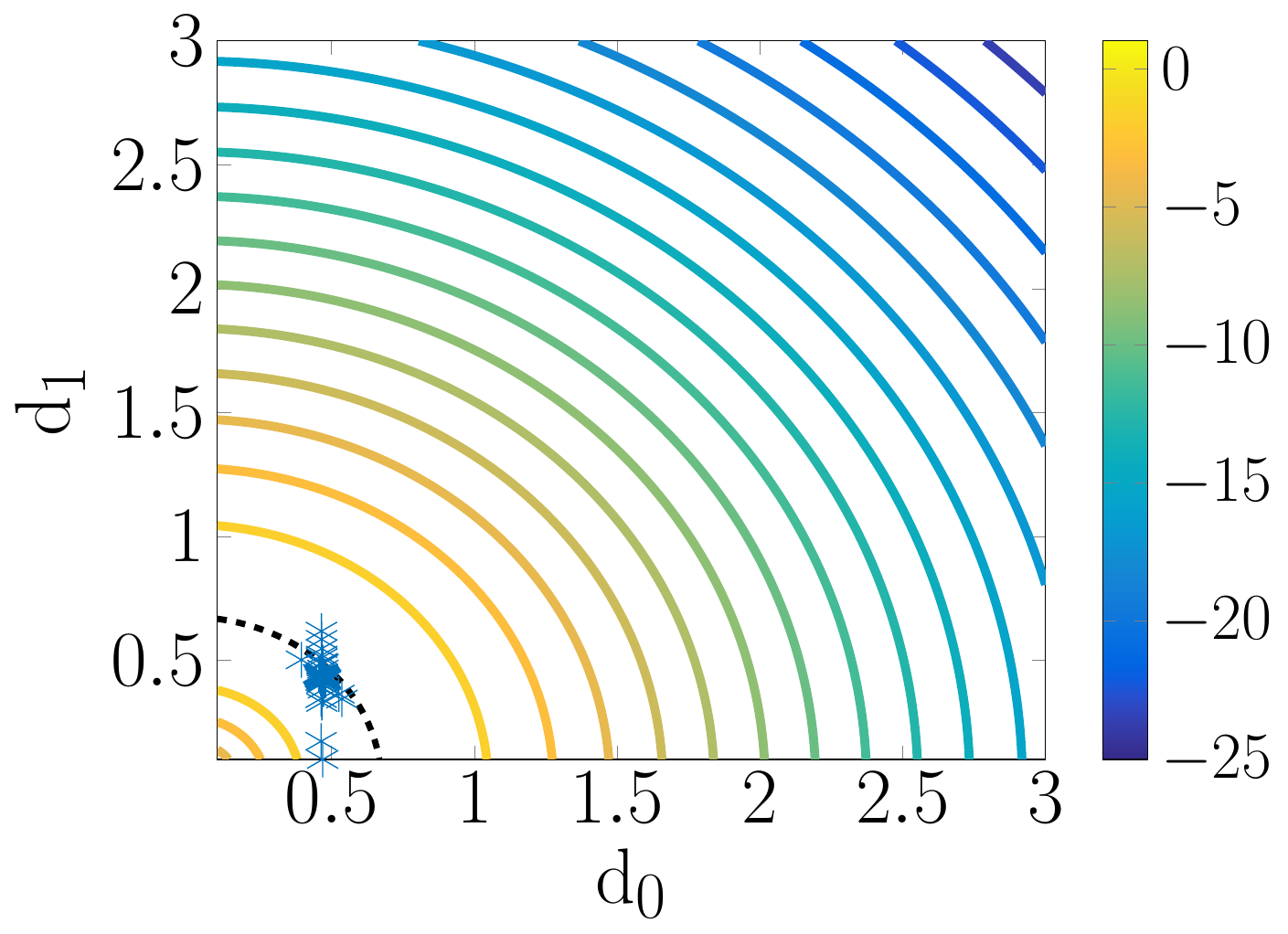}
    }
  }
  % \mbox{\subfigure[Map method]
  %   {\includegraphics[width=0.32\textwidth]{figures/linear_gaussian/d_pairs_VFA_iter0_linear_gaussian_map_exploration_map_only_perturb.pdf}
  %     \includegraphics[width=0.32\textwidth]{figures/linear_gaussian/d_pairs_VFA_iter1_linear_gaussian_map_exploration_map_only_perturb.pdf}
  %     \includegraphics[width=0.32\textwidth]{figures/linear_gaussian/d_pairs_VFA_iter2_linear_gaussian_map_exploration_map_only_perturb.pdf}
  %   }
  %  }
  \caption{Linear-Gaussian problem: scatter plots of design pairs
    $(d_0,d_1)$ from 1000 simulated trajectories, superimposed on
    contours of the exact expected reward function. The black dotted
    line is the locus of exact optimal designs, obtained
    analytically. The left, middle, and right columns correspond to
    $\ell=1$, 2, and 3, respectively.}
  \label{f:linear_gaussian_d_pairs}
\end{figure}

Note that the optimal policy is not unique for this problem, as there
is a natural notion of exchangeability between the two designs $d_0$
and $d_1$. With no stage cost, the overall objective of this problem
is the sum of the expected KL divergence and the expected distance of the final
log-variance to the target log-variance. Both terms only depend on the
final variance, which is determined exactly by the chosen values of
$d_k$ through \eqref{e:linear_gaussian_analytic_inference}---i.e.,
independently of the observations $y_k$. This phenomenon is particular to
the linear-Gaussian problem with constant $\sigma_{\epsilon}^2$: the
design problem is in fact \textit{deterministic}, as the final variance is
independent of the realized values of $y_k$. The optimal policy is
then reducible to a choice of scalar-valued optimal designs
$d_0^{\ast}$ and $d_1^{\ast}$. In other words, batch design will
produce the same optimal designs as sOED for such deterministic
problems since feedback does not add any design-relevant information. Moreover, we can find the exact optimal designs and expected
reward function \textit{analytically} in this problem; a derivation
is given in Appendix B of~\cite{Huan2015}.
% , with
% \begin{eqnarray}
%   d_0^{\ast 2} + d_1^{\ast 2} =
%   \frac{1}{9}\[\exp\(\frac{18014398509481984 \ln 3 - 
%     5117414861322735}{9007199254740992}\) -
%   1\],
% \end{eqnarray}
% and
% \begin{eqnarray}
%   U(d_0^{\ast},d_1^{\ast}) \approx 0.783289.
% % 0.78328869838813691695023067040843
% \end{eqnarray}
% Indeed, there is a ``front'' of optimal designs, as there are
% different combinations of $d_0$ and $d_1$ that together lead to the
% underlying optimal final variance.  
Figure~\ref{f:linear_gaussian_d_pairs} plots the exact
expected reward function over the design space, and overlays pairwise scatter plots of $(d_0,d_1)$ from 1000
trajectories simulated with the numerically-obtained sOED policy. The
dotted black line represents the locus of exact optimal designs. The
symmetry between the two experiments is evident, and the optimizer may
hover around different parts of the optimal locus in different
cases. The expected reward is relatively flat around the optimal
design contour, however, and thus all these methods perform
well.
%% \todo{Something to check when you read through: we alternate
%%   between talking about expected utility and expected reward. For
%%   consistency, should we stick with reward? Also, is it clear to the
%%   reader why $\tJ_1$ is an expected utility/reward, in this problem?}
% XH: yes let's switch to expected reward then, but I'll check at
% formulation to make sure we also use expected utility once so that
% the connection to decision theory is established.
% The expected
% utility surface also appears flat around the optimal
% front, thus we expect all these methods to have performed fairly
% well. 
Histograms of total reward and their means from the 1000 trajectories
are presented in Figure~\ref{f:linear_gaussian_rewards} and
Table~\ref{t:linear_gaussian_rewards}. The exact optimal reward value
is $U(\pi^\ast) \approx 0.7833$; the numerical sOED approaches agree with this
value very closely, considering the Monte Carlo standard error. 
% , with grid and map methods
% exhibiting slightly lower mean values but all 
In contrast, the exploration policy produces a much lower expected
reward of $U(\pi^{\textrm{explore}}) = -8.5$.

\begin{table}[htb]
  \caption{Linear-Gaussian problem: expected reward values (mean values of
    histograms in Figure~\ref{f:linear_gaussian_rewards}) from 1000
    simulated trajectories. Monte Carlo standard errors are all $\pm
    0.02$. The true optimal expected reward is $\approx 0.7833$.}
\label{t:linear_gaussian_rewards}
\centering
\begin{tabular}{c|ccc}
  \hline  & $\ell = 1$ & $\ell = 2$ & $\ell = 3$ \\ \hline
  Analytical & 0.77 & 0.78 & 0.78 \\
  Grid & 0.74 & 0.76 & 0.75 \\
%  Map & 0.77 & 0.75 & 0.75 \\
  \hline
\end{tabular}
\end{table}

% The pairwise and marginal kernel density estimates (KDEs) from samples
% used to construct the joint exploration map, and samples generated
% from the resulting map, are shown in
% Figure~\ref{f:linear_gaussian_exploration_map}. Excellent agreement is
% observed between the two sets of KDEs, implying the map is capturing
% the joint distribution well.  As evident by, for example, the pairwise
% KDEs between $d_k$ and $y_k$, the joint distribution is not Gaussian
% even for a linear-Gaussian problem with a Gaussian exploration measure
% on $d_k$.
% % ; this is discussed in Example~\ref{ex:d} form a
% % theoretic perspective.

% \begin{figure}[htb]
%   \centering
%   \mbox{\subfigure[Samples used for map construction]
%     {\includegraphics[width=0.45\textwidth]{figures/linear_gaussian/samples_build_map_exploration_linear_gaussian_map_exploration_map_only_perturb.pdf}
%     }
%   }
%   \mbox{\subfigure[Samples generated from resulting map]
%     {\includegraphics[width=0.45\textwidth]{figures/linear_gaussian/samples_from_map_exploration_linear_gaussian_map_exploration_map_only_perturb.pdf}
%     }
%   }
%   \caption{Linear-Gaussian problem: pairwise KDEs from samples used to
%     construct the exploration joint map and samples generated from the
%     resulting map.}
%   \label{f:linear_gaussian_exploration_map}
% \end{figure}

In summary, we have shown agreement between numerical sOED results and
the true optimal design. Furthermore, we have demonstrated agreement
between the analytical and grid-based methods of representing the
belief state and performing inference, thus establishing credibility
of the grid-based method for the subsequent nonlinear and non-Gaussian
example, where an exact representation of the belief state is no
longer possible.

\subsection{Contaminant source inversion problem}
\label{s:source_inversion_1D}

Consider a situation where a chemical contaminant is accidentally
released into the air. The contaminant plume diffuses and is advected
by the wind. It is crucial to infer the location of the contaminant
source so that an appropriate response can be undertaken. Suppose that
an aerial or ground-based vehicle is dispatched to measure contaminant
concentrations at a sequence of different locations, under a fixed
time schedule. We seek the optimal policy for deciding \textit{where}
the vehicle should take measurements in order to maximize expected
information gain in the source location. Our sOED problem will also
account for hard constraints on possible vehicle movements, as well
movement costs incorporated into the stage rewards.

We use the following simple plume model of the contaminant
concentration at location $z$ and time $t$, given source location
$\theta$:
\begin{eqnarray}
  G(\theta, z, t) = \frac{s}{\sqrt{2 \pi} \(\sqrt{1.2+4Dt}\)}
  \exp \(-\frac{\norm{\theta+d_w(t)-z}{}^2}{2(1.2+4Dt)}\), \label{e:contaminant_profile}
\end{eqnarray}
%% \todo{What does the $(4)$ in parentheses in the
%%   denominator mean? Is it really there?}
% XH: Yes, it is trying to view the term ``2 sqrt(0.3 + Dt)'' in the
% denominator outside the exponential as a standard deviation (sigma),
% so then the denominator inside the exponential would be 2 sigma^2 =
% 2 (4) (0.3+Dt). I'll absorb it inside the square-root to make it
% look nicer.
where $s$, $D$, and $d_w(t)$ are the \textit{known} source intensity,
diffusion coefficient, and cumulative net displacement due to wind up
to time $t$, respectively. The displacement $d_w(t)$ depends on the
time history of the wind velocity. (Values of these coefficients will
be specified later.)  A total of $N$ measurements are performed, at
uniformly spaced times given by $t=k+1$. (While $t$ is a continuous
variable, we assume it to be suitably scaled so that it corresponds to
the experiment index in this fashion; hence, observation $y_0$ is taken at $t=1$, $y_1$ at $t=2$,
etc.) The state $x_k$ is a combination of a belief state and a
physical state. Because an exact parametric representation of the
posterior is not available in this nonlinear problem, the belief state is
represented by an adaptive discretization of the posterior probability density
function, using 100 nodes.
The relevant physical state is the current location of the vehicle,
i.e., $x_{k,p} = z$. Inclusion of physical state is necessary since
the optimal design is expected to depend on the vehicle position as
well as the belief state. 
Here we will consider the source inversion problem in one spatial
dimension, where $\theta$, $d_k$, and $x_{k,p}$ are scalars (i.e., the
plume and vehicle are confined to movements in a line).
The design variables themselves correspond to the spatial displacement
of the vehicle from one measurement time to the next. To introduce
limits on the range of the vehicle, we use the box constraint
$d_k\in [-d_L, d_R]$, where $d_L$ and $d_R$ are bounds on the
leftwards and rightwards displacement. The physical state dynamics
then simply describe position and displacement:
$x_{k+1,p} = x_{k,p} + d_k$.

The concentration measurements are corrupted by additive Gaussian noise:
\begin{eqnarray}
  y_k=G(\theta,x_{k+1,p},k+1)+\epsilon_k(x_k,d_k),\label{e:contaminant_likelihood}
\end{eqnarray}
where the noise $\epsilon_k\sim\CN(0, \sigma_{\epsilon_k}^2(x_k,d_k))$
may depend on the state and the design. When simulating a trajectory,
the physical state must first be propagated before an observation
$y_k$ can be generated, since the latter requires the evaluation of
$G$ at $x_{k+1,p}$. Once $y_k$ is obtained, the belief state can then
be propagated forward via Bayesian inference. 

The reward functions used in this problem are
\begin{eqnarray*}
  g_k(x_k, y_k, d_k) &=& -c_{b} - c_{q}\vert {d_k} \vert^2, \ \text{and} \\ g_N(x_N) &=&
  \DKL( f_{\theta|I_N} || f_{\theta|I_0} ), \label{e:source_inversion_1D_terminal}
\end{eqnarray*}
for $k=0,\ldots,N-1$. The terminal reward is simply the KL
divergence, and the stage reward consists of a base cost of
operation plus a penalty that is quadratic in the vehicle displacement.

We study three different cases of the problem, as described at the start of
Section~\ref{ch:numerical_results}. Problem and algorithm settings
common to all cases can be found in
Tables~\ref{t:source_inversion_problem_settings}
and~\ref{t:source_inversion_algorithm_settings}, and additional
variations will be described separately.

\begin{table}[htb]
\begin{footnotesize}
\caption{Contaminant source inversion problem: problem settings.}
\label{t:source_inversion_problem_settings}
\begin{center}
\begin{tabular}{c|c}
  \hline
%  Number of experiments $N$ & 2  \\ 
  Prior on $\theta$ & $\CN(0, 2^2)$  \\
  Design constraints & $d_k \in \[-3, 3\]$ \\
  Initial physical state & $x_{0,p} = 5.5$ \\
  Concentration strength & $s=30$  \\
  Diffusion coefficient & $D = 0.1$  \\
  Base operation cost & $c_b = 0.1$  \\
  Quadratic movement cost coefficient & $c_q = 0.1$ \\ \hline
\end{tabular}
\end{center}
\end{footnotesize}
\end{table}

\begin{table}[htb]
\begin{footnotesize}
\caption{Contaminant source inversion problem: algorithm settings.}
\label{t:source_inversion_algorithm_settings}
\begin{center}
\begin{tabular}{c|c}
  \hline
  Number of grid points & 100 \\
  Design measure for exploration policy & $d_k \sim \CN(0, 2^2)$ \\
  Total number of regression points & 500  \\
  \% of regression points from exploration & 30\% \\
  Maximum number of optimization iterations & 50  \\
  Monte Carlo sample size in stochastic optimization & 100 \\
%  Robbins-Monro harmonic gain sequence multiplier & 5 \\
  \hline
\end{tabular}
\end{center}
\end{footnotesize}
\end{table}

\subsubsection{Case 1: comparison with greedy (myopic) design}

This case highlights the advantage of sOED over greedy design, which
is accentuated when future factors are important to the design of the
current experiments. sOED will allow for coordination between
subsequent experiments in a way that greedy design does not. We
illustrate this idea via the wind factor: the air is calm initially,
and then a constant wind of velocity 10 commences at $t=1$, leading to
the following net displacement due to the wind up to time $t$:
\begin{eqnarray}
  d_w(t) = \left\{ \begin{array}{cc} 0, & t < 1 \\ 10 (t-1), & t \geq
      1 \end{array} \right..\label{e:1D_wind}
\end{eqnarray}
Consider $N=2$ experiments. Greedy design, by construction, chooses
the first design to yield the single best experiment at $t=1$. This
experiment is performed before plume has been advected. sOED, along
with batch design, can take advantage of the fact that the plume will
have moved by the second experiment, and make use of this knowledge
even when designing the first experiment.

Details of the problem setup and numerical solution are as
follows. The observation noise variance is set to
$\sigma_{\epsilon_k}^2 = 4$ for all experiments.  Features for the representation of the
value function are analogous to those used in the previous example but
now include the physical state as well: that is, we use polynomials up to
total degree two in the posterior mean, posterior log-variance, and
physical state (including cross-terms). Posterior moments are
evaluated by trapezoidal rule integration on the adaptive grid. The
terminal KL divergence is approximated by first estimating the mean
and variance, and then applying the analytical formula for KL
divergence between the associated Gaussian approximations.
%% ; this is a reasonable approximation in
%% the unimodal case, and much easier to evaluate than the full
%% divergence. \todo{Is this justification useful, or should we skip it?
%%   Actually, see later comment on multi-modal dists.}
% XH: I was debating between either skipping it, or note something
% like ``much easier to evaluate than full KL, reasonable approx for
% unimodal, but becomes less accurate for
% multimodal/nonGausian''. Skip for now.
sOED uses $L=3$ policy updates,
with a design measure for exploration of $d_k\sim\CN(0, 2^2)$.
%% \todo{Does this
%%   mean we use regression points generated by the exploration policy?
%%   In any case, we should specify the choice of exploration design
%%   measure here.}
% XH: Done. Exploration policy is listed in Table 3 but we'll repeat
% it here. I did do the sOED with up to L=3, but didn't include them
% before because (1) they are similar, and
% even the L=1 (ie no update) is sufficient to illustrate the sOED
% advantage over batch and greedy so I just used that and didn't need
% to show results from other iterations, and (2) there's no iteration
% concept for batch/greedy, so *maybe* it would be more fair if we
% didn't do that for sOED (this is very arguable though).
%
% But excluding might raise questions too. So I'll just present the L=3
% results now, and verbally say the results are similar for different
% l's and we only present the l=3 plots.
%
Policies are compared by applying each to 1000 simulated trajectories,
as summarized in Algorithm~\ref{a:linear_gaussian_fair_evaluator}. The
common assessment framework for this problem is a high-resolution grid
discretization of the posterior probability density function with 1000
nodes.

% \begin{algorithm}[h]
%  \SetAlgoLined
%   \textbf{Select ``native'' belief state representation to generate
%     policy:} for example: grid or map\; 
%   \textbf{Construct policy:} use the native belief state
%   representation, and the numerical methods developed in this thesis
%   for solving the sOED problem\;
%   \For{$q=1,\ldots,n_{\textrm{trajectories}}$}
%   {
%     \textbf{Apply policy:} generate a trajectory using the native
%     belief state representation: sample $\theta$ from prior,
%     evaluate $d_k$ by applying the constructed policy, sample $y_k$
%     from the likelihood, for $k=0,\ldots,N-1$\;

%     \textbf{Evaluate rewards via a ``common'' evaluation framework:}
%     perform inference on the 
%     $d_k$ and $y_k$ values from this trajectory and evaluate all
%     rewards, using a high-resolution 1000-node grid state representation\;
%   }
%   \caption{Procedure for evaluating policies by simulating
%     trajectories.}
%   \label{a:linear_gaussian_fair_evaluator}
% \end{algorithm}

\begin{figure}[htb]
  \centering
  \mbox{\subfigure[Physical state and plume profiles]
    {\includegraphics[width=0.45\textwidth]{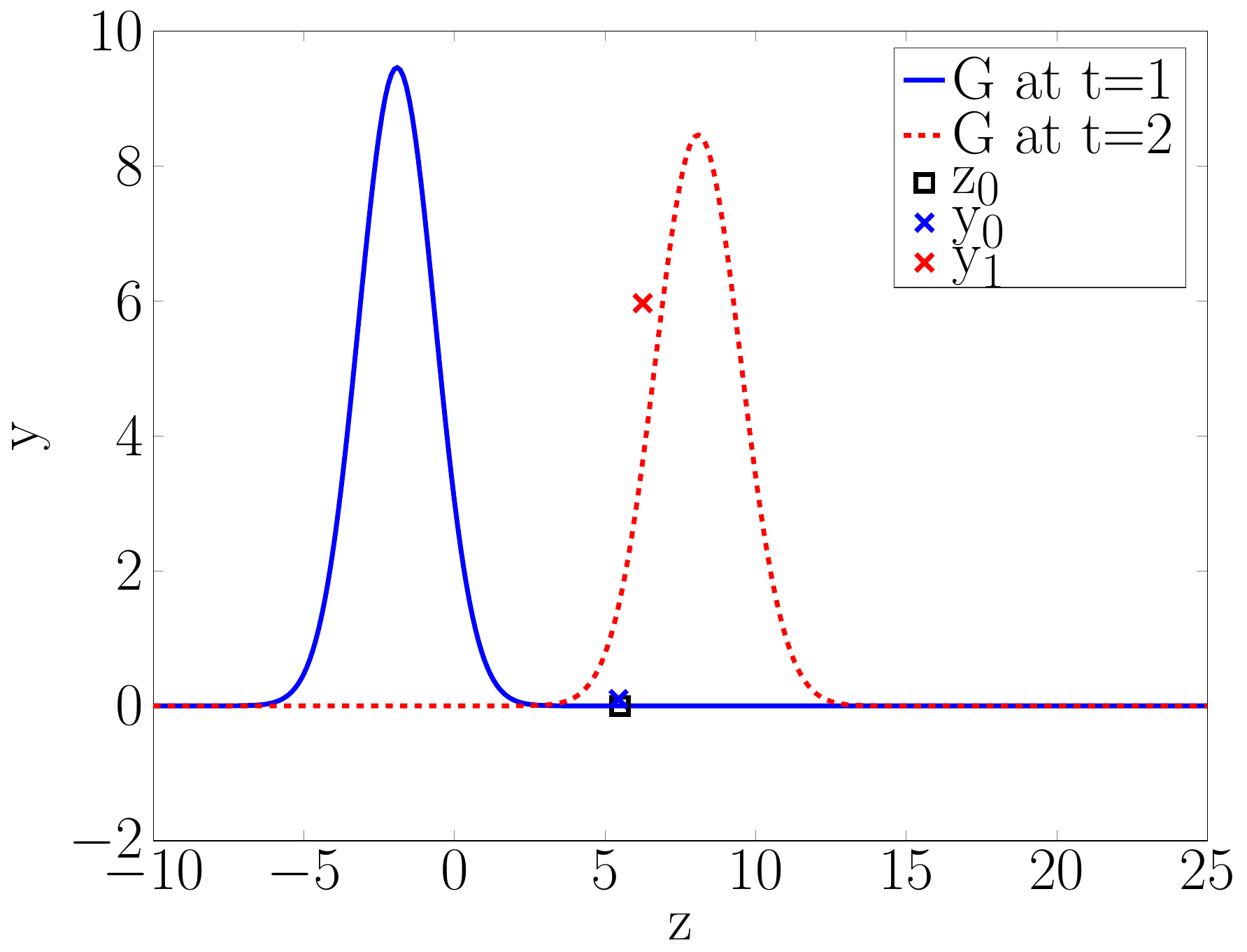}
    }
  }
  \mbox{\subfigure[Belief state, i.e., probability densities of $\theta$.]
    {\includegraphics[width=0.45\textwidth]{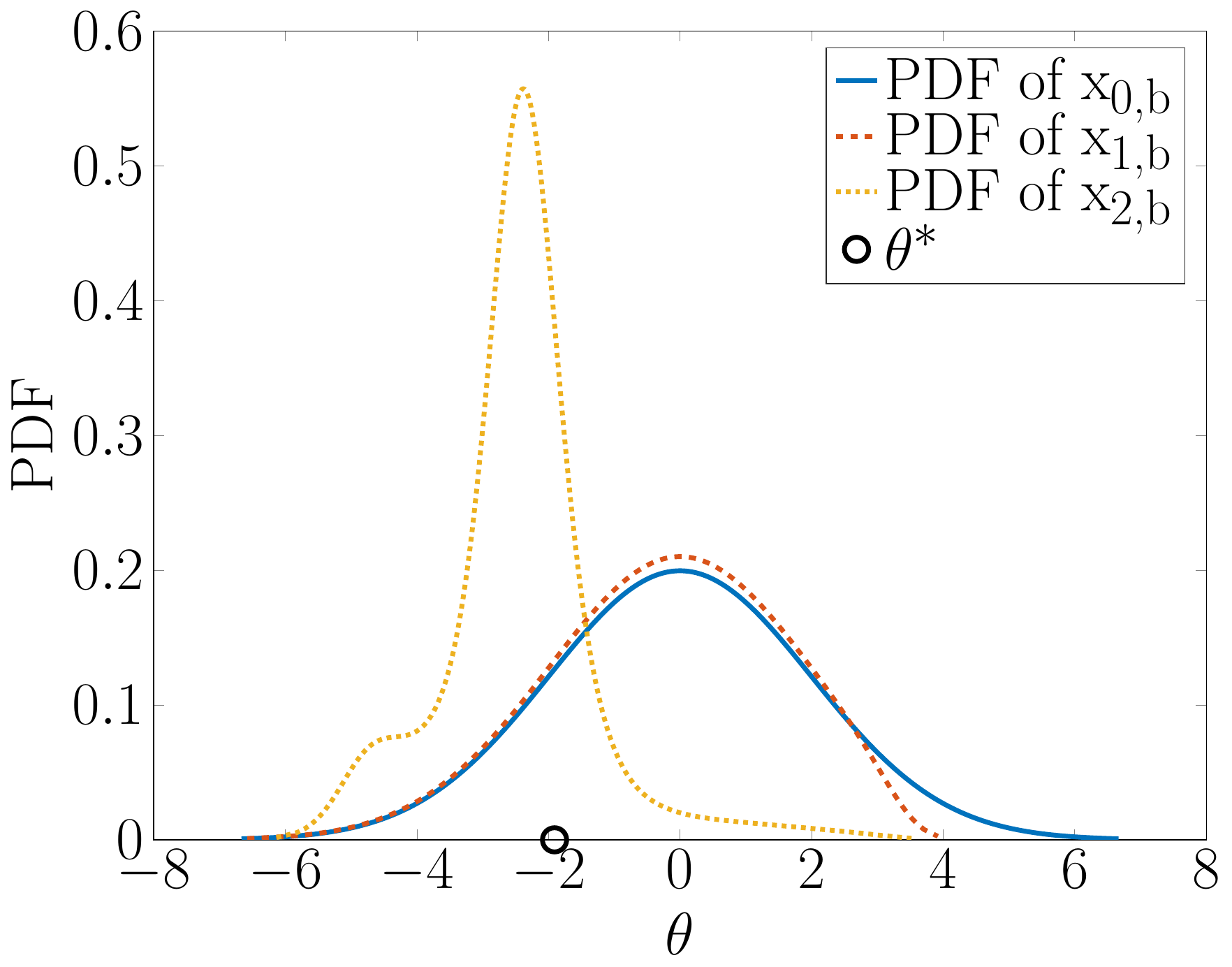}
    }
  }
  \caption{Contaminant source inversion problem, Case 1: progression
    of the physical state and belief state for a sample trajectory.}
  \label{f:source_inversion_1D_case3_sample_progression}
\end{figure}

Before presenting the results, we first provide some intuition for the
progression of a single example trajectory, as shown in
Figure~\ref{f:source_inversion_1D_case3_sample_progression}. Suppose
that a true value $\theta^\ast$ of the source location has been fixed
(or sampled from the prior). The horizontal axis of the left figure
corresponds to the physical space, with the vehicle starting at the
location marked by the black square. To perform the first experiment,
the vehicle moves to a new location and acquires the noisy observation
indicated by the blue cross,
%% \todo{Would it be ``safer'' if the blue
%%   cross were at a positive value of $G$?}
% XH: updated with a trajectory that gave positive y's, and is in tail
% for y0 and informative in y1.
with the solid blue curve
indicating the plume profile $G(\theta^\ast, z, t=1)$ at that time. For
the second experiment, the vehicle moves to another location and
acquires the noisy observation indicated by the red cross; the dotted
red curve shows the plume profile $G(\theta^\ast, z, t=2)$,
which has diffused slightly and been carried to the right by the
wind. The right figure shows the corresponding belief state (i.e.,
posterior density) at each stage. Starting from the solid blue prior
density, the posterior density after the first experiment (dashed red)
narrows only slightly, since the first observation is in a region
dominated by measurement noise. The posterior after both experiments
(dotted yellow), however, becomes much narrower, as the second
observation is in a high-gradient region of the concentration
profile and thus carries significant information for identifying
$\theta$. The posteriors in this problem can become quite non-Gaussian
and even multimodal.
%% \todo{True. So maybe my earlier addition on a
%%   good approximation of the KL divergence should be dropped!}
The
black circle in Figure
\ref{f:source_inversion_1D_case3_sample_progression}(b) indicates the
true value of $\theta$; the posterior mode after the second experiment is
close to this value but should not be expected to match it exactly,
due to noisy measurements and the finite number of observations.

\begin{figure}[htb]
  \centering
  \mbox{\subfigure[Greedy design]
    {\includegraphics[width=0.45\textwidth]{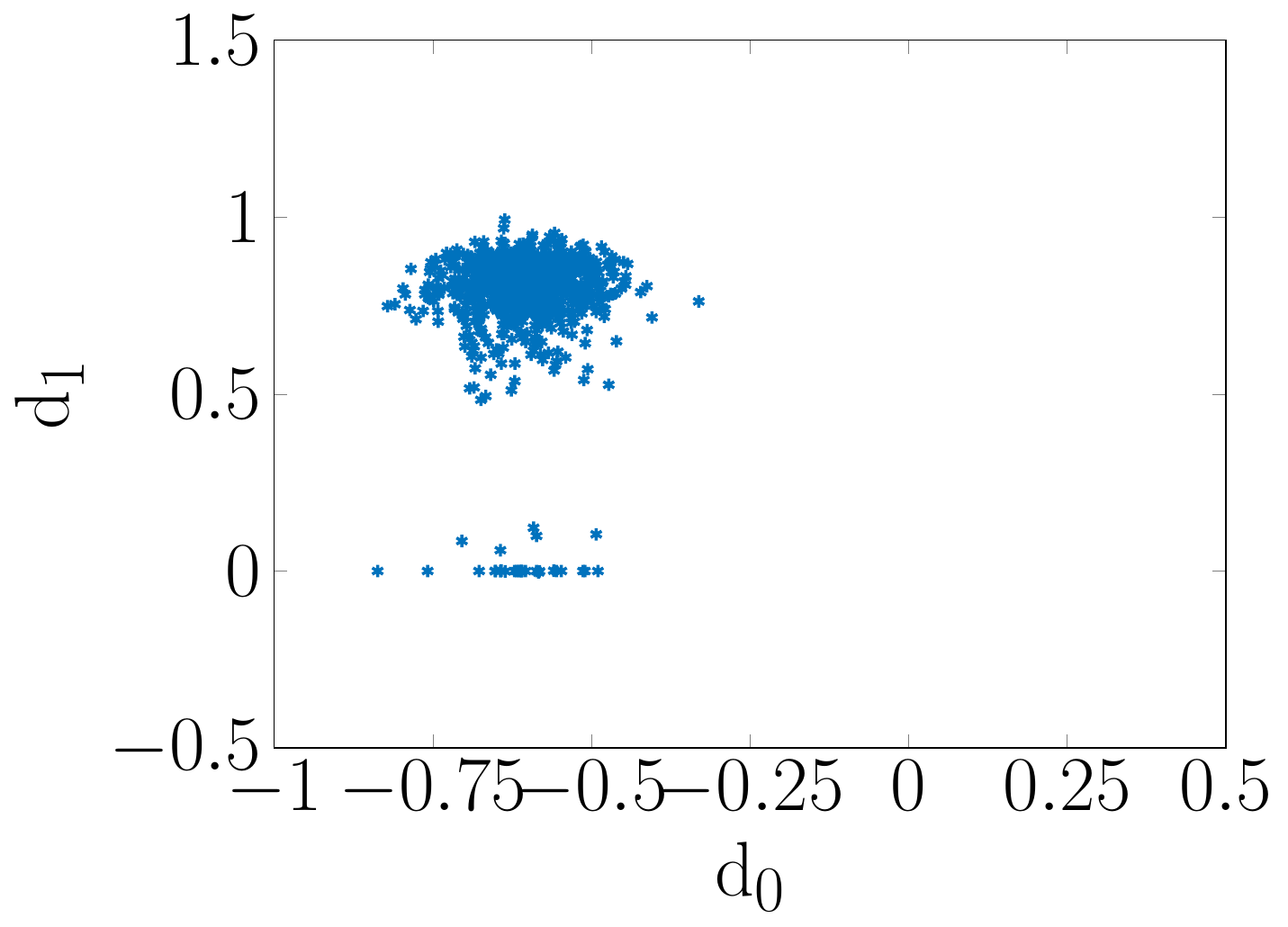}
    }
  }
  %% \mbox{\subfigure[sOED]
  %%   {\includegraphics[width=0.45\textwidth]{figures/source_inversion_1D/case1/d_pairs_VFA_iter0_source_inversion_1D_case1_grid_perturb.pdf}
  %%   }
  %% }
  %% \mbox{\subfigure[sOED]
  %%   {\includegraphics[width=0.45\textwidth]{figures/source_inversion_1D/case1/d_pairs_VFA_iter1_source_inversion_1D_case1_grid_perturb.pdf}
  %%   }
  %% }
  \mbox{\subfigure[sOED]
    {\includegraphics[width=0.45\textwidth]{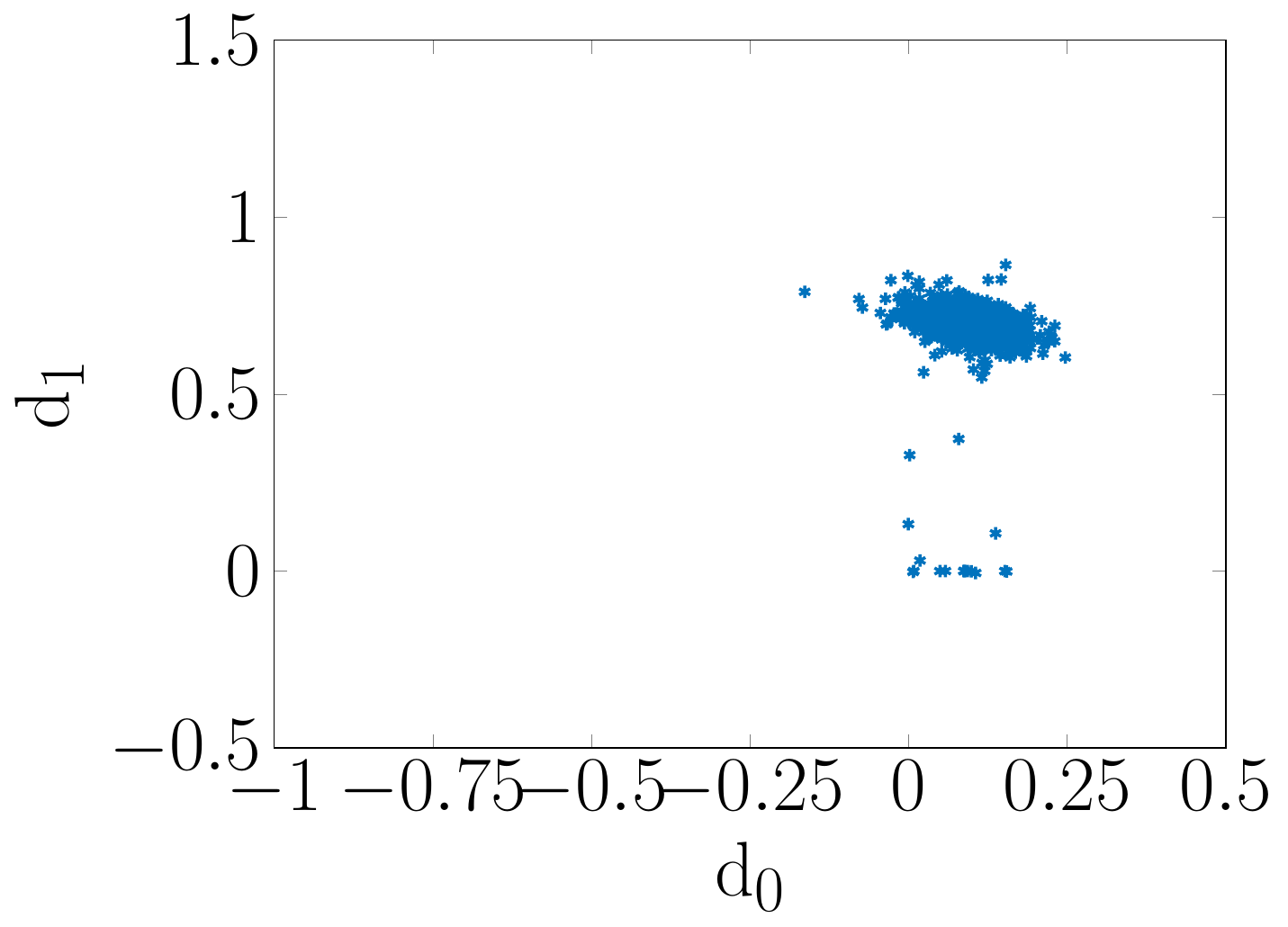}
    }
  }
  \caption{Contaminant source inversion problem, Case 1:
  scatter plots of $(d_0,d_1)$ from 1000 simulated trajectories
  for the greedy and sOED policies.}
  \label{f:source_inversion_1D_case1_d_pairs}
\end{figure}

Now we compare results of the greedy and sOED policies. Only data 
from $\ell=3$ are shown for sOED, as other iterations produced similar results.
Pairwise
scatter plots of the generated designs $(d_0,d_1)$ for 1000 simulated
trajectories are shown in
Figure~\ref{f:source_inversion_1D_case1_d_pairs}.  Greedy designs
generally move towards the left for the first experiment (negative
values of $d_0$) since for almost all prior realizations of $\theta$,
the bulk of the plume is to the left of the initial vehicle
location.
%% \todo{The first design in both greedy and sOED should
%%   actually be deterministic, right? Or am I missing something?}
% XH: Yes. Deterministic in theory, but due to algorithmic uncertainty
% there's some spread which is what is in the plot (imagine a marginal
% on the x-axis). The ``generally moves towards the left'' is directly
% in reference what we see in the plot, which is actually due
% to this algorithmic uncertainty, I don't mean to imply that there's
% adaptivity in the first experiment. The second part of the sentence
% is to explain why they would generally move towards the left: it is
% because the in most prior samples, the plume would be to the
% vehicle's left.
% Let me know if that is clear; we can discuss in person.
% 
%
When designing the first experiment, greedy design by construction
does not account for the fact that there will be a second experiment
and that the wind will eventually blow the plume to the right; thus it
moves the vehicle to the left in order to acquire information
\textit{immediately}. Similarly, when designing the second experiment,
the greedy policy chases after the plume, which is now to the right of
the vehicle; hence we see positive values of $d_1$.  sOED, however,
generally starts moving the vehicle to the right for the first
experiment, so that it can arrive in the regions of highest
information gain (which correspond to high expected gradient of the
concentration field) in time for the second experiment, after the
plume has been carried by the wind. Both policies produce a few cases
where $d_1$ is very close to zero, however. These cases correspond to
samples of $\theta$ drawn from the right tail of the prior, making the
plume much closer to the initial vehicle location. As a result, a
large amount of information is obtained from the first
observation. The plume is subsequently carried 10 units to the right
by the wind, and the vehicle cannot reach regions that yield
sufficiently high information gain from the second experiment to
justify the movement cost for $d_1$. The best action is then to simply
stay put, leading to these near-zero $d_1$ values. Note that these
small-$d_1$ cases are very much the result of \textit{feedback} from
the first experiment, which would not be available in a batch
experimental design setting.
%% \todo{I thought this last sentence could
%%   be an interesting addition.}
% XH: nice!

\begin{figure}[htb]
  \centering
  \mbox{\subfigure[Greedy design]
    {\includegraphics[width=0.45\textwidth]{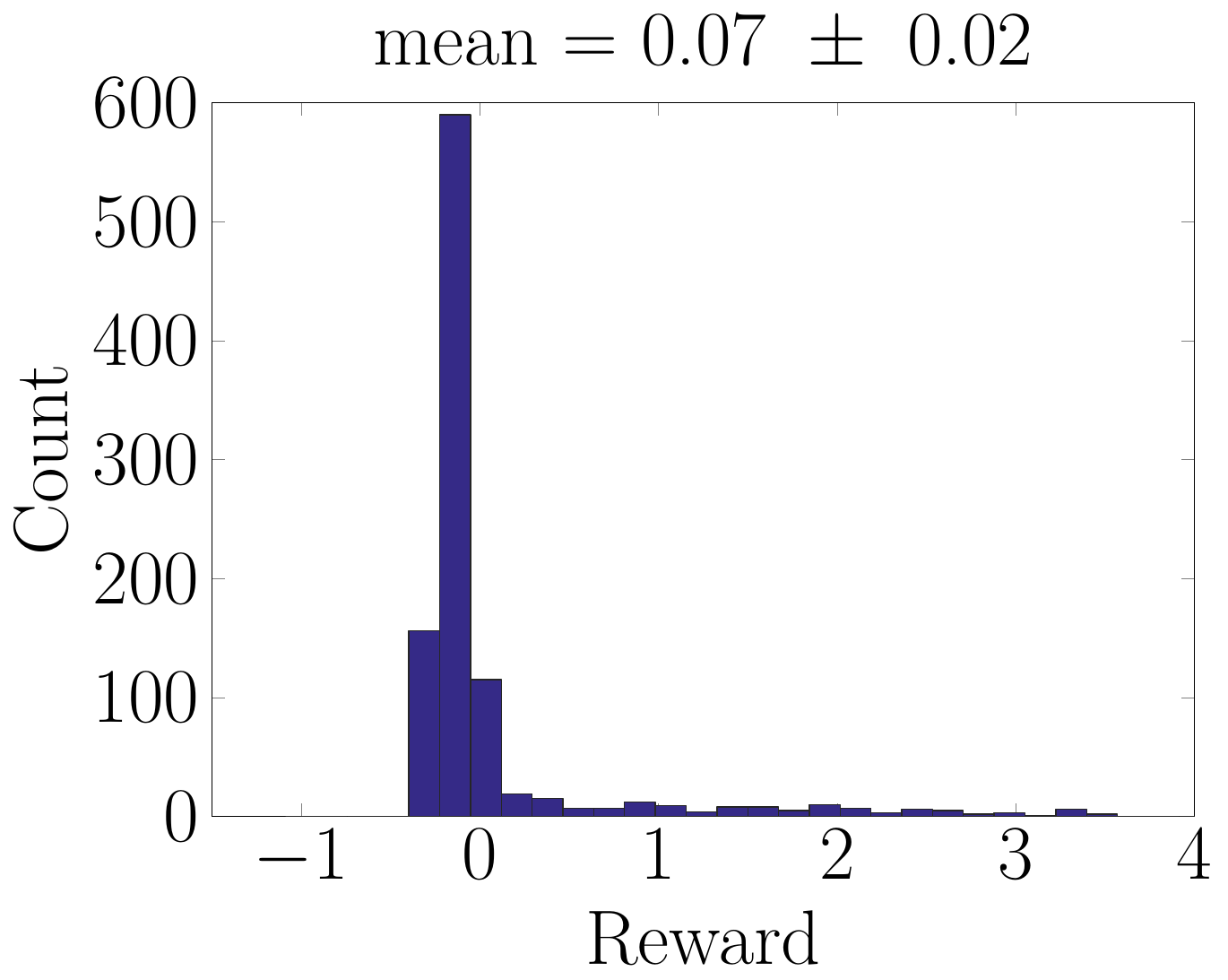}
    }
  }
  %% \mbox{\subfigure[sOED]
  %%   {\includegraphics[width=0.45\textwidth]{figures/source_inversion_1D/case1/rewards_VFA_iter0_source_inversion_1D_case1_grid_perturb.pdf}
  %%   }
  %% }
  %% \mbox{\subfigure[sOED]
  %%   {\includegraphics[width=0.45\textwidth]{figures/source_inversion_1D/case1/rewards_VFA_iter1_source_inversion_1D_case1_grid_perturb.pdf}
  %%   }
  %% }
  \mbox{\subfigure[sOED]
    {\includegraphics[width=0.45\textwidth]{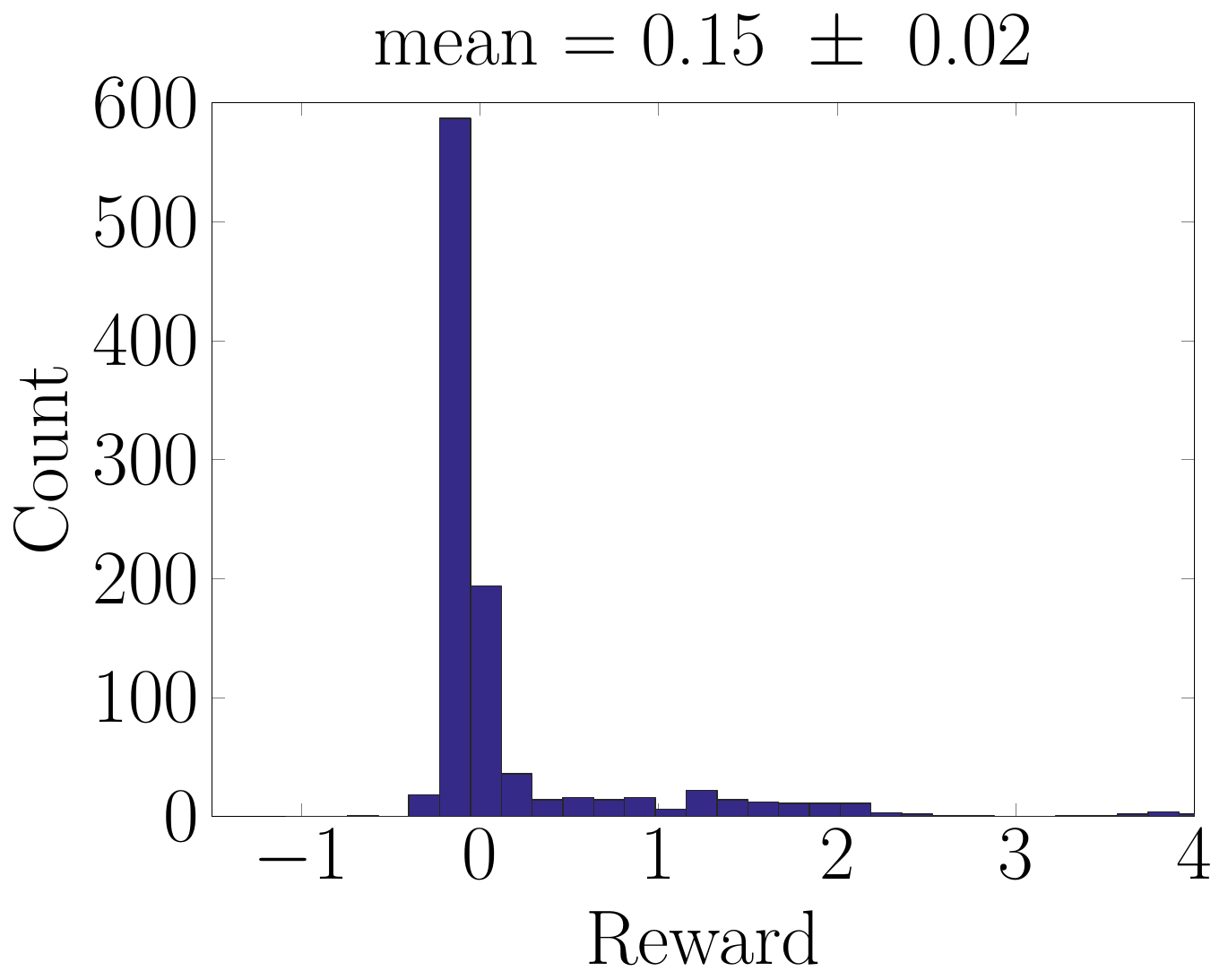}
    }
  }
  \caption{Contaminant source inversion problem, Case 1: histograms of
    total reward from 1000 simulated trajectories for greedy
    design and sOED.}
  \label{f:source_inversion_1D_case1_rewards}
\end{figure}

Overall, the tendency of the greedy design policy to ``chase''
high-information experimental configurations turns out to be costly
due to the quadratic movement penalty. This cost is reflected in
Figure~\ref{f:source_inversion_1D_case1_rewards}, which shows
histograms of total reward from the trajectories.  sOED yields an
expected reward of $U(\pi^L) = 0.15\pm0.02$, whereas greedy produces
% XH: updated results for l=3 iteration (before was l=1).
a lower value of $U(\pi^{\text{greedy}}) = 0.07\pm0.02$; the
plus-minus quantity is the Monte Carlo standard error.

\subsubsection{Case 2: comparison with batch (open-loop) design}
\label{ss:source_inversion_1D_case2}

This case highlights the advantage of sOED over batch design, which is
accentuated when useful information for designing experiments can be
obtained by performing some portion of the experiments first (i.e., by
allowing feedback). Again consider $N=2$ experiments. We illustrate the
impact of feedback via a scenario involving two different measurement
devices: suppose that the vehicle carries a ``coarse'' sensor with an
observation noise variance of $\sigma^2_{\epsilon_k}=4$ and a
``precise'' sensor with $\sigma^2_{\epsilon_k}=0.25$.
%
% The precise device is much more expensive to operate. Fortunately, the
% device cost is charged to the funding agency, and not reflected in our
% reward functions.
%
Suppose also that the precise sensor is a scarce resource, and can
only be utilized when there is a good chance of localizing the source;
in particular, rules require it to be used if and only if the current
posterior variance is below a threshold of 3.
%\footnote{Which sensor to use is not a design decision.}  
(Recall that the prior variance
is 4.) The observation noise standard deviation thus has the form
\begin{eqnarray}
  \sigma_{\epsilon_k}^2(x_{k,b}) = \left \{ \begin{array}{cc} 
    0.25, & \textrm{if \{variance corresponding to } x_{k,b} \} < 3 \\ 
         4, & \textrm{otherwise} \end{array} \right . .\label{e:1D_two_tier_device}
\end{eqnarray}
The same wind conditions as in \eqref{e:1D_wind} are applied.  Batch
design is expected to perform poorly for this case, since it cannot
include the effect of feedback in its assessment of different
experimental configurations, and thus cannot plan for the use of the
precise sensor. % \todo{Wording.}

The same numerical setup as Case 1 is used. Policies are assessed
using the same procedure but with one caveat. Batch design selected
$d_0$ and $d_1$ without accounting for feedback from the result of the
first experiment. But in assessing the performance of the optimal
batch design, we \textit{will} allow the belief state to be updated
between experiments; this choice maintains a common assessment
framework between the batch and sOED design strategies, and is in fact
more favorable to batch design than a feedback-free assessment. In
particular, updating the belief state permits the use of the precise
sensor. In other words, while batch design cannot strategically plan
for the use of the precise sensor, we make this sensor available in
the policy assessment stage if the condition in
\eqref{e:1D_two_tier_device} is satisfied. %\todo{Wording.}

%since the measurement noise depends on the belief state and hence the
%numerical belief state representation and associated inference
%methodology.
% YMM: removed this, since the same numerical belief state
% representation (grid) is used in all cases here. This comment may
% have been left over from a maps version?

%
% the noise standard deviation is also recorded as the observations are
% generated.  The correct corresponding standard deviation is used when
% inference is performed on the common assessment framework. In other
% words, we would know which device is used to obtain any particular
% observation.

\begin{figure}[htb]
  \centering
  \mbox{\subfigure[Batch design]
    {\includegraphics[width=0.45\textwidth]{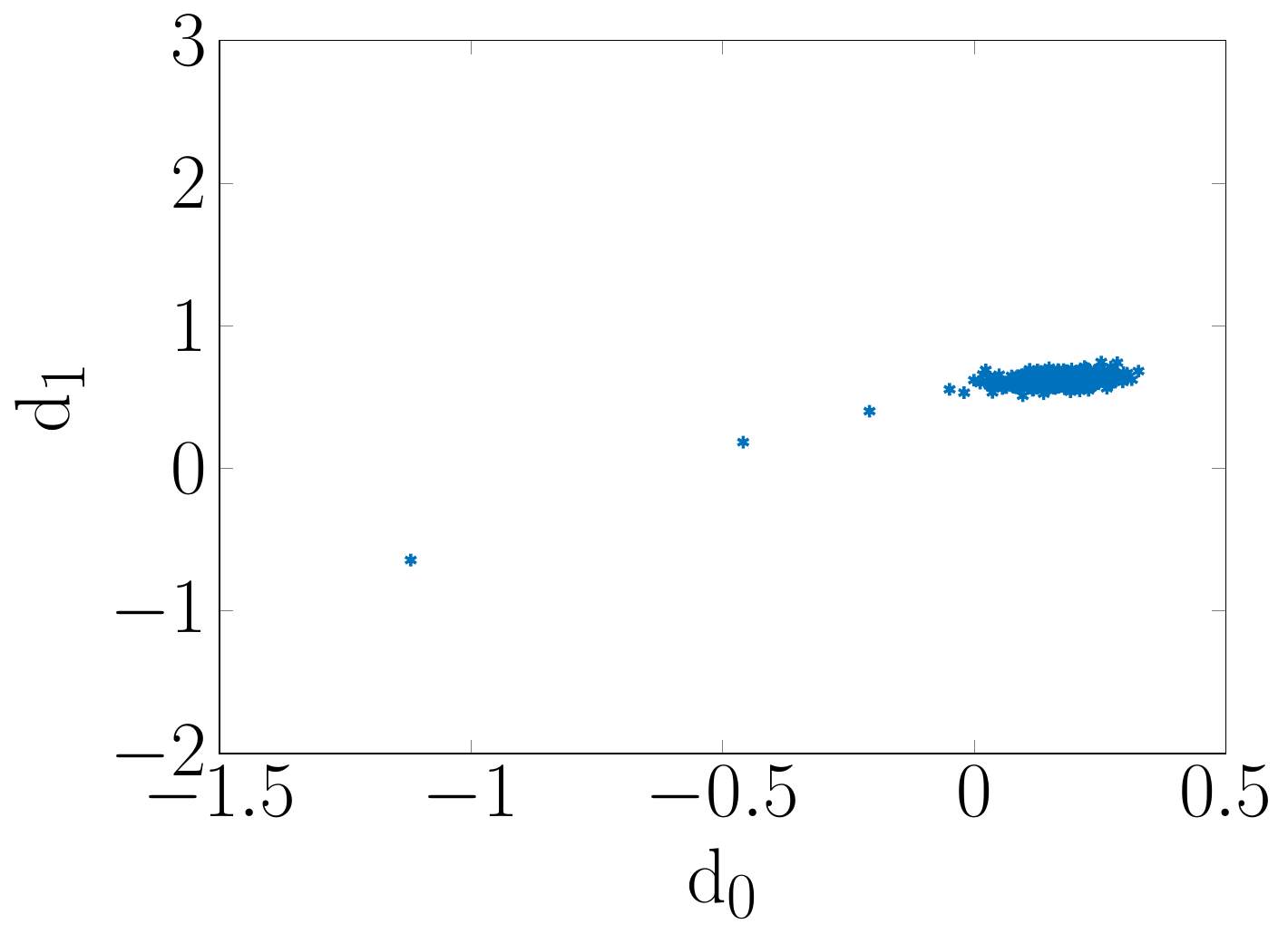}
    }
  }
  \mbox{\subfigure[sOED]
    {\includegraphics[width=0.45\textwidth]{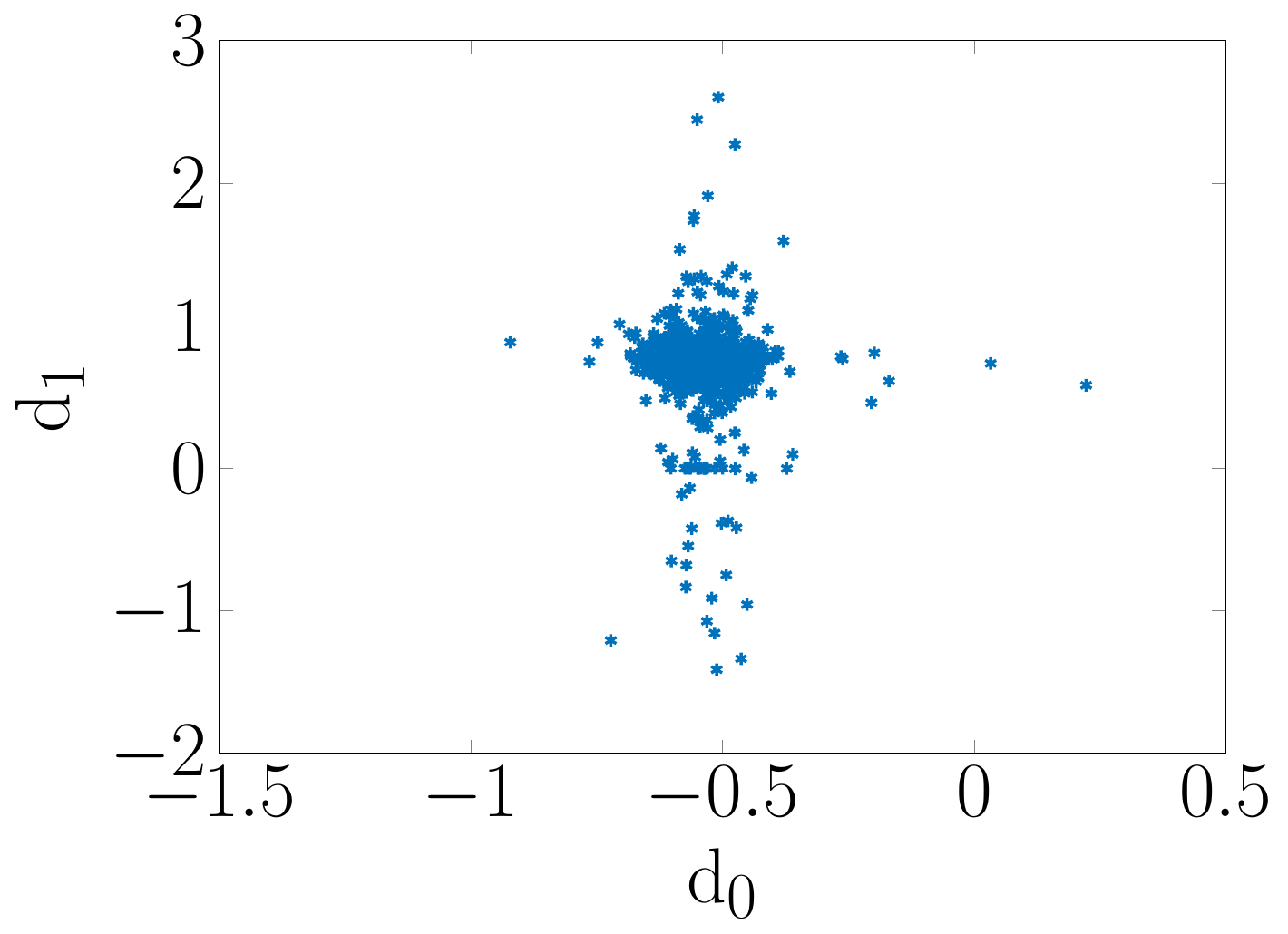}
    }
  }
  %% \mbox{\subfigure[sOED]
  %%   {\includegraphics[width=0.45\textwidth]{figures/source_inversion_1D/case2/d_pairs_VFA_iter0_source_inversion_1D_case2_grid_perturb.pdf}
  %%   }
  %% }
  %% \mbox{\subfigure[sOED]
  %%   {\includegraphics[width=0.45\textwidth]{figures/source_inversion_1D/case2/d_pairs_VFA_iter1_source_inversion_1D_case2_grid_perturb.pdf}
  %%   }
  %% }
  \caption{Contaminant source inversion problem, Case 2: scatter plots of $(d_0,d_1)$ from 1000 simulated trajectories
  of batch design and sOED.} 
% Roughly 55\% of the sOED trajectories qualify for the precise device
% in the second experiment. However, there is no particular pattern or
% clustering of these designs, thus we do not separately color-code them
% in the scatter plot. 
% YM: no reason to bring this up?
  \label{f:source_inversion_1D_case2_d_pairs}
\end{figure}

Again, only data from $\ell=3$ for sOED are shown, with other
iterations producing similar results.
Pairwise scatter plots of $(d_0,d_1)$ for 1000 simulated trajectories
are shown in Figure~\ref{f:source_inversion_1D_case2_d_pairs}. As
expected, batch design is able to account for change in wind velocity
at $t=1$ and immediately starts moving the vehicle to the right for the first
experiment so that it can reach locations of higher information gain
for the second
experiment, after the plume is advected to the right. sOED, however,
realizes that there is the possibility of using the precise device in
the second experiment if it can reduce the posterior variance below
the threshold using the first observation. Thus it moves to the left
in the first experiment (towards the more likely initial plume
locations) to obtain an observation that is expected to be more
informative, even though the movement cost is higher. This behavior is
in stark contrast with that of the sOED policy for Case 1. 
Roughly 55\% of the resulting sOED trajectories achieve the threshold
for using the precise device in the second experiment compared to only
8\% from batch design. This subset of
the sOED trajectories has an expected reward of $U=0.51$, in contrast to
$U= -0.01$ for the subset of trajectories that fail to qualify.
Effectively, the sOED policy creates an opportunity for a much larger
reward with a slightly higher cost of initial investment.
%% takes a risk
%% \todo{Should we say `gamble' instead of `risk?'}
% XH: hmm, 'gamble' and 'risk' both sound somewhat negative. How about
% this?
%% in order to achieve an overall
%% higher expected reward.
%
% The notion of risk is currently not in the problem formulation, but
% it certainly should be considered in practice, especially for crucial
% missions where there is perhaps only one chance to ensure public
% safety.
% YMM: too digressive!
%
Histograms of the total rewards from all 1000 trajectories are shown in
Figure~\ref{f:source_inversion_1D_case1_rewards}. Indeed, the reward
distribution for the sOED policy sees more mass in higher values.
%\todo{Is there really more mass at lower values?}
% XH: if you zoom in on the figures, the first 2 bars just left of 0:
% the left most bar for sOED is higher than batch, though the
% second-most-left bar is higher for batch than sOED. They should be
% plotted with the same bins. The difference is small and probably
% affected by histogram binning ad plotting etc so I took that part out.
Overall the risk taken by sOED pays off as it produces an expected
reward of $U(\pi^L) = 0.26 \pm 0.02$, while batch design produces a much
lower value of $U(\pi^{\text{batch}}) = 0.15 \pm 0.02$.

\begin{figure}[htb]
  \centering
  %% \mbox{\subfigure[Batch design]
  %%   {\includegraphics[width=0.45\textwidth]{figures/source_inversion_1D/case2/rewards_VFA_iter0_source_inversion_1D_case2_open_loop_grid.pdf}
  %%   }
  %% }
  \mbox{\subfigure[Batch design]
    {\includegraphics[width=0.45\textwidth]{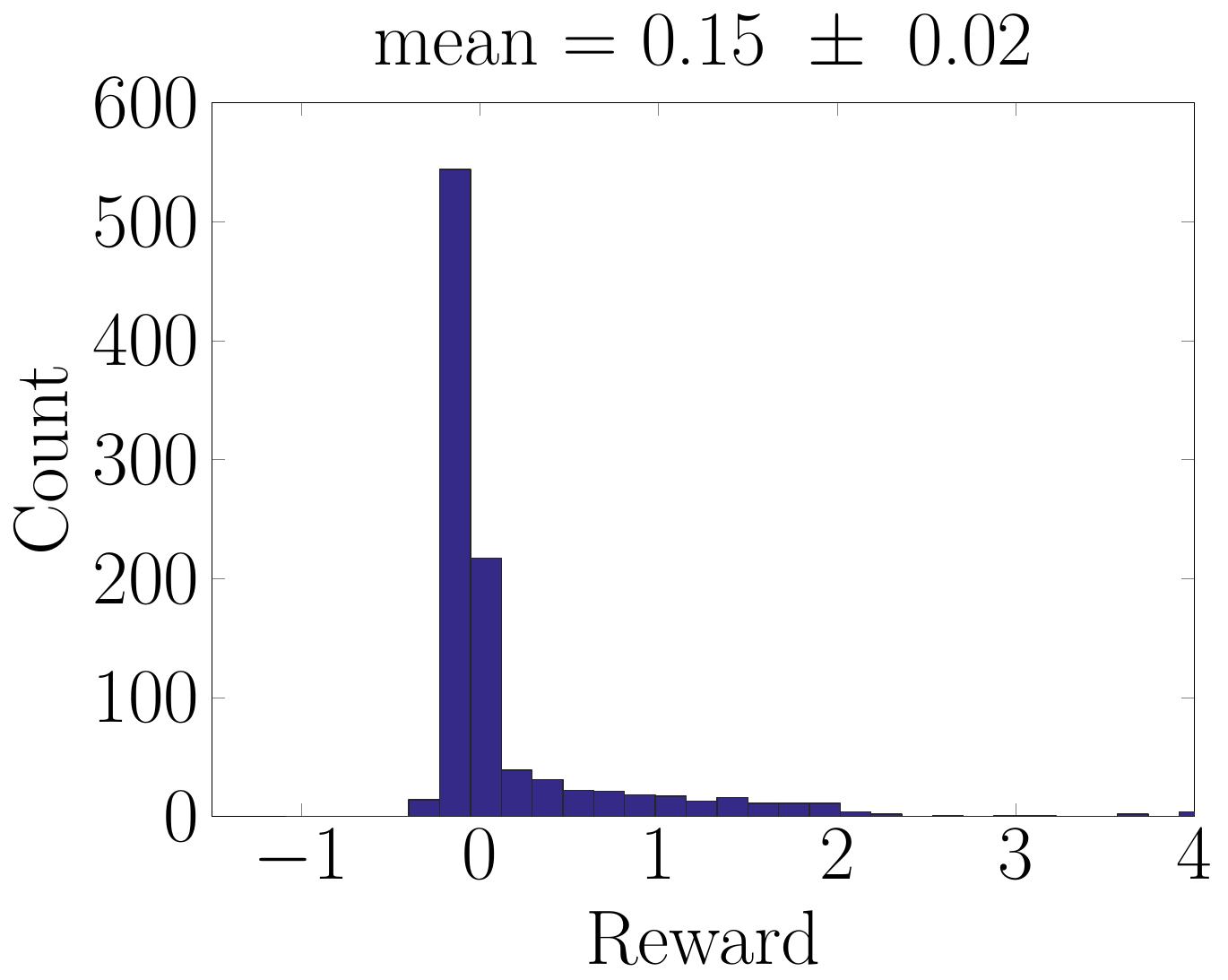}
    }
  }
  \mbox{\subfigure[sOED]
    {\includegraphics[width=0.45\textwidth]{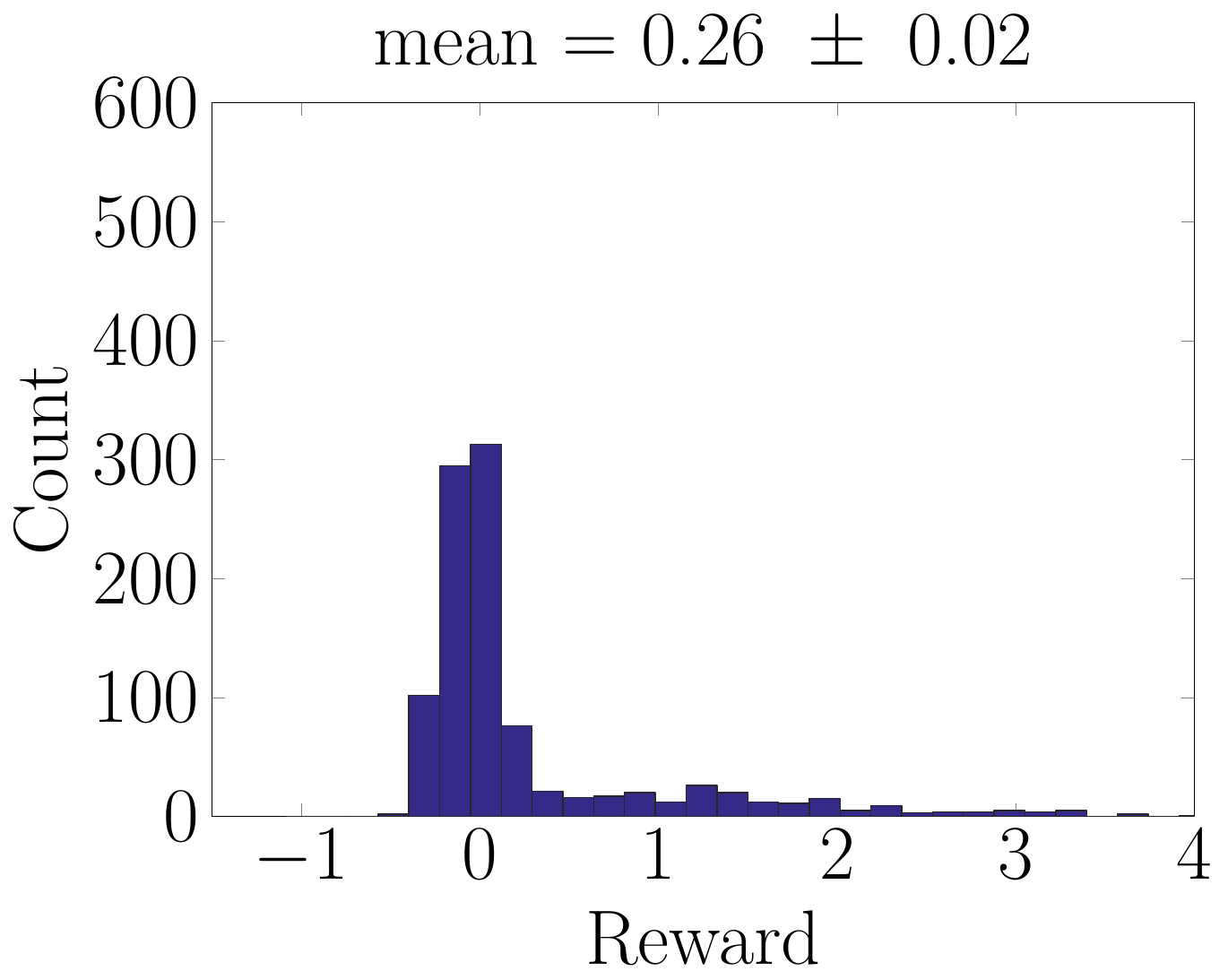}
    }
  }
  %% \mbox{\subfigure[sOED]
  %%   {\includegraphics[width=0.45\textwidth]{figures/source_inversion_1D/case2/rewards_VFA_iter0_source_inversion_1D_case2_grid_perturb.pdf}
  %%   }
  %% }
  %% \mbox{\subfigure[sOED]
  %%   {\includegraphics[width=0.45\textwidth]{figures/source_inversion_1D/case2/rewards_VFA_iter1_source_inversion_1D_case2_grid_perturb.pdf}
  %%   }
  %% }
  \caption{Contaminant source inversion problem, Case 2: histograms of
    total reward from 1000 simulated trajectories of batch design
%    with and without allowing belief state update,
    and sOED. }
  \label{f:source_inversion_1D_case2_rewards}
\end{figure}

\subsubsection{Case 3: additional experiments and policy updates}
This case demonstrates sOED with a larger number of experiments, and
also explores the ability of our policy update mechanism 
to improve the policy resulting from a poor initial choice
of design measure for exploration. We consider $N=4$ experiments, with
the wind condition
\begin{eqnarray}
  d_w(t) = \left\{ \begin{array}{cc} 0, & t < 1 \\ 5 (t-1), & t \geq
      1 \end{array} \right.,\label{e:1D_wind_case4}
\end{eqnarray}
and a two-tiered measurement system similar to that of Case 2:
\begin{eqnarray}
  \sigma_{\epsilon_k}^2(x_{k,b})= \left\{\begin{array}{cc} 0.25, & \textrm{if
    \{variance corresponding to } x_{k,b} \} < 2.5 \\ 4, &
  \textrm{otherwise} \end{array} \right..
\end{eqnarray}
A total of $L=10$ policy updates are conducted,
starting with a design measure for exploration of $\CN(-2.5,0.1)$, which
targets a particular point in the left part of design space and
ignores the remaining regions. Subsequent iterations use a mix of 5\%
exploration and 95\% exploitation trajectories as regression
points. 100-node grids are used both to construct the policy and to
drive the common assessment framework. All other settings remain the
same as in Cases 1 and 2. For this case, we intuitively expect the
most informative data (and the most design ``activity,'' i.e.,
variation in policy outputs) to occur at the second and third
experiments, as the plume is carried by the wind through the vehicle
starting location ($x_{0,p} = 5.5$) at those times. Additionally, we
anticipate that the advantage of the precise instrument might be less
prominent compared to Case 2, since more experiments are performed overall.

\begin{figure}[htb]
  \centering
  \includegraphics[width=0.45\textwidth]{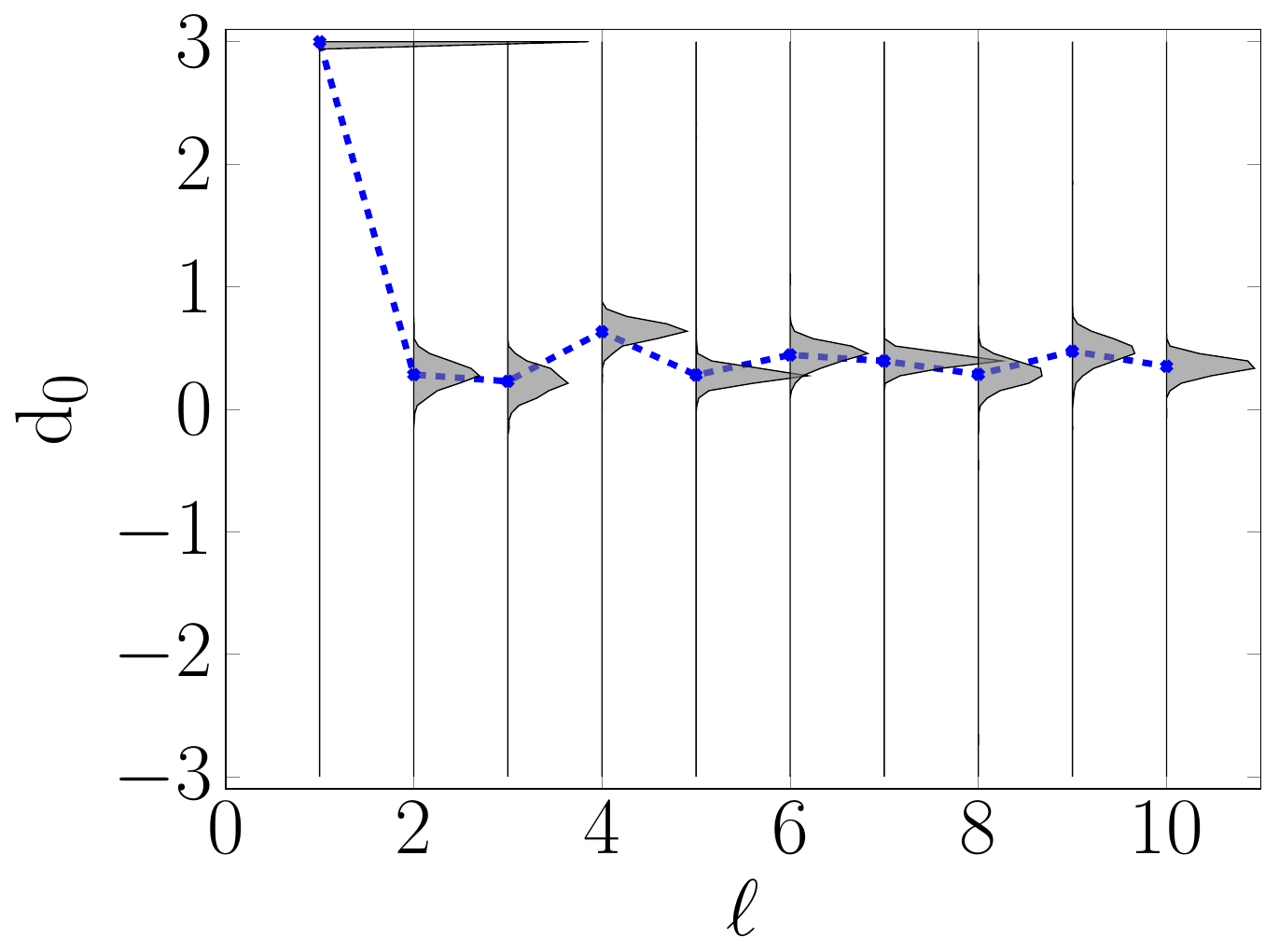}
  \includegraphics[width=0.45\textwidth]{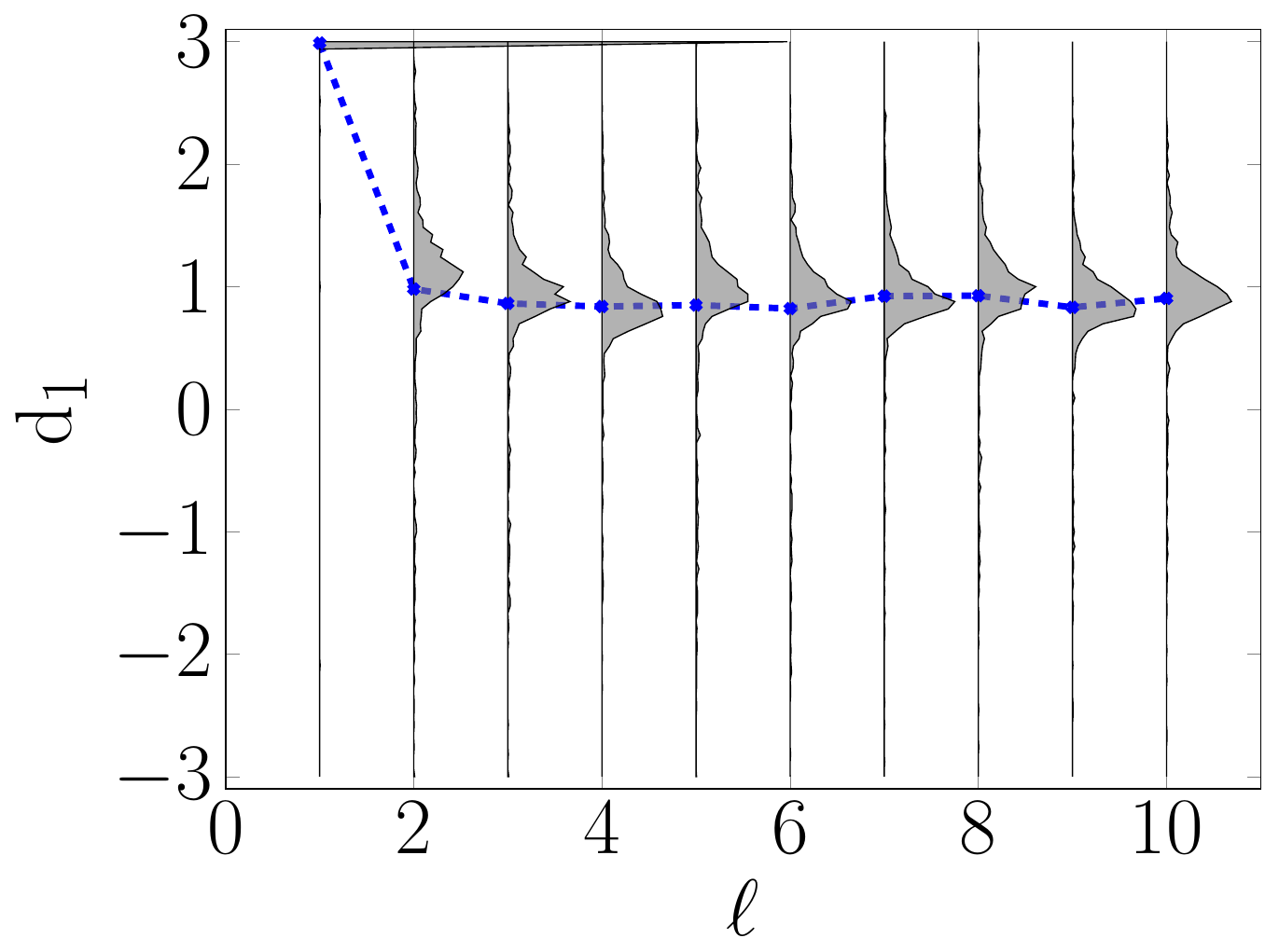}
  \includegraphics[width=0.45\textwidth]{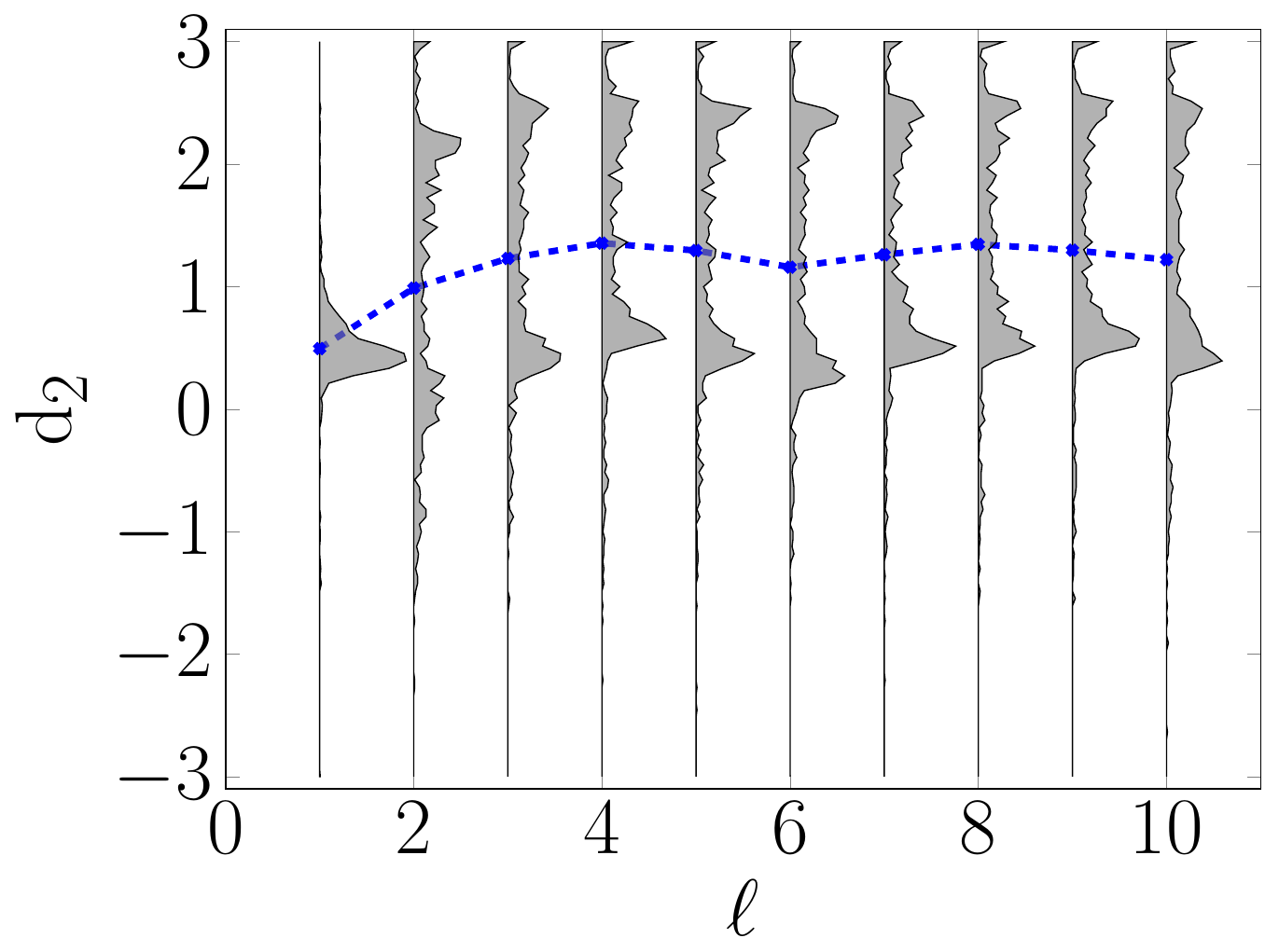}
  \includegraphics[width=0.45\textwidth]{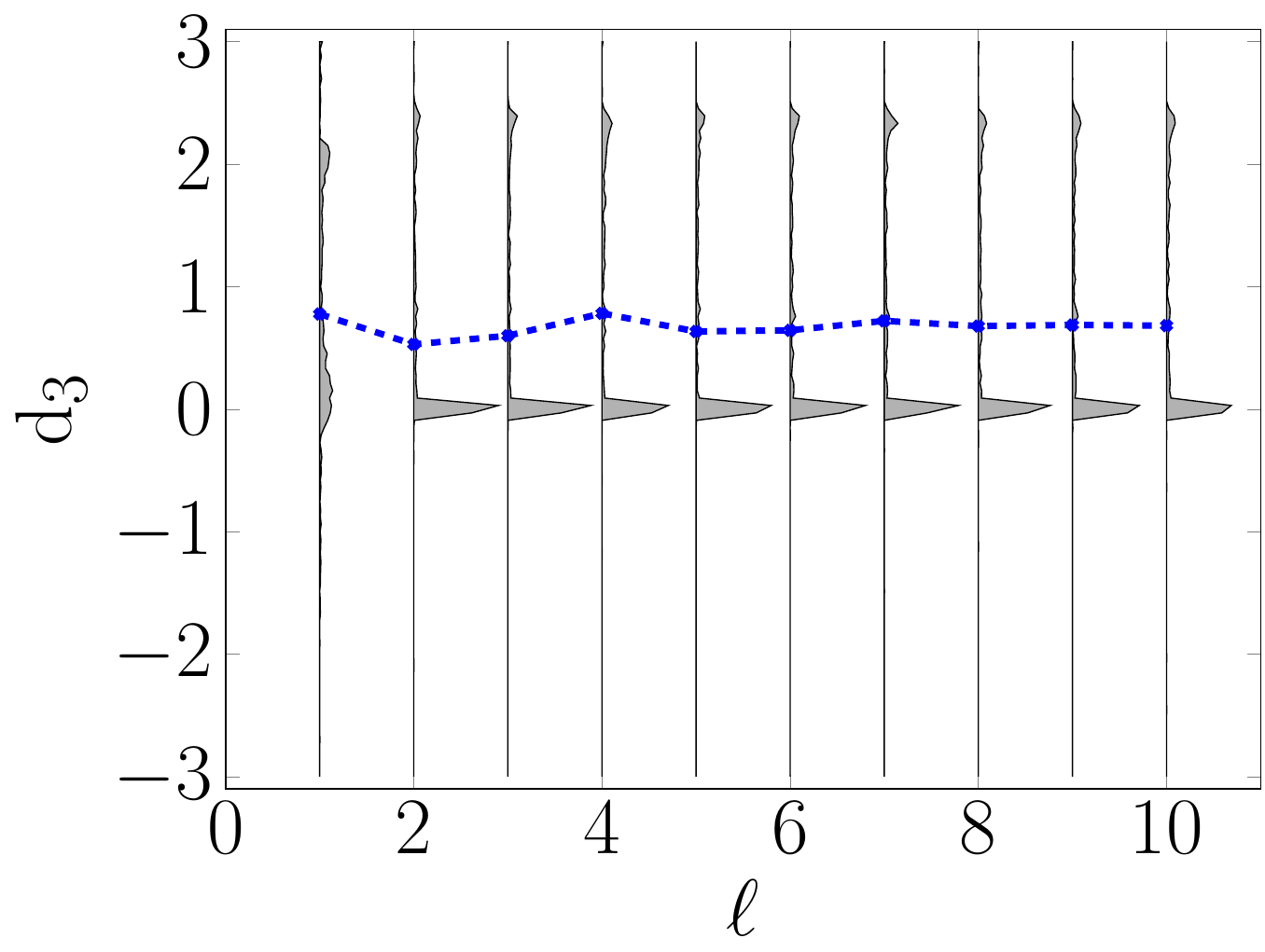}
  \caption{Contaminant source inversion problem, Case 3: design
    distributions for $d_0$, $d_1$, $d_2$, and $d_3$, from 1000
    simulated trajectories over successive steps $\ell$ of policy
    update. The blue dashed line connects
    the means of the distributions.}
  \label{f:source_inversion_1D_case4_dk}
\end{figure}

\begin{figure}[htb]
  \centering
  \includegraphics[width=0.6\textwidth]{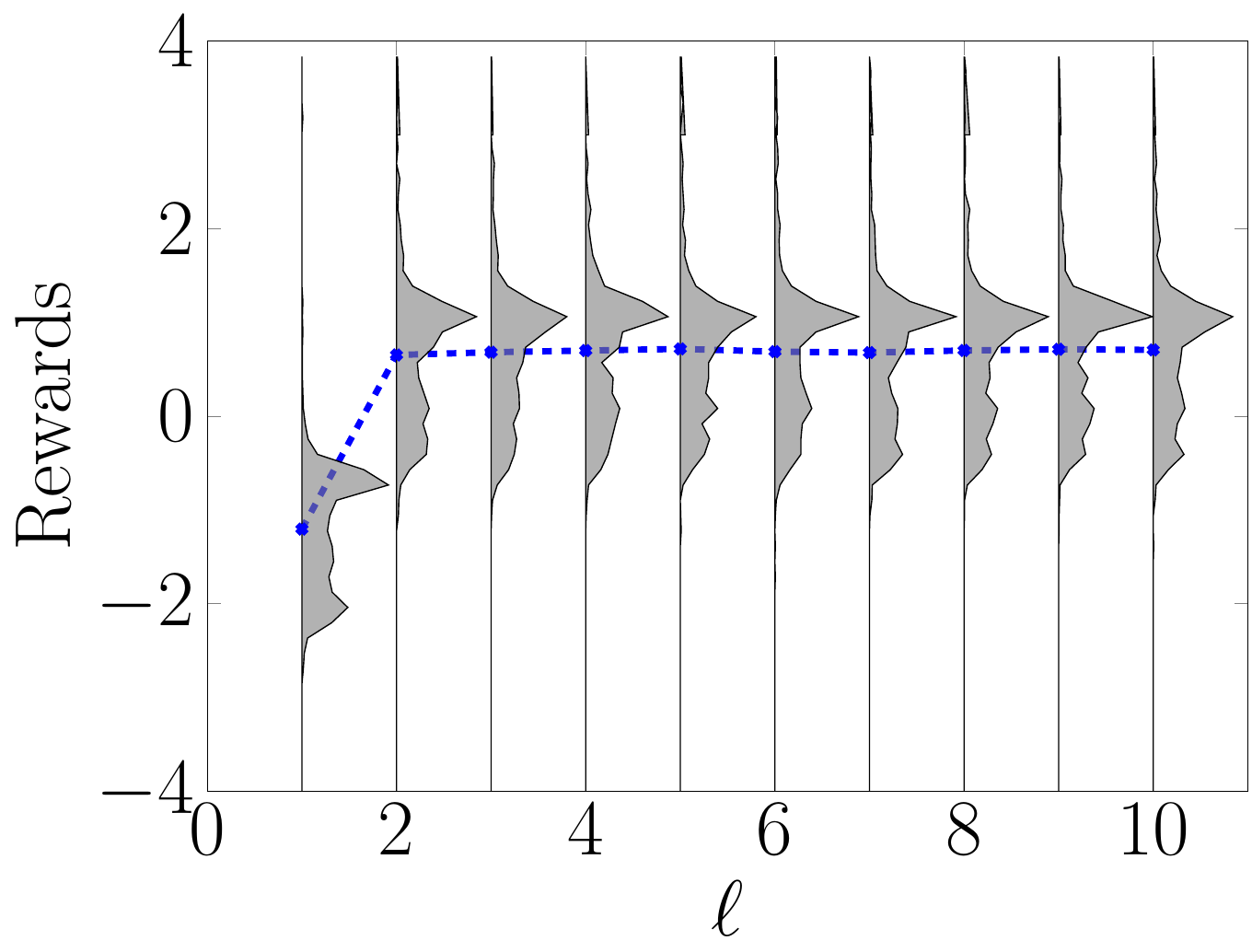}
  \caption{Contaminant source inversion problem, Case 3: distributions
    of total reward following $N=4$ experiments, from 1000 simulated
    trajectories over successive steps $\ell$ of policy update. The
    blue dashed line connects the means of
    distributions. Mean values are also given in
    Table~\ref{t:source_inversion_1D_case4_rewards_values}. }
  \label{f:source_inversion_1D_case4_rewards}
\end{figure}

Figure~\ref{f:source_inversion_1D_case4_dk} presents distributions of
each design $d_k$ (drawn sideways) from 1000 simulated trajectories
versus the policy update index, $\ell$. The blue dashed line
connects the mean values of successive distributions.
Figure~\ref{f:source_inversion_1D_case4_rewards} shows a similar
plot for the total reward, with expected rewards also reported in
Table~\ref{t:source_inversion_1D_case4_rewards_values}. We can
immediately make two observations.
First, the policy at $\ell=1$ appears to be quite poor, producing an
expected reward that is significantly lower than that of subsequent
iterations. This low value is to be expected, given the poor initial
choice of design measure for exploration.  Starting at $\ell=2$, exploitation
samples become available for regression, and they contribute to a
dramatic improvement in the expected reward. (Under the hood, this
jump corresponds to a significant improvement in the value function
approximations $\tJ_k$.)
%\todo{Check wording.}
% XH: looks good.
Large differences between the $d_k$ distributions of these first two iterations can also be observed.  In
comparison, subsequent iterations show much smaller changes; this is
consistent with the intuition described in Section~\ref{ss:ell}.

Second, a small oscillatory behavior can be seen between iterations,
most visibly in $d_0$.  This is likely due to some interaction between
the limitations of the selected features in capturing the landscape of
the true value function, and the ensuing locations of the regression points
generated by
% peculiar 
%% \todo{check wording; I removed the word `peculiar' as it seemed kinda
%%   negative (and I wasn't sure what was peculiar about our exploration
%%   measure)}
% XH: looks good.
exploration. In other words, the regression points might oscillate
slightly around the region of the state space visited by the optimal
policy.
%% In other words, the regression points might
%% not fully align with [[the optimal region of the value function
%% approximation.]] \todo{Bracketed part: do you mean the region of the
%%   state space visited by the optimal policy? Overall, is the wording
%%   of the previous sentence on the right track? Feel free to make it
%%   better! ALT VERSION: ``In other words, the regression points might oscillate
%% slightly around the region of the state space visited by the optimal policy.''}
%
% XH: sounds good, ALT adopted.
Nonetheless, the design oscillations are quite small compared to the
first change following $\ell=1$. The oscillations are also small
relative to the range of the overall design space and the width of the
design distributions. More importantly, an oscillation is \textit{not}
visible in the reward distributions of
Figure~\ref{f:source_inversion_1D_case4_rewards} and in the values of
the expected reward, implying a sense of ``flatness'' around the
optimal policy.

We also performed numerical tests on this problem using a more
reasonable design measure for exploration of $\CN(0,4)$, with 30\% exploration
samples; results are reported in
Table~\ref{t:source_inversion_1D_case4_better_exploration_rewards_values}. The
expected reward values from $\ell=1$ and from the pure exploration
policy are much better compared to their counterparts in
Table~\ref{t:source_inversion_1D_case4_rewards_values}.

Overall, these results indicate that the first iteration of policy
update (from $\ell =1$ to $\ell = 2$) can be very helpful,
especially when it is unclear what a suitable design measure for exploration
should be. This first iteration can help recover from extremely poor
initial choices of exploration policy. Subsequently, it seems that
only a small number of policy updates (e.g., $L=3$) are
needed in practice.

\begin{table}[htb]
  \caption{Contaminant source inversion problem, Case 3: expected
    rewards (mean values of histograms in
    Figure~\ref{f:source_inversion_1D_case4_rewards}) estimated from 1000 
    simulated trajectories, using a ``poor'' design measure for exploration of
    $\CN(-2.5, 0.1)$.}
  \label{t:source_inversion_1D_case4_rewards_values}
  \centering
  \begin{tabular}{c|r||c|r}
    \hline
    $\ell$ & Expected reward & $\ell$ & Expected reward \\ \hline
    1 & $-1.20 \pm 0.02$ & 6 & $0.69 \pm 0.03$ \\
    2 & $0.65 \pm 0.03$ & 7 & $0.68 \pm 0.03$ \\
    3 & $0.68 \pm 0.02$ & 8 & $0.70 \pm 0.03$ \\
    4 & $0.70 \pm 0.02$ & 9 & $0.71 \pm 0.02$ \\
    5 & $0.72 \pm 0.03$ & 10 & $0.71 \pm 0.03$ \\
    \hline \hline
    Exploration & $-2.00 \pm 0.03$\\ \hline
  \end{tabular}
\end{table}

\begin{table}[htb]
  \caption{Contaminant source inversion problem, Case 3: expected
    rewards, estimated from 1000 
    simulated trajectories, using a ``better'' design measure for
    exploration of
    $\CN(0, 4)$.}
  \label{t:source_inversion_1D_case4_better_exploration_rewards_values}
  \centering
  \begin{tabular}{c|r}
    \hline
    $\ell$ & Expected reward \\ \hline
    1 & $0.67 \pm 0.02$ \\
    2 & $0.65 \pm 0.03$ \\
    3 & $0.68 \pm 0.03$ \\
    \hline \hline
    Exploration & $-0.70 \pm 0.03$\\ \hline
  \end{tabular}
\end{table}

% Through the three contaminant source inversion cases in
% Section~\ref{s:source_inversion_1D}, we have demonstrated the
% advantages of sOED over batch and greedy designs in realistic
% situations involving nonlinear models and non-Gaussian posteriors, and
% the ability of our numerical methods to handle multiple experiments.
%
% YM: Seems unnecessary in the paper, since the conclusions section
% will follow immediately...

\section{Conclusions}
\label{ch:conclusions}

We have developed a rigorous formulation of the sequential optimal
experimental design (sOED) problem, in a fully Bayesian and
decision-theoretic setting. 
The solution of the sOED problem is not a set of designs, but rather a
feedback control \textit{policy}---i.e., a decision rule that
specifies which experiment to perform as a function of the current
system state. The latter incorporates the current posterior
distribution (i.e., belief state) along with any other design-relevant
variables. More commonly-used batch and greedy experimental design
approaches are shown to result from simplifications of this sOED
formulation; these approaches are in general suboptimal. Unlike batch
or greedy design alone, sOED combines \textit{coordination} among
multiple experiments, which includes the ability to account for future
effects, with \textit{feedback}, where information derived from each
experiment influences subsequent designs.

% framework and a set of numerical tools for performing sOED in a
% computationally feasible manner. This framework is formulated from a
% decision-theoretic perspective, with a Bayesian treatment of
% uncertainty and an information measure objective. It is capable of
% accommodating the sequential design of a finite number of experiments,
% with nonlinear models and non-Gaussian distributions, and under
% continuous parameter, design, and observation spaces.  Popular design
% approaches such as batch OED and greedy design are shown
% mathematically to be special cases of sOED, and thus suboptimal. sOED
% seeks the optimal \textit{policy}, a decision rule that makes use of
% newly acquired information and takes into account of future effects.

Directly solving for the optimal policy is a challenging task,
particularly in the setting considered here: a finite horizon of
experiments, described by continuous parameter, design, and
observation spaces, with nonlinear models and non-Gaussian posterior
distributions. Instead we cast the sOED problem as a dynamic program
and employ various approximate dynamic programming (ADP) techniques to
approximate the optimal policy. Specifically, we use a one-step
lookahead policy representation, combined with approximate value
iteration (backward induction with regression). Value functions are
approximated using a linear architecture and estimated via a series of
regression problems obtained from the backward induction procedure, with regression
points generated via both exploration and exploitation. In obtaining
good regression samples, we emphasize the notion of the \textit{state
  measure} induced by the current policy; our algorithm
incorporate an update to adapt and refine this measure as better
policy approximations are constructed.

We apply our numerical ADP approach to several examples of
increasing complexity. The sOED policy is first verified on a
linear-Gaussian problem, where an exact solution is available
analytically. The methods are then tested on a nonlinear and
time-dependent contaminant source inversion problem. %, with a forward
                                %model that includes the effects of
                                %both advection and diffusion. 
Different test cases demonstrate the advantages of sOED
over batch and greedy design, and the ability of policy update
to improve the initial policy resulting from a poor choice
of design measure for exploration. 

There remain many important avenues for future work. One challenge
involves the representation of the belief state $x_{k,b}$, and the
associated inference methodologies. The ADP approaches developed here
are largely \textit{agnostic} to the representation of the belief
state: as long as one can evaluate the system dynamics
$x_{k+1} = \CF_k(x_k,y_k,d_k)$ and the features $\phi_{k,i}(x_k)$,
then approximate value iteration and all the other elements of our ADP
algorithm can be directly applied. In the case of parametric
posteriors (e.g., Bayesian inference with conjugate priors and
likelihoods) the representation and updating of $x_{k,b}$ is
trivial. But in the general case of continuous and non-Gaussian
posteriors, finding an efficient representation of $x_{k,b}$---one
that allows repeated inference under different realizations of the
data---can be difficult. In this paper, we have employed an adaptive
discretization of the posterior probability density function, but this
choice is impractical for higher parameter dimensions. A companion
paper will explore a more flexible and scalable alternative, based on
the construction of transport maps over the joint distribution of
parameters and data, from which conditionals can efficiently be
extracted \cite{Parno2015a,Marzouk2016}.
Alternative ADP approaches are also of great interest, including
model-free methods such as $Q$-learning.

% Another major challenge involves the expression of the belief state,
% which are non-Gaussian continuous posterior random variables.  The
% current implementation involves using a grid to capture the
% probability density function, and the grid nodes is redistributed when
% inference is performed in order to better capture the resulting
% posterior. A grid approach is suitable for one dimensional parameter
% space, but becomes impractical for higher dimensions. A more flexible
% alternative, based on transport maps, will be presented in a future
% paper.

% Interesting and promising avenues of future work include developing
% efficient belief state representation and Bayesian inference and
% alternative approximate dynamic programming techniques.

\section*{Acknowledgments}

The authors would like to acknowledge support from the Computational
Mathematics Program of the Air Force Office of Scientific Research.
% and by the National Science Foundation under award number ECCS-1128147.

\bibliography{library}
\bibliographystyle{siamplain_modified}

\end{document}